\documentclass[prd,floatfix,preprint,nofootinbib,eqsecnum,12pt,superscriptaddress,tightenlines]{revtex4}
\usepackage{amsmath}
\usepackage{graphics, graphicx}
\usepackage{ulem}

\renewcommand{\Im}{{\rm Im}}
\renewcommand{\Re}{{\rm Re}}
\def\bh{\hat b}
\def\ch{\hat\chi}
\def\mb{\bar\mu}
\def\ah{\hat a}
\def\Nh{\hat N}
\def\phih{\hat\phi}
\def\cS{{\cal  S}}
\def\cV{{\cal V}}

\def\qf{}
\def\rf{}
\def\sf{}
\def\uf{}
\def\vf{}

\def\be{\begin{equation}}
\def\ee{\end{equation}}

\begin{document}

\title{The Classical Universes of  the \\ No-Boundary Quantum State}

\vspace{.5cm}

\author{James B. Hartle}
\email{hartle@physics.ucsb.edu}

\affiliation{Department of Physics,
 University of California,
 Santa Barbara, CA 93106-9530}
 
\author{S.W. Hawking}
\email{S.W.Hawking@damtp.ac.uk}

\affiliation{DAMTP, CMS, Wilberforce Road, CB3 0WA Cambridge, UK}

\author{Thomas Hertog}
\email{thomas.hertog@apc.univ-paris7.fr}
\affiliation{Laboratoire APC, 10 rue A.Domon et L.Duquet, 75205 Paris, France, and\\
International Solvay Institutes, Boulevard du Triomphe, ULB -- C.P. 231, 1050 Brussels, Belgium}

\date{\today}

\begin{abstract}

We analyze the origin of the quasiclassical realm from the no-boundary proposal for the universe's quantum state in a class of minisuperspace models. The models assume homogeneous, isotropic, closed spacetime geometries, a single scalar field moving in a quadratic potential, and a fundamental cosmological constant. The allowed classical histories and their probabilities are calculated to leading semiclassical order. We find that for the most realistic range of parameters analyzed a minimum amount of scalar field is required, if there is any at all, in order for the universe to behave classically at late times. If the classical late time histories are extended back, they may be singular or bounce at a finite radius. The ensemble of classical histories is time symmetric although individual histories are generally not. The no-boundary proposal selects inflationary histories, but the measure on the classical solutions it provides is heavily biased towards small amounts of inflation. However, the probability for a large number of efoldings is enhanced by the volume factor needed to obtain the probability for what we observe in our past light cone, {\qf given our present age}. Our results emphasize that it is the quantum state of the universe that determines whether or not it exhibits a quasiclassical realm and what histories are possible or probable within that realm.


\end{abstract}


\maketitle

\tableofcontents

\eject
\section{Introduction} 

The inference is inescapable from the physics of the last eighty years that we live in a quantum  mechanical universe. If so, the universe has a quantum state. A theory  of that state is as important a challenge for fundamental physics as a theory of the dynamics. Providing that theory, and testing its observational predictions, are the goals of quantum cosmology. 

A central prediction of the universe's quantum state is the classical spacetime that is a manifest fact of the present universe. {\qf Predicting classical spacetime}  is a  constraint on the theory of the state  because we can no more expect to find classical predictions following from a general state in quantum gravity than we can in the non-relativistic quantum mechanics of a particle. Histories exhibit classical correlations in time only when they are suitably {\it coarse-grained} and then only for {\it particular} kinds of states (e.g \cite{Har94b}). 

{\uf The probabilities for the alternative classical histories of a  quantum universe answer questions such as the following:   Is  approximate homogeneity and isotropy likely?}
Is the probability high for sufficient inflation to explain the present spatial flatness? Is a homogeneous thermodynamic arrow of time likely?  What is the probability that the universe bounced at a minimum radius above the Planck scale in the past? 

This paper is concerned with the classical histories predicted by  the no-boundary wave function of the universe (NBWF) \cite{Haw84a} in  homogeneous, isotropic minisuperspace models with a fundamental cosmological constant and a single scalar field moving in a quadratic potential. Many of our results and conclusions together with speculations concerning {\uf their} extensions  to other models have been summarized in 
\cite{HHH07}. This paper presents the detailed derivations of these and deals exclusively  with quadratic potentials for the scalar field. 

By way of introduction we now briefly sketch the standard procedure for classical prediction in quantum cosmology.  More details will be found in Section II, and derivations in the context of generalized quantum theory in \cite{HHer08}.  

States in quantum cosmology are represented by wave functions on the superspace of three-geometries and spatial matter field configurations. For the homogeneous, isotropic, spatially closed, minisuperspace models with one scalar field that are the subject of this paper, wave functions depend on the scale factor $b$ determining the size of the spatial geometry and the value $\chi$ of the homogeneous scalar field. Thus, $\Psi = \Psi(b,\chi)$. 

A class of states of particular interest are those whose wave functions can be approximated {\uf to leading order in $\hbar$ in some region of superspace}  by the semiclassical form (or superpositions of such forms)
\begin{equation}
\Psi(b,\chi) \approx  \exp\{[-I_R(b,\chi) +i S(b,\chi)]/\hbar\} 
\label{semiclass}
\end{equation}
 with both $I_R$ and $S$ real. {\rf When $S$ varies rapidly and $I_R$  varies slowly}  such wave functions  predict an ensemble of suitably coarse-grained Lorentzian histories  with high probabilities for correlations in time governed by classical deterministic laws for spacetime geometry and {\uf the matter field}. {\uf This requirement on the gradients of $I_R$ and $S$ is called the {\it classicality condition}. When it is satisfied}  the action $S$ determines the ensemble as in familiar Hamilton-Jacobi theory. Classical histories not contained in the ensemble have zero probability in this approximation. The classical histories that are members of the ensemble have probabilities proportional to $\exp[-2 I_R(b,\chi)]/\hbar]$. {\uf In this way particular states can predict classical spacetime. } 
 
The no-boundary wave function {\qf for these models}  is defined by the sum-over-histories
\begin{equation}  
\Psi(b,\chi) =  \int_{\cal C} \delta a \delta \phi \exp(-I[a(\tau),\phi(\tau)]/\hbar) .
\label{nbwf}
\end{equation}
Here, $a(\tau)$ and $\phi(\tau)$ are the histories of the scale factor and matter field and $I[a(\tau),\phi(\tau)]$ is their Euclidean action. The sum is over cosmological geometries that  are regular  on a disk with only one boundary at which $a(\tau)$ and $\phi(\tau)$ take the values $b$ and $\chi$. The integration is carried out along a suitable complex contour ${\cal C}$ which ensures the convergence of \eqref{nbwf} and the reality  of the result \cite{HH90}. 

For some ranges of $b$ and $\chi$ it may happen that the integral in \eqref{nbwf} can be approximated by the method of steepest descents. Then the wave function will be well approximated by a sum of terms of the form \eqref{semiclass} --- one for each extremizing history $(a(\tau),\phi(\tau))$ matching $(b,\chi)$ on the boundary of the manifold and regular elsewhere. 
In general these solutions  will be complex --- ``fuzzy instantons''. 
For each contribution $I_R(b,\chi)$ is the real part of the action $I[a(\tau),\phi(\tau)]$ evaluated at the extremizing history and $-S(b,\chi)$ is the imaginary part. When the classicality condition is satisfied the no-boundary wave function predicts an ensemble of classical histories as described above. 

Two key points should be noted about this prescription for classical prediction:
(1) The NBWF provides probabilities for entire classical {\it histories}. It therefore supplies a classical {\qf history} measure, i.e. {\qf a measure on classical phase space  that is conserved along the classical trajectories}. (2) The histories in the classical ensemble are not the same as the extremizing histories that provide the steepest descents approximation to the integral \eqref{nbwf} defining the NBWF.  

We apply this prescription for classical prediction to the NBWF in minisuperspace models with a cosmological constant $\Lambda$  and a quadratic potential of the form $V(\Phi) = (1/2)m^2 \Phi^2$. Our main aim is determining the ensemble of classical cosmologies predicted by  the no-boundary proposal and the probabilities of its members. From these {\uf we  calculate the probabilities for whether universe  bounces or is singular, for whether it expands forever or recollapses,} for the magnitude of any time asymmetries, for different amounts of matter content, and for different amounts of inflation.

 Our detailed conclusions are given in Section \ref{conc} , but were it necessary  to single out  just two they would be the following:   (1)  Not all classical behaviors of the universe are allowed by  the no-boundary proposal and for some ranges of {\qf model} parameters no classical behavior is predicted at all. The manifest existence of the quasiclassical realm in this universe is therefore an important, non-trivial, constraint on theories of its initial quantum state. (2) The probability for significant inflation depends sensitively on the {\uf limitations of the classical ensemble arising from the}  classicality condition and also on the limited scale of our observations in a large universe and the values of cosmological parameters such as the present age.

The predictions of the no-boundary quantum state were extensively analyzed in minisuperspace models in the '80s and '90s (see e.g. \cite{Halbib}).  We are unable to give anything like a complete survey of this work, but relevant papers include the following: Classicality conditions were studied in similar models in \cite{Haw84a,GR90,UJ98} especially in connection with the amount of inflation predicted by the NWBF.  For related work on this question see \cite{HP88,GT06} and \cite{BK90}  for the influence of higher order quantum corrections which are neglected here. Complex solutions were  studied in detail by  \cite{Lyo92,HLL93,LS87}; this paper relies heavily on these works. 

 What is new  is the following: 
(1)  A better understanding of the prescription for classical prediction attained by providing a firmer foundation through the probabilities for histories provided by generalized quantum theory \cite{Har95c,HHer08}.
(2)  A more complete analysis of the complex solutions that provide the semiclassical approximation to the no-boundary proposal and their connection to the probabilities for classical cosmologies. 
(3) The inclusion of a fundamental cosmological constant which both generalizes the discussion and simplifies it. 
(4) A derivation of the probabilities for what we observe within our past light cone from the probabilities for the ensemble of entire classical histories predicted by the NBWF.

This paper is organized as follows:  In Section \ref{sec2} we review the prescription for extracting the predictions for classical cosmologies from a wave function of the universe.  Section \ref{sec3} describes the no-boundary wave function and its semiclassical approximation. The homogenenous, isotropic minisuperspace models that are the focus of this paper are laid out in detail in Section \ref{sec4}.
Section \ref{sec5}  analyzes the complex fuzzy instantons that provide the semiclassical approximation to the NBWF. The {\uf predicted ensemble of } classical ensemble and the probabilities of its members are discussed in Section \ref{sec6}.  Section \ref{sec8} discusses conditional (top down) probabilities relevant for the amount of inflation of the universe given our present observations. Section \ref{sec7} discusses the arrow of time in those  histories that bounce at a minimum radius in our past. Section \ref{sec9} contains our conclusions. 

\section{Classical Prediction in Quantum Cosmology}
\label{classicalpred}\label{sec2}

A quantum system behaves classically when the probability is high for histories of its motion that exhibit patterns of classical correlation in time governed by deterministic dynamical laws. The Moon can be said to move on a classical orbit when the quantum mechanical  probability is high for histories of  coarse grained positions of the Moon's center of mass that obey Newton's laws. The universe behaves classically when the quantum probability is high for histories of coarse grained geometry and matter fields that are correlated in time by the Einstein equation. 

In this section we summarize the {\qf prescription} for classical prediction in quantum cosmology. The context is the minisuperspace models with homogeneous, isotropic geometries and  homogeneneous scalar field studied in this paper.  
These rules can be derived in a quantum  framework that predicts probabilities for sets of alternative, coarse-grained histories of cosmological geometries and matter fields whether or not they behave classically  \cite{Har95c,HM97,CH04}. Alternatively the rules can be motivated as a simple extension of the analogous algorithm derived in non-relativistic quantum mechanics. For brevity here we defer both of these arguments to a separate paper \cite{HHer08}.

A detailed discussion of our minisuperspace models will be given in Section \ref{minimodels}. But for the present discussion it is sufficient  to note that they lie in the class specified by a classical action $\cS$ of the form:
\begin{equation}
\cS[N(\lambda),q^A(\lambda)] = K \int d\lambda \Nh  \left[\frac{1}{2} G_{AB} \left(\frac{1}{\Nh}\frac{dq^A}{d\lambda}\right) \left(\frac{1}{\Nh}\frac{dq^B}{d\lambda}\right) -\cV (q^A) \right] .
\label{mini-act}
\end{equation}
Here, the $q^A$ are a set of coordinates for the minisuperspace of homogeneous, isotropic  three-geometries and homogeneous three-dimensional field configurations. {\uf {\uf For our models the  scale factor $b$ and the field value $\chi$ are the $q^A$. $\Nh$ is a multiplier ensuring reprametrization invariance.} (The hat is for consistency with later notation.)} Histories are {\qf curves in minisuperspace} specified by giving these coordinates as a function of a parameter $\lambda$, viz $q^A(\lambda)$. The metric\footnote{Up to factors this is the inverse DeWitt metric.}  on superspace $G_{AB}(q^A)$ and the potential $\cV(q^A)$ specify the model. 
The constant $K$ is fixed by scaling conventions for the variables and the conventions for $\cS$.

The action  \eqref{mini-act} is invariant under reparametrizations of the histories $q^A(\lambda)$. 
As a consequence there is a constraint relating  the coordinates $q^A$ and their conjugate momenta $p_A$. This can be found by varying \eqref{mini-act} with respect to $N(\lambda)$ and expressing the result in terms of $q^A$ and $p_A$. The result can be put in the form:
\begin{equation}
H(p_A, q^B) \equiv \frac{1}{2} G^{AB} p_A p_B   +\cV(q^A) = 0 .
\label{mini-constraint}
\end{equation}

In quantum cosmology the state of the universe is specified by giving a wave function on superspace. For minisuperspace models this is $\Psi(q^A)$. In this paper that is the no-boundary wave function \eqref{nbwf}. All  wave functions satisfy an operator implementation of the classical constraint \eqref{mini-constraint} 

\begin{equation}
H\left(-i \hbar\frac{\partial}{\partial q^A}, q^B \right) \Psi(q^A)=\left( -\frac{\hbar^2}{2} \nabla^2  +\cV (q^A)\right)\Psi(q^A)=0 .
\label{mini-wdw}
\end{equation}
(We retain the factors of $\hbar$ for later convenience in discussing classicality.)
This is the Wheeler-DeWitt equation for these models.  
There is a conserved current associated with the Wheeler-DeWitt equation 
\begin{equation}
J_A \equiv -\frac{i\hbar}{2} \Psi^*\frac{\stackrel{\leftrightarrow}{\partial}}{\partial q^A} \Psi .
\label{mini-current}
\end{equation}
This will play an important role in defining the probabilities for histories.

Suppose that in some region of superspace the wave function of the universe has the approximate semiclassical form (or is a sum of such forms)
\begin{equation}
\Psi(q^A) \approx A(q^A) e^{\pm iS(q^A)/\hbar}
\label{mini-scform}
\end{equation}
where $S(q^A)/\hbar$ varies rapidly over the region and $A(q^A)$ varies slowly. 
Under these circumstances the Wheeler-DeWitt equation \eqref{mini-wdw} requires that $S(q^A)$ satisfies the classical Hamilton-Jacobi equation {\uf to a good approximation}
\begin{equation}
H\left(\frac{\partial S}{\partial q^A}, q^B\right) \equiv \frac{1}{2} G^{AB}\frac{\partial S}{\partial q^A}   \frac{\partial S}{\partial q^B}+\cV(q^A) = 0 .
\label{mini-hamjac}
\end{equation}

In a suitable coarse-graining the only  histories that have significant probability are the classical histories corresponding to the integral curves of $S(q^A)/\hbar$ (e.g. \cite{HHer08}). These are  curves $q^A(\lambda)$  which satisfy 
\begin{equation}
p_A \equiv  G_{AB} \frac{1}{N}\frac{dq^B}{d\lambda} = \frac{\partial S}{\partial q^A} .
\label{mini-momenta}
\end{equation}
In short,  wave functions of the form \eqref{mini-scform} with rapidly varying $S(q^a)/\hbar$ and slowly varying $A(q^A)$ predict an ensemble of  classical Lorentzian cosmological histories.

Consider any surface in minisuperspace that is spacelike with respect to the metric $G_{AB}$ and has unit normal $n_A$.  We assume that the {\it relative} probability density $\wp$  of classical histories passing through this  surface is the component of the conserved current  \eqref{mini-current} along the normal if  it is positive. In leading order in $\hbar$ this is
\begin{equation}
\wp(q^A) \equiv J \cdot n  = |A(q^A)|^2 \nabla_n S(q^A) 
\label{mini-probs}
\end{equation}
in any region in which it is positive.  The first order in $\hbar$ implications of the Wheeler-DeWitt equation  \eqref{mini-wdw} ensure that these probabilities are constant along classical trajectories. Thus the formula \eqref{mini-probs}  could be evaluated on any spacelike surface with the same result for the probabilities of the classical histories {\uf that intersect it.}\footnote{This expression for probability of classical histories has been advocated by other authors, see e.g.  \cite{Vil88,HTsum}.}

Several key points should be noted about this prescription for classical prediction:
\begin{itemize}

\item The no-boundary wave function provides probabilities for entire four-dimensional classical {\it histories}. When the strong energy condition is not satisfied (as for the present models)  these may bounce at a minimum radius. If that radius is large enough we expect a classical extrapolation from present data to be a good approximation over the whole history of the universe --- from an infinite volume in the past to an  infinite volume in the future. Alternatively when extrapolated from present data  some classical histories assigned probabilities may have initial or final singularities or both. 

\item {\uf Singularities in the extrapolation of classical solutions}  do not signal the break down of quantum mechanics but rather of the approximation. In particular the NBWF predicts probabilities for the classical description of late time observables such as CMB fluctuations whatever happens to an extrapolation of that classical description. That is because the NBWF predicts probabilities for histories not their initial data. In this sense the NBWF {\it resolves} classical singularities. 
 
\item The histories in the classical ensemble are not the same as the extremizing histories that provide the steepest descents approximation to the integral \eqref{nbwf} {\uf defining the NBWF}. The classical histories are real and Lorentzian. The extrema are generally complex --- neither Euclidean nor Lorentzian except in very special cases. The classical histories may contract from an infinity in the past and reexpand to another one in the future; the no-boundary extrema can have only one infinity. Indeed, in the no-boundary case the classical histories and extremizing histories are on different manifolds. 
 This clean separation into real classical histories and complex extremizing ones helps to clarify the meaning of both,  and resolves issues that arise from their identification such {\uf as those dicsussed in} \cite{UJ98}.  

\end{itemize}

\noindent From this point of view, a wave function of the universe is best thought of, not as an initial condition, but rather in a four dimensional sense as giving probabilistic weight to the possible four-dimensional histories of a quantum universe.

\section{The No-Boundary Wave Function and its Semiclassical Approximations}
\label{nbwf_sec}\label{sec3}

This section considers the steepest descents approximation to the NBWF for homogeneous isotropic minisuperspace models. It describes when this leads to a semiclassical form like \eqref{mini-scform} which  predicts an ensemble of classical Lorentzian histories with probabilities for each.

\subsection{Steepest descents approximation}
\label{steepest}

The NBWF is defined by a path integral over homogeneous field configurations and homogeneous isotropic metrics of the form
\begin{equation}
ds^2=(3/\Lambda)\left[N^2(\lambda) d\lambda^2 +a^2(\lambda) d\Omega^2_3\right] .
\label{eucmetric}
\end{equation} 
Here, $ d\Omega^2_3$ is the round metric on the unit three-sphere and the factor in front is a convenient normalization. The defining path integral has the specific form [cf. \eqref{nbwf}]
\begin{equation}  
\Psi(b,\chi)\equiv \Psi(q^A) =  \int_{\cal C} \delta N \delta x  \exp(-I[N(\lambda), x^A(\lambda)]/\hbar) 
\label{mini-nbwf}
\end{equation}
where $x^A(\lambda)=(a(\lambda), \phi(\lambda))$ are histories of the scale factor and scalar field. 
The integral is over all ($x^A(\lambda), N(\lambda)$) that {\qf define regular geometries on a disk}  which  match the values of $q^A=(b,\chi)$ on its boundary.  
The functional $I$ is the Euclidean action which for our class of models is [cf. \eqref{mini-act}]
\begin{equation}
I[N(\lambda),x^A(\lambda)] = K \int^1_0 d\lambda N   \left[\frac{1}{2} G_{AB}(q^A) \left(\frac{1}{N}\frac{dx^A}{d\lambda}\right) \left(\frac{1}{N}\frac{dx^B}{d\lambda}\right) +\cV (q^A) \right] .
\label{mini-eucact}
\end{equation}
where the parameter values labeling the  endpoints have been  conventionally chosen to  be $0$ and $1$. 
The {\uf integration}  measure in \eqref{mini-nbwf} contains the usual apparatus of gauge fixing terms and their associated determinants made  necessary by  reparametrization invariance. We have left all of this unspecified because it will not be important in the  leading steepest descents approximation. From now on in this paper  $\Psi(q^A)$ should be understood  to be this no-boundary wave function.

The steepest descents approximation to the path integral \eqref{mini-nbwf} defining the NBWF starts with those paths that extremize the action {\uf in the class integrated over}. 
Such paths $(N_{\rm ext}(\lambda), x^A_{\rm ext}(\lambda))$ are solutions the equations of motion 
\begin{equation}
\frac{\delta I}{\delta x^A(\lambda)} = 0,  \quad  \frac{\delta I}{\delta N(\lambda)}=0 
\label{mini-eqmotion}
\end{equation}
that are regular on the disk and match the values $q^A$ on its boundary The explicit form of the equations  \eqref{mini-eqmotion} will be displayed in Section \ref{minimodels}. 

The steepest descents approximation to the NBWF is then given by 
\begin{equation}
\Psi(q^A) \approx \sum_{\rm ext} \exp[-{\cal A}_{\rm ext}(q^A)/\hbar]
\label{mini-scapprox}
\end{equation} 
where the sum is over all extrema that contribute to the integral.  The exponent has an expansion in powers of $\hbar$ of the form
\begin{equation}
{\cal A}_{\rm ext}(q^A) = I_{\rm ext}(q^A) + \hbar I^{(1)}_{\rm ext}(q^A) + \cdots. 
\label{mini-scexpn}
\end{equation}
The leading order in this expansion is the Euclidean action evaluated at the extremum:
\begin{equation}
I_{\rm ext}(q^A) \equiv I[N_{\rm ext}(\lambda), x^A_{\rm ext}(\lambda)] .
\label{mini-clasact}
\end{equation}
Next order corrections in $\hbar$ include terms like 
\begin{equation}
-(1/2) Tr \log (\delta^2 I) 
\label{mini-det}
\end{equation}
where $\delta^2I$ is the operator resulting from the second variation of the action\footnote{This correction leads to a well known prefactor which can be written as the inverse square root of the  determinant of $\delta^2 I$}. Factors arising from the measure would also contribute at this order. 

The Hamiltonian-Jacobi equation for the Euclidean action is the order $\hbar^0$ consequence of the Wheeler-DeWitt equation \eqref{mini-wdw}. The order $\hbar^1$ implication is the conservation of the probabilities \eqref{mini-probs}. We therefore should retain both orders in the steepest descents approximation to be consistent with these features.
Traditonally the order $\hbar$ contributions are written as a prefactor to the exponential. 
Thus we have to write for the contribution to the wave function of one extremum
\begin{equation}
\Psi_{\rm ext}(q^A) \approx P_{\rm ext}(q^A) \exp[-I_{\rm ext}(q^A)/\hbar]. 
\label{scwavfn}
\end{equation}

From now on we will consider the extrema one at a time and drop the subscript ``ext'' that distinguished one from the other. 

\subsection{Classicality}
\label{classicality}\label{sec2b}
There is no reason to assume that the leading steepest descents approximation will be given by a real path $(N(\lambda), x^A(\lambda))$. The reality of the NBWF only means that the extrema must come in complex conjugate pairs. (See Section \ref{timesym} below for more on the consequences of this.) Indeed, we will show in Section \ref{mini-extrema} how the extremizing paths are necessarily complex. 
The action at an extremum will therefore have both real and imaginary parts which we write:
\begin{equation}
I(q^A) = I_R(q^A) - i S(q^A) .
\label{mini-realimag}
\end{equation}
The first two orders in steepest descents approximation to the wave function therefore take the form \eqref{mini-scform}
\begin{equation} 
\Psi(q^A) = A(q^A) e^{iS(q^A)/\hbar} , 
\label{mini-scform1}
\end{equation}
with $A$  given by 
\begin{equation}
A(q^A) \equiv P(q^A)e^{-I_R(q^A)/\hbar} .
\label{mini-A}
\end{equation}

As reviewed in Section \ref{classicalpred}, a wave function with the semiclassical form \eqref{mini-scform1} in some region of minisuperspace predicts an ensemble of classical trajectories provided that $S(q^A)/\hbar$ is rapidly varying and $A(q^A)$ is slowly varying. {\rf Assuming that  $P(q^A)$ is slowly varying}, a necessary  condition for this is that the gradient of $I_R(q^A)$ be small compared to the gradient of $S(q^A)$ {\rf in the coordinates $q^A$ {\sf that enter into the defining path integral} and for which we expect to have classical equations of  motion e.g $(b,\chi)$ \cite{HHer08}.} That is, 
\begin{equation}
|\nabla_A I_R| \ll |\nabla_A S|  .
\label{classcond}
\end{equation}
This is the {\it classicality condition} which plays a central role in our work. 

Further, as we saw in Section \ref{sec2}, when the action $S(q^A)$ satisfies the classical Hamilton-Jacobi equation
\begin{subequations}
\label{loreqs}
\begin{equation}
\frac{1}{2} (\nabla S)^2 + \cV(q^A) = 0,
\label{classHJ}
\end{equation}
the Lorentzian histories in the classical ensemble are the integral curves of $S(q^A)$. Specifically, choosing $N=i$ so the metrics \eqref{eucmetric} are Lorentizian, the integral curves obey the equations of motion  
\begin{equation}
p_A(q^A) \equiv G_{AB}\frac{dx^B}{d\lambda }= \nabla_A S(q^A) . 
\label{momenta}
\end{equation}
\end{subequations}

Defined in \eqref{mini-realimag}  as (minus) the imaginary part of the Euclidean action evaluated at an extremum, there is no reason to believe that eqs \eqref{loreqs}  hold in all regions of minisuperspace, and indeed they do not. Rather the Euclidean action satisfies its own `Euclidean Hamilton-Jacobi' equation 
\begin{subequations}
\begin{equation}
-\frac{1}{2} (\nabla I)^2 + \cV(q^A) = 0 ,
\label{eucHJ}
\end{equation}
with its own equation of motion
\begin{equation}
{G_{AB}}\frac{1}{N}\frac{dx^B}{d\lambda} = \nabla_A I(q^A). 
\label{euceqmotion}
\end{equation}
\end{subequations}
These follow directly from the definition \eqref{mini-eucact}. 

Using \eqref{mini-realimag} the real and imaginary parts of the Euclidean Hamilton-Jacobi relation \eqref{eucHJ} can be  written as
\begin{subequations}
\label{realimaghj}
\begin{align}
-\frac{1}{2} (\nabla I_R)^2 + \frac{1}{2} (\nabla S)^2 + \cV(q^A) &= 0, \label{realjhj1}\\
\nabla I_R \cdot \nabla S &= 0.
\label{maghj2}
\end{align}
\end{subequations}
Eq \eqref{realjhj1} shows that the Hamilton-Jacobi equation \eqref{classHJ}  for $S(q^A)$ follows from the Euclidean Hamiltonian-Jacobi equation when the  classicality condition \eqref{classcond} holds  because this implies
\begin{equation}
|(\nabla I_R)^2| \ll |(\nabla S)^2| .
\label{oldclasscond}
\end{equation}
But \eqref{oldclasscond} is not enough to guarantee classicality. It is sufficient for the Lorentizian Hamilton-Jacobi equation \eqref{classHJ}. But it is  not necessarily enough to ensure that the Euclidean equations of motion \eqref{euceqmotion} reduce to their Lorentzian forms \eqref{momenta}. For that the stronger condition \eqref{classcond} is necessary. 

The other consequence of the Euclidean Hamilton-Jacobi equation is \eqref{maghj2}. 
This implies that $I_R(q^A)$ is constant along  the integral curves of $S(q^A)$. That is, each classical Lorentzian history is associated with a value of $I_R$.  Indeed, in a two-dimensional minisuperspace like that of this paper, \eqref{maghj2} implies that curves of constant $I_R$ {\it are} integral curves of  $S$.

In principle the probability for any history can be calculated from the wave function $\Psi(q^A)$ without a semiclassical approximation. The semiclassical form is only a {\it sufficient} criterion for classicality. {\it We will assume that once classical histories have been identified in a region of minisuperspace where the classicality condition holds they may be extended to regions where it does not hold using the classical equations of motion unless they become classically singular.} It is plausible, for instance, that a bouncing universe whose radius never falls below the Planck length will remain classical throughout its history even if it can only be identified by a steepest descents approximation in some regions of minisuperspace. That is an assumption which can in principle be checked in the full quantum mechanical theory.

We next turn to the probabilities predicted by the NBWF for the individual histories in the classical ensemble. According to \eqref{mini-probs}, the relative probability density $\wp$  for classical histories passing through a spacelike surface in minisuperspace with unit normal $n_A$  is given by 
\begin{equation}
\wp(q^A) = |P(q^A)|^2 e^{-2I_R(q^A)} \nabla_n S
\label{probhist}
\end{equation}
where this expression is positive. For this  to be a probability for histories it must be constant along the integral curves of $S$. As discussed  in Section \ref{classicalpred},  the order $\hbar^0$ approximation to the Wheeler-DeWitt equation \eqref{mini-wdw} ensures this. But since $I_R$ is already constant along classical trajectories the rest of the measure \eqref{probhist} must be separately constant.

{\it In this paper we will consider only the lowest order semiclassical approximation to the probabilities and ignore the prefactor $P$ which arises in the next order. }  That is we discuss only  the $\exp(-2I_R)$ contribution to the probabilities which is possible because it is conserved along classical histories.  
This  `approximation' is forced on us  by our limited ability to compute the corrections to the leading term at this time. Higher order corrections can be crucial for some questions (e.g \cite{BK90}) but for the diagnosis of classicality we expect that the this lowest approximation is enough. 

\subsection{Extrema of the Euclidean Action and Fuzzy Instantons} 
\label{mini-extrema}
We now return to a more detailed examination of the conditions determining the complex paths that extremize the action and the differential equations \eqref{mini-eqmotion} which are necessary conditions for an extremum.
The set of equations  \eqref{mini-eqmotion} consists of two second order differential equations for $a(\lambda)$ and $\phi(\lambda)$
 together with a constraint involving only their first derivatives.  There are thus four real second order differential equations and two real constraint conditions. 

We first show that there are no free parameters in the boundary conditions that determine the solutions {\uf to these equations}. 
The domain on which the equations are to be solved ranges from the center of symmetry of the geometry on the 3-disk at $\lambda =0$ (called the `South Pole' (SP)) to the boundary {\uf where $b$ and $\chi$ are specified} at $\lambda=1$. The conditions for the geometry and field to be regular at the SP are 
\begin{subequations}
\label{bcs}
\begin{equation}
a(0)=0, \quad \phi'(0)=0 
\label{SPconds}
\end{equation}
where a prime denotes a derivative with respect to $\lambda$. The conditions at the boundary are 
\begin{equation}
a(1)= b, \quad \phi(1)=\chi
\label{Bconds}
\end{equation}
\end{subequations}
where $b$ and $\chi$ are real. Eqs \eqref{bcs} constitute eight real conditions for the four real second order equations for $a(\lambda)$ and $\phi(\lambda)$.

With a suitable choice of parametrization the multiplier $N(\lambda)$ can be taken to be a complex constant $N$.  For each $N$ solve the second order differential equations with the boundary conditions \eqref{bcs} for $a(\lambda)$ and $\phi(\lambda)$. Then find the real and imaginary parts of $N$ so that the two real constraint conditions are satisfied. Were the equations linear the solutions would be determined. There may be more than one solution to this non-linear set of equations with the boundary conditions \eqref{bcs} but there are no free parameters to specify. Hence whether the solutions are real or complex is not up to us; it is determined by the equations and the boundary conditions. 

In the case of a cosmological constant and no scalar field there is a real solution consisting of a real Euclidean instanton with the geometry of half a round 4-sphere joined smoothly onto de Sitter space through a sphere of minimum radius \cite{HH83}. But as shown by a number of authors (see e.g. \cite{UJ98}), and, as will be verified here, there are no real extrema when the scalar field is non-zero  except for special ``false-vacuum" potentials. In general the extrema are necessarily complex. 
There is thus generally no meaningful notion of a Euclidean instanton nucleating the universe. But, as we will see in Section \ref{sec5},  for a wide range of parameters the imaginary parts of extremizing  solutions  that satisfy the classicality condition \eqref{classcond} are small.
In this regime we can therefore think of these extremizing geometries as ``fuzzy instantons'' in which there is a transition from a real Euclidean geometry at the South Pole to an asymptotically real Lorentzian geometry at large volume. The transition is not sharp as in the zero scalar field case, but rather spread out over a region.  We will give explicit examples in Section \ref{complexsolutions}. 

We stress again, however, that the complex fuzzy instantons that provide the semiclassical approximation to the NBWF are distinct from the Lorentzian histories in the classical ensemble for which they provide probabilities through the real part of their complex  action.

\subsection{Time Symmetry of the Classical Ensemble}
\label{timesym}

The metric $G_{AB}(q^A)$ and the potential $\cV(q^A)$  that define the Euclidean action
\eqref{mini-eucact} are real analytic functions of their arguments for the models we will consider. So, therefore, are the  coefficients in the equations \eqref{mini-eqmotion} which are the necessary conditions for an extremum of the action. The boundary conditions for their solutions \eqref{bcs} are real. 

Therefore, for every extremum $(N(\lambda),a(\lambda),\phi(\lambda))$, there is also a complex conjugate extremum $(N^*(\lambda),a^*(\lambda),\phi^*(\lambda))$. If
\begin{subequations}
\label{mini-realimag-all}
\begin{equation}
I(b,\chi) = I_R(b,\chi) - i S(b,\chi)
\label{mini-realimag1}
\end{equation} 
is the action for the first solution then the action for the second will be
\begin{equation}
I(b,\chi) = I_R(b,\chi) + i S(b,\chi).
\label{mini-realimag2}
\end{equation}
\end{subequations}
The real part of both actions is the same. Both extrema therefore count equally in their contribution to the steepest descents approximation to the NBWF. This shows explicitly that the NBWF is real in the semiclassical approximation. 

But the opposite signs for $S$  in \eqref{mini-realimag-all} means that the momenta of a classical history passing through $q^A$ will be opposite in the two cases [cf. \eqref{mini-momenta}]. The classical ensembles of each extremum consist of histories that are time reversals of one another.
Both will have the same probability because $I_R(b,\chi)$ is the same for both.  
The individual histories need not be time symmetric \cite{HLL93} and indeed we will find that they are not [cf. Figure \ref{Lor-ex}]. But the ensemble of predicted classical histories is time symmetric in the sense that for any history in it,  its time reversed is also {\uf a member} with the same probability.

\subsection{Measures on Classical Phase Space}
\label{measures}
A classical history measure is any function on phase space that is conserved along classical trajectories because of the classical equations of motion. Classical history measures on the phase space of classical minisuperspace models have been adroitly employed by Gibbons and Turok \cite{GT06} to analyze the probability of inflation in the absence of a theory of the universe's state. 

The predictions of the NBWF for an ensemble of classical histories provides a history measure on the classical phase space of minisuperspace models. The condition \eqref{momenta} between the coordinates $q^A$ and the momenta $p_A$ shows the the NBWF measure is concentrated on a surface in classical phase space of half its dimension. One could think of the  NBWF measure as a $\delta$-function on this slice through phase space that assigns zero probability to those points not on it. 

As  we will see in Section \ref{complexsolutions}, the NBWF surface does not slice through the whole of classical phase space. Quantum mechanics assigns probabilities generally to decoherent sets of alternative histories of the universe. But only in special circumstances are the probabilities high for the correlations in time that define classical histories. A classical history therefore cannot be expected to pass through every point $q^A$.  The classicality condition \eqref{classcond} generally specifies a {\it boundary} to the surface in phase space on which points corresponding to classical histories lie. 

This restriction of the ensemble of possible classical histories to a  surface  with boundary  in phase space is already a powerful prediction of the NBWF whatever relative probabilities are predicted for histories in it. 
Eq. \eqref{probhist} gives the predictions of the NBWF for the probabilities of the histories within the surface in phase space defining the classical ensemble. These probabilities define a classical history measure because they are conserved along the classical trajectories (recall \eqref{maghj2}.)

\section{Homogeneous Isotropic Mini-Superspace Models}
\label{minimodels}\label{sec4}

\subsection{Euclidean Action and Equations for its Extrema}

From now on, we use Planck units where $\hbar=c=G=1$. 
The Euclidean action $I[g,\Phi]$ is a sum of a  curvature part $I_C$ and a part $I_\Phi$ for the scalar field $\Phi$.  The general form for the curvature action is: 
\begin{equation}
I_C[g] = -\frac{1}{16\pi}\int_M d^4 x (g)^{1/2}(R-2\Lambda) +\text{(surface terms)} .
\label{curvact}
\end{equation}
The general form for the matter action for a scalar field moving in a quadratic potential is:
\begin{equation}
I_{\Phi}[g,\Phi]=\frac{1}{2} \int_M d^4x (g)^{1/2}[(\nabla\Phi)^2 +m^2 \Phi^2] . 
\label{mattact}
\end{equation}
The integrals in these expressions are over the manifold $M$ with one boundary defining the NBWF (cf. eq \eqref{nbwf}). 
With a convenient overall scale,  the homogeneous, isotropic metrics are defined as in \eqref{eucmetric}.
With that normalizing factor  the scale factor $a(\lambda)$, nor  the lapse $N(\lambda)$,  nor any of the coordinates carry  dimensions\footnote{The scaling of the metric used here is different from that employed in \cite{Lyo92}, as are others in this paper, but they prove convenient for simplifying the numerical work.}. 

It proves convenient to introduce dimensionless measures $H$,   $\phi$, and $\mu$  of  $\Lambda$, $\Phi$, and $m$ respectively as follows: 
\begin{subequations}
\begin{equation}
H^2 \equiv \Lambda/3 ,
\label{defH}
\end{equation}
\begin{equation}
\phi \equiv (4\pi/3)^{1/2} \Phi   ,
\label{defphi}
\end{equation}
\begin{equation}
\mu \equiv  (3/\Lambda)^{1/2}m .
\label{defmu}
\end{equation}
\end{subequations}
The scaling for $H$ was chosen so that the scale factor of a classical inflating universe is proportional to $\exp(H t)$ --- the usual definition of $H$. The other scalings were chosen to make the action simple.
In these variables the Euclidean action takes the following simple form:
\begin{equation} 
I[a(\lambda),\phi(\lambda)] = \frac{3\pi}{4 H^2}\int^1_0 d\lambda N \left\{ -a \left(\frac{a'}{N}\right)^2 -a +a^3 +a^3\left[\left(\frac{\phi'}{N}\right)^2 + \mu^2\phi^2\right]\right\}
\label{eucact}
\end{equation}
where $'$ denotes $d/d\lambda$ and the surface terms in \eqref{curvact}  have been chosen to eliminate second derivatives.  The center of symmetry SP  and the boundary of the manifold $M$ have arbitrarily been labeled by coordinates $\lambda =0$ and $\lambda=1$ respectively.

Three equations follow from extremizing the action with respect to $N$, $\phi$, and $a$. 
They imply the following equivalent relations:
\begin{subequations}
\label{euceqns_N}
\begin{equation}
\left(\frac{a'}{N}\right)^2 -1 +a^2 +a^2\left[-\left(\frac{\phi'}{N}\right)^2 + \mu^2\phi^2\right]=0,
\label{eucconstraint_N}
\end{equation}
\begin{equation}
\frac{1}{a^3N}\left(a^3\frac{\phi'}{N}\right)' - \mu^2 \phi = 0 , 
\label{eucphieqn_N}
\end{equation}
\begin{equation}
\frac{1}{N}\left(\frac{a'}{N}\right)' +2a\left(\frac{\phi'}{N}\right)^2 + a(1+\mu^2\phi^2) = 0 \ .
\label{eucaeqn_N}
\end{equation}
\end{subequations}
These three equations are not independent. The first of them is the Hamiltonian constraint. From it, and any of the other two, the third follows.

From \eqref{eucact} we can read off the explicit forms of the factors in the general form of the actions \eqref{mini-act} and \eqref{mini-eucact}. We have $K=3\pi/2 H^2$ and 
\begin{equation}
G_{AB}= {\rm diag}(-a,a^3), \quad  {\cal V}= (1/2)(-a +\mu^2\phi^2) .
\label{GV}
\end{equation}

\subsection{Complex Contours for the Action}
\label{complexgauge} \label{sec4b}
The extremizing solutions $a(\lambda)$, $\phi(\lambda)$, and $N(\lambda)$  will generally be complex. Assuming they are analytic functions, the integral \eqref{eucact} can be thought of as taken over a real contour in the complex $\lambda$ plane between $0$ and $1$.  Following Lyons \cite{Lyo92} it is then useful to introduce a new complex variable $\tau$ defined by 
\begin{equation}
\tau(\lambda) \equiv \int_0^\lambda d\lambda' N(\lambda').
\label{deftau}
\end{equation} 
The function $\tau(\lambda)$  defines a contour in the complex $\tau$-plane for each lapse function $N(\lambda)$. Conversely {\uf for each  contour starting at $\tau=0$, \eqref{deftau}  defines a multiplier  $N(\lambda)\equiv d\tau(\lambda)/d\lambda$. }The action \eqref{eucact} can be rewritten as an integral over the countour ${C}(0,\upsilon)$ in the complex $\tau$-plane corresponding to the $N(\lambda)$ in \eqref{eucact} and connecting $\tau=0$ with an endpoint we denote by $\upsilon$. Specifically, 
\begin{equation} 
I[a(\tau),\phi(\tau)] = \frac{3\pi}{4 H^2}\int_{{C}(0,\upsilon)}  d\tau \left[ -a {\dot a}^2 -a +a^3 +a^3\left({\dot\phi}^2 + \mu^2\phi^2\right)\right]
\label{eucact_tau}
\end{equation}
and  $\dot f$ denotes $df/d\tau$. 

The equations \eqref{euceqns_N} also simplify in the new variable, viz:
\begin{subequations}
\label{euceqns}
\begin{equation}
{\dot a}^2 -1 +a^2 +a^2\left(-{\dot \phi}^2 + \mu^2\phi^2\right)=0,
\label{eucconstraint}
\end{equation}
\begin{equation}
\ddot\phi + 3({\dot a}/a)\dot\phi  - \mu^2 \phi = 0 , 
\label{eucphieqn}
\end{equation}
\begin{equation}
\ddot a  +2a {\dot\phi}^2 + a(1+\mu^2\phi^2) = 0 \ .
\label{eucaeqn}
\end{equation}
\end{subequations}
These are the equations we will use to calculate the complex extremizing geometries and matter field configurations.

{\uf Using these equations the value of the action \eqref{eucact_tau} on a solution can be rexpressed as}
\be
I[a(\tau),\phi(\tau)] = \frac{3\pi}{2H^2}\int_{{C}(0,\upsilon)}  d\tau a \left[ a^2\left(1 + \mu^2\phi^2\right)-1\right],
\label{eucact_sol}
\ee

Two contours that connect the same endpoints in the $\tau$-plane give the same value  for the  action provided they can smoothly be distorted into one another. They are different representations of  the same extremum as far as the semiclassical approximation to the NBWF is concerned and we count their contributions only once.   Another way of saying this is that \eqref{deftau} defines a complex  transformation of the coordinates in the formula for the action under which it is invariant if the contours can be smoothly distorted into one another. It should not, however, be thought of as a transformation of the coordinates on the manifold which remain real throughout. 

This suggests that a {\it solution} to equations \eqref{euceqns} should be considered as a pair of complex analytic {\it functions} $a(\tau)$ and $\phi(\tau)$. We can evaluate the action with these functions by picking  any convenient contour in $\tau$ connecting the center of symmetry to the boundary. We will exploit this in what follows.

\subsection{Lorentzian Equations}

For the semiclassical approximation to the NBWF we will be interested in complex solutions to equations \eqref{euceqns}. But the ensemble of histories to which these solutions supply probabilities will be real, Lorentzian metrics of the form
\begin{equation}
d{\hat s}^2=(3/\Lambda)\left[-\Nh^2(\lambda) d\lambda^2 +\ah^2(\lambda) d\Omega^2_3\right].
\label{lormetric}
\end{equation}
We will use hats to distinguish Lorentzian quantities that are always real from the complex metrics that extremize the Euclidean action. 

Both the Lorentizian action $\cal S$ and the equations locating its extrema can be obtained from the complex relations by substituting  $N=\pm i \Nh$, $a=\ah$ and $\phi=\phih$ into \eqref{eucact}.  Adhering to the usual convention that the kinetic energy term of the matter be positive, the Lorentzian action is
\begin{equation} 
{\cal  S}[\ah(\lambda),\phih(\lambda)] = \frac{3\pi}{4 H^2}\int d\lambda \Nh \left\{ -\ah \left(\frac{\ah'}{\Nh}\right)^2 +\ah -\ah^3 +\ah^3\left[\left(\frac{\phih'}{\Nh}\right)^2 - \mu^2\phih^2\right]\right\}.
\label{loract}
\end{equation}
We quote the consequent Lorentzian equations in terms of $dt=\Nh d\lambda$,
\begin{subequations}
\label{loreqns}
\begin{equation}
\left(\frac{d\ah}{dt}\right)^2 +1 -\ah^2 -\ah^2\left[\left(\frac{d\phih}{dt}\right)^2 + \mu^2\phih^2\right]=0,
\label{lorconstraint}
\end{equation}
\begin{equation}
\frac{1}{a^3}\frac{d}{dt}\left(a^3\frac{d\phih}{dt}\right)+ \mu^2 \phih = 0 , 
\label{lorphieqn}
\end{equation}
\begin{equation}
\frac{d^2\ah}{dt^2} +2\ah\left(\frac{d\phih}{dt}\right)^2 -a(1+\mu^2\phih^2) = 0 \ .
\label{loraeqn}
\end{equation}
\end{subequations}
The energy density in the scalar field $\rho_\Phi$ is a useful quantity for analyzing Lorentzian solutions. For example, if its exceeds the Planck density we can consider the solution classically singular. An expresson for it  can be derived from the action \eqref{loract} or from the form of the constraint equation \eqref{lorconstraint}. The result is
\begin{equation} 
\rho_\Phi = \left(\frac{3 H^2}{8\pi} \right)\left[\left(\frac{d\phih}{dt}\right)^2 + \mu^2 \phih^2\right] \  .
\label{rho}
\end{equation}

\subsection{The Classical Ensemble of a Complex Extremum}
\label{classens}

{\qf As discussed in Section \ref{sec2b}, the semiclassical approximation to the NBWF given by  a solution to \eqref{euceqns} corresponds to a solution to the Lorentzian equations \eqref{loreqns} when the classicality condition \eqref{classcond} are satisfied. The Lorentzian solutions obtained this way are the integral curves of 
\begin{equation}
S(b,\chi) = -\Im [I(b,\chi)]. 
\label{Sdef}
\end{equation}
To calculate this ensemble explicitly for these models we proceed as follows: Choose a {\it matching surface}  of constant $b=b_*$ in a region of minisuperspace where the classicality condition is satisfied --- typically for large values of $b_*$. The integral curves of $S$ can be labeled by the value of $\chi=\chi_*$ where they intersect this matching surface. The Lorentzian and Euclidean momenta there are given by gradients of $S$ and  $I$ respectively [cf. \eqref{momenta}]. Their explicit forms in terms of scale factor and field can be found from the actions \eqref{eucact}, and \eqref{loract}. From \eqref{Sdef} we have on the matching surface:
\begin{subequations}
\label{match}
\begin{align}
{\hat b} = b_*, \quad & {\hat p}_b = -\Im(p_b)|_{b_*},  \label{matcha} \\
{\hat \chi} = \chi_*, \quad &{\hat p_\chi} = -\Im(p_\chi)|_{b_*}. \label{matchb}
\end{align}
\end{subequations}
These relations show how a complex extremum specifies  Cauchy data for solving the equations \eqref{loreqns} to find the complete Lorentzian history labeled by $\chi_*$. The value of $\exp[-2I_R(b_*,\chi_*)]$  gives its relative probability. The classical ensemble is generated as $\chi_*$ varies across the matching surface. 
}
\section{Complex Solutions}
\label{complexsolutions}\label{sec5}

 {\vf As the first step in determining the probabilities of the ensemble of classical cosmologies predicted by the NBWF we begin  by evaluating it semiclassically in this section. The classical ensemble implied by this approximation will be determined in the following section.  

We find the semiclassical approximation to the NBWF by} numerically solving eqs \eqref{euceqns} for $a(\tau)$ and $\phi(\tau)$ along a suitable contour in the complex $\tau$-plane connecting the South Pole $\tau=0$ with an 
endpoint $\upsilon=X+iY$ where $a$ and $\phi$ take real values $b$ and $\chi$. 
The (no) boundary conditions \eqref{SPconds} of regularity {\vf at the SP mean {\vf that} the complex value of $\phi$ at the origin is the only free parameter there}. To reach the prescribed values $(b,\chi)$ at the boundary, we will adjust {\vf both} this and the endpoint of integration $\upsilon$. This gives four real adjustable parameters
to meet four real conditions at $\upsilon$. Hence for each $b$ and $\chi$ there is a unique solution. {\vf These solutions can be found analytically in the limits when $\phi(0)$ is very large and very small. We discuss these limiting cases first as they will motivate our numerical search procedure.}

When the scalar field is large (but well below the Planck density) the classical dynamics should be governed {\vf only} by the scalar field and its backreaction on the geometry. The background cosmological constant should be largely irrelevant. In this regime therefore we expect to recover the {\vf approximate}  complex solutions found by Lyons \cite{Lyo92} {\vf for} a scalar field model with a quadratic potential and $\Lambda=0$ . Indeed, following Lyons one can show that for $|\phi(0)|  \gg 1$ the complex Einstein equations \eqref{euceqns} admit the approximate `slow roll' solution \be
 \label{ansol}
\phi_+(\tau)  \approx \phi(0) +i \frac{\mu \tau}{3},  \qquad a_+(\tau)  \approx \frac{i}{2\mu \phi (0)} e^{-i\mu \phi(0)\tau+\mu^2 \tau^2/6}.
\ee
{\vf There is a similar approximate solution  $(\phi_{-}(\tau), a_{-}(\tau))$ found by changing $i$ to $-i$ in \eqref{ansol}. }

{\vf These solutions are the complex analogs of the standard `slow roll' inflationary solutions. They can be found up to a constant multiplicative normalization  of the scale factor by neglecting the cosmological constant, spatial curvature, and $\ddot \phi$ terms in equations \eqref{euceqns} and solving the resulting simple set of equations. The results are not good approximations everywhere in the complex $\tau=x+iy$ plane. They hold when  when $y$ is not so large that the slow role assumption breaks down, and only in regions where $|a(\tau)|>>1$ so that the spatial curvature is exponentially negligble as was assumed in deriving them. The existence and location of such a region depends on the value of $\Re[\phi(0)]$. 
When $\Re[\phi(0)]>0$ the solution $(a_+(\tau),\phi_+(\tau))$ is valid in a region in the $y>0$ half-plane, and $(a_-(\tau),\phi_-(\tau))$ holds in a region in the lower half-plane. }

{\vf When $|\phi(0)|$ is very large and $\tau$ is sufficiently small that any change in $\phi$ is negligible these solutions must match the `no-roll' solution
\be
\phi(\tau) \approx \phi (0), \qquad a(\tau)  \approx \frac{\sin [\mu \phi (0) \tau]}{\mu \phi (0)}.
\label{noroll}
\ee
This is regular at the origin and valid in regions in both half-planes. Matching with this solution determines the multiplicative normalization factor in \eqref{ansol}. }

{\vf Following Lyons we  now make the further approximation that $\Re[\phi(0)]\gg \Im[\phi(0)]$. 
Then in the  solution \eqref{ansol}} the scalar field is approximately real along a vertical line $\tau = -3 \phi_{I}(0) /\mu+iy$ where $\phi_{I}(0) \equiv {\rm Im}[\phi(0)] $.  Eliminating $\tau$ from the solution \eqref{ansol} for $a$ gives
\be
a \approx \frac{i}{2\mu \phi_R (0)} e^{3[\phi (0))^2 - \phi(\tau)^2]/2}
\ee
where $\phi_R (0) \equiv {\rm Re}[\phi(0)]$. Therefore, if one takes
\be \label{imSP}
\phi_I (0) = - \frac{\pi}{6 \phi_R (0)},
\ee
one obtains vertical lines given by 
\be\label{turningpt}
\tau = \frac{\pi}{2 \mu \phi_R (0)} +iy
\ee
along which both $a$ and $\phi$ are approximately real. It is clear that {\vf progressively} finer  tuning of $\phi_I (0)$ will yield approximately vertical curves of exactly real $a(y)$ and $\phi(y)$.  Notice also that the condition \eqref{imSP} at the SP that fixes $a$ and $\phi$ to be approximately real does not depend on the time parameter along the vertical line. 

The complex action of a solution to the equations of motion {\vf can be}  obtained from \eqref{eucact_sol}.
For the solutions \eqref{ansol} the main contribution to the real part $I_R$ comes from the integral over real $\tau$ from the SP to $X=\pi/(2\mu \phi_R (0))$. In this regime the solutions are approximately given by the no-roll solution \eqref{noroll} with $\phi(0)\approx \phi_R(0)$. 
When $\phi$ is large this yields
\be\label{act}
I_R \approx - \frac{\pi}{2(H\mu \phi_R (0))^2} \approx - \frac{\pi}{2(H\mu \chi)^2} 
\ee 
where we are assuming that at the boundary $\chi \sim \phi_R(0)$.
This is the action of Euclidean de Sitter space with effective cosmological constant 
$3m ^2(\phi_R (0))^2$. The main contribution to $S$ comes from the integration over the vertical line to an endpoint $\upsilon$. It is given by \cite{Lyo92}
\be
S \approx \frac{i\mu \chi b^3}{3}.
\label{lyonsS}
\ee
This can be used to verify whether the solutions satisfy the classicality condition \eqref{classcond} at large scale factor. One has
\begin{subequations}
\be
(\nabla I_R)^2 \equiv -\frac{1}{b} \left( \frac{\partial I_R}{\partial b} \right)^2 + \frac{1}{b^3} \left( \frac{\partial I_R}{\partial \chi} \right)^2 \approx \frac{1}{b^3} \left( \frac{\partial I_R}{\partial \chi} \right)^2 
\approx \frac{1 }{\mu^4 b^3 \chi^6}
\label{gradIR}
\ee
and 
\be
(\nabla S)^2  \approx -\mu^2 \chi^2 b^3 -\mu^2  b^3.
\label{gradS}
\ee
\end{subequations}
This means $\vert (\nabla_b I_R)^2 \vert / \vert (\nabla_b S)^2 \vert \approx 0 $. More particularly from \eqref{act} and \eqref{lyonsS} we have  $\vert (\nabla_{\chi} I_R)^2 \vert / \vert (\nabla_{\chi} S)^2 \vert \approx 1/ (\mu \chi b)^6 $ and 
$\vert (\nabla I_R)^2 \vert / \vert (\nabla S)^2 \vert \approx 1/ (\mu b)^6 \chi^8 $. Hence
$| (\nabla_A I_R)^2| <<|(\nabla_A S)^2| $, provided $\chi$, and hence $\phi_R (0)$, is sufficiently large (which we assumed from the outset). The complex solutions \eqref{ansol} therefore tend to solutions of the Lorentzian Hamilton-Jacobi equation along the vertical lines where both $a$ and $\phi$ are real, and they do so in the ``inflationary"  slow roll regime where $\chi$ is still large. {\rf The classicality condition is satisfied.}

Small values {\vf of the scalar field} are another regime for which analytic approximations are possible. These can be calculated by perturbation theory which is the subject of the Appendix. {\vf  There we find that in the leading approximation of vanishing scalar field the contours where $a(\tau)$ is real are exactly vertical in the complex $\tau$-plane. In the linear approximation the tuning of $\gamma$ to give a real value of $\phi(\tau)$ along this contour can be carried out explicitly with the result given in Figure \ref{gam}. }

{\vf These analytic approximations for large and small scalar field do not, of course, give us the complete ensemble of classical histories in these models.} To find the complete ensemble we now solve eqs \eqref{euceqns} numerically in a systematic manner.
Guided by the analytic solutions \eqref{ansol} we begin by taking the scalar field at the origin to be large and approximately real. {\vf Define $\phi_0$ and $\theta$ by}
\begin{equation} 
\phi(0)=|\phi(0)|e^{i\theta} \equiv \phi_0 e^{i\theta}. 
\label{defphi0}
\end{equation} 
Then with $\phi_0 \gg 1$ and $\theta$ small, we integrate \eqref{euceqns} along a broken contour $C_B(X)$  that runs along the real axis to {\vf a point}  $X$, and then up the imaginary $y$-axis. We are able to adjust both the turning point 
$X$ and the phase angle $\theta$  so that $a$ and $\phi$ tend to real functions $b(y)$ and 
$\chi(y)$ along the vertical line given by $\tau = X + iy$ in the complex $\tau$-plane. 

We find that for all $\mu$ and for {\uf each}  large $\phi_0$ there exists a {\it unique} combination $X$ and $\theta$   for which the fields become real at large $y$. Furthermore for these pairs $(X,\theta)$, the ratio of the gradients of the real to the imaginary part of the action in different directions all tend to zero at large $y$. Hence the classicality conditions \eqref{classcond} hold for this set of solutions, which therefore specifies a one-parameter set of classical histories as described in Section \ref{sec3}. It will prove convenient to label the members of this set by $\phi_0$. 
\begin{figure}[t]
\includegraphics[width=3.2in]{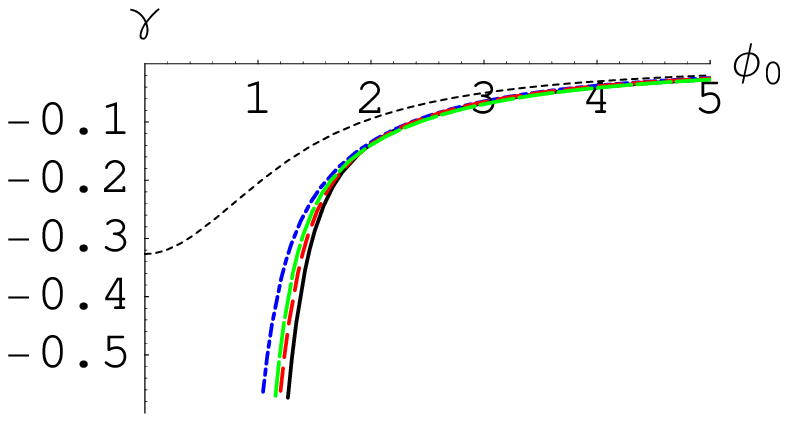} \hfill
\includegraphics[width=3.2in]{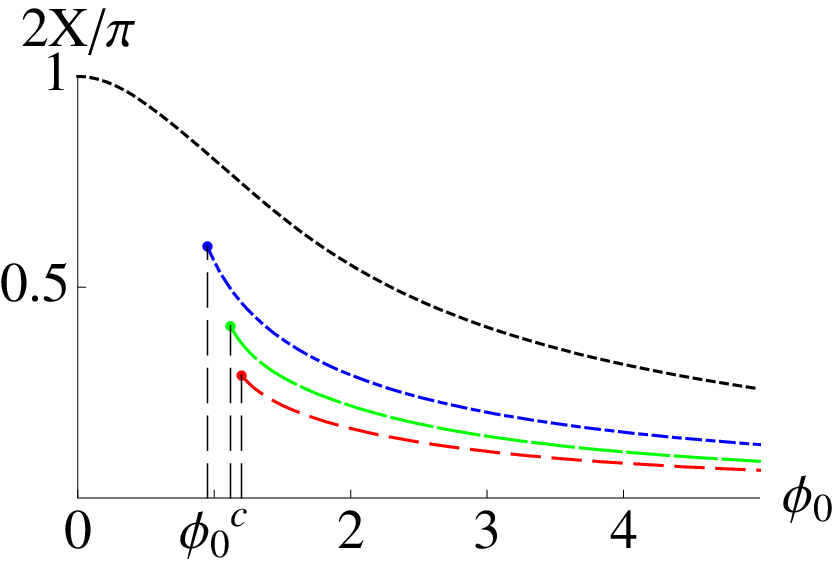}
\caption{The complex solutions that provide the steepest descents approximation to the NBWF are found by integrating the field equations along a broken contour $C_B(X)$ in the complex $\tau$-plane. In order for the solutions to behave classically at late times one ought to tune the tangent $\gamma$ of the phase of $\phi$ at the South Pole and the turning point $X$ of the contour. We show the tangent $\gamma$ (left) and the turning point $X$ (right) here as a function of the absolute value $\phi_0$ of $\phi$ at the South Pole. There is a qualitative difference between $\mu<3/2$ models on the one hand where $\gamma$ remains finite for all $\phi_0$ (dotted curve), and $\mu >3/2$ models on the other hand where $\gamma$ diverges at a critical value $\phi_0^c$ (remaining curves). In the latter case there is no combination $(X,\gamma)$ for which the classicality conditions at large scale factor hold when $\phi_0 < \phi_0^c$: the ensemble of possible classical histories is restricted to a bounded surface in phase space. The right panel shows the critical value $\phi_0^c$ increases slightly with $\mu$, for fixed $m$, and tends to $1.27$ as $\Lambda \rightarrow 0$, independently of the value of $m^2$. From top to bottom, the different curves show $\gamma$ and $X$ for $\mu=3/4,\  33/20,\ 9/4, 3$ and (in the left panel) for a scalar field model with $m^2=.05$ and $\Lambda=0$.}
\label{angle}
\end{figure}

However, when we decrease $\phi_0$ to $\phi_0 \sim {\cal O}(1)$ there is an important qualitative difference between $\mu<3/2$ models and models with $\mu>3/2$.  We illustrate this in Figure \ref{angle} where we plot the values for the turning point $X$ of the contour $C_B(X)$  and for the tangent $\gamma \equiv \tan \theta$ of $\phi$ at the SP, both as a function of $\phi_0$. The values given there have been fine-tuned so that the complex solutions behave classically at large scale factor. One sees $\gamma$ remains finite for all $\phi_0$ when $\mu<3/2$. By contrast  in all $\mu >3/2$ models --- and for quadratic scalar field potentials with $\Lambda=0$ --- we find $\gamma$ diverges as $\phi_0$ decreases to a critical value $\phi_0^c \sim {\cal O}(1)$. 
\begin{figure}[t]
\includegraphics[width=3.2in]{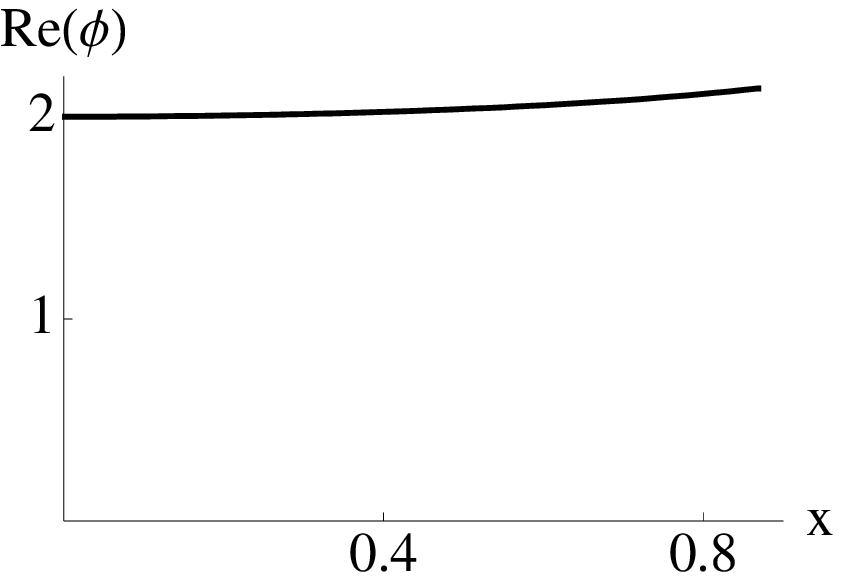} \hfill
\includegraphics[width=3.2in]{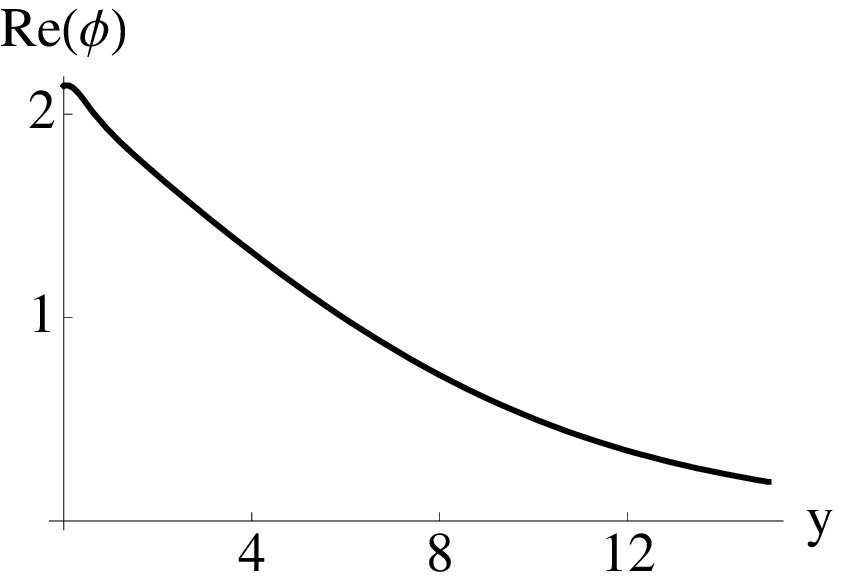} \hfill
\includegraphics[width=3.2in]{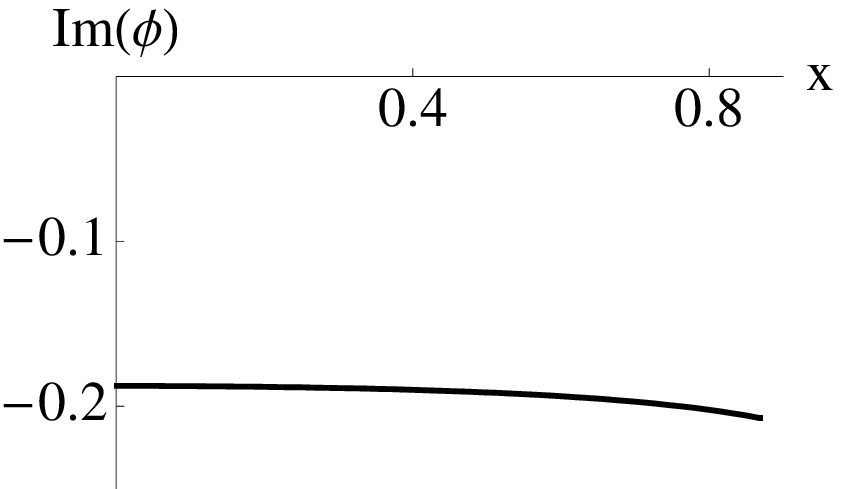} \hfill
\includegraphics[width=3.2in]{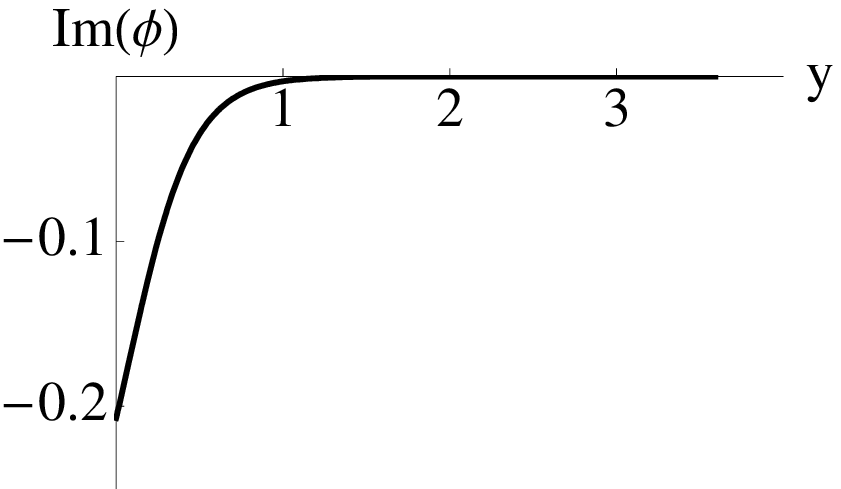} \hfill
\caption{The real and imaginary part of the scalar field $\phi$ for a typical complex solution that provides the semiclassical approximation to the NBWF. This solution has  $\mu=3/4$ and $\phi_0=2$.  
It is shown along a broken contour $C_B(X)$  in the complex $\tau=(x,y)$ plane that runs first along the $x$-axis from the South Pole at $x=0$ to a value $x=X$, and then vertically in the $y$-direction. {\rf The turning point $X$ is the largest value of $x$ plotted in the left hand two figures. } It  and the imaginary part of $\phi$ at the SP are determined by the requirement that the imaginary part of the action becomes constant with increasing $y$ (cf. Figure \ref{angle}) . This is necessary for classicality at late times and it implies that the imaginary part of $\phi$ decays rapidly to zero with increasing $y$.}
\label{ex}
\end{figure}

The classicality conditions \eqref{classcond} do not impose a constraint on $\phi_0$ when $\mu < 3/2$. 
For $\mu=3/4$, as $\phi_0 \rightarrow 0$, it is clear from Figure \ref{angle} that $\gamma \rightarrow -.32$ and $X \rightarrow \pi/2$. These limiting values agree with the predictions of the perturbation theory for small values of $\phi$ around empty de Sitter space, as  discussed in Appendix A, Fig \ref{gam}.

\begin{figure}[t]
\includegraphics[width=3.2in]{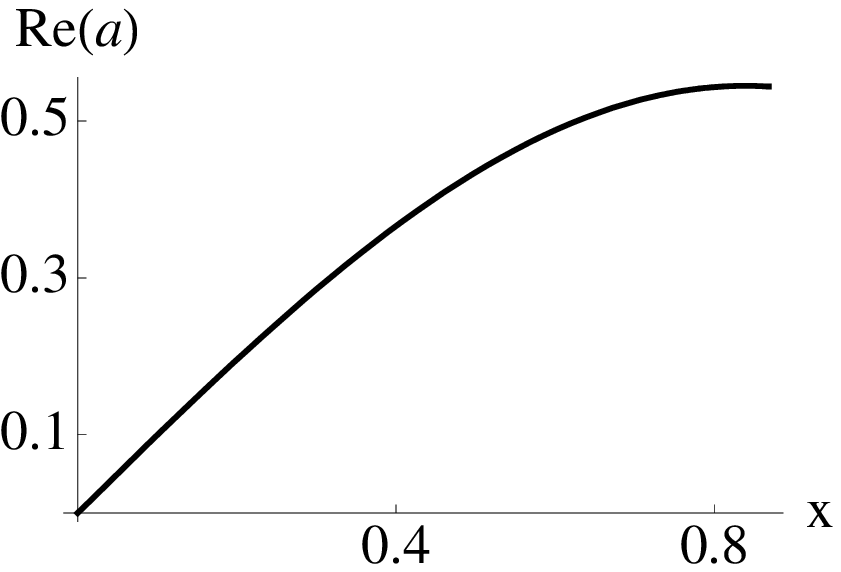} \hfill
\includegraphics[width=3.2in]{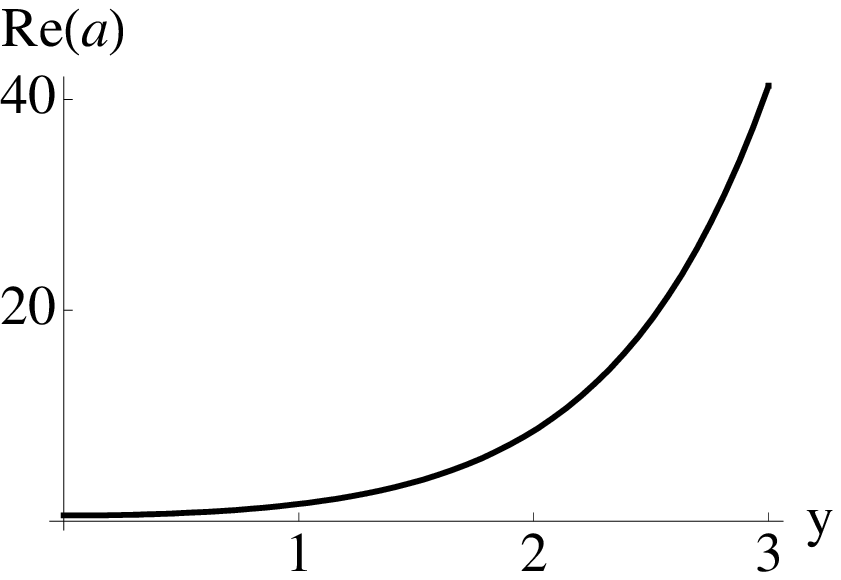} \hfill
\includegraphics[width=3.2in]{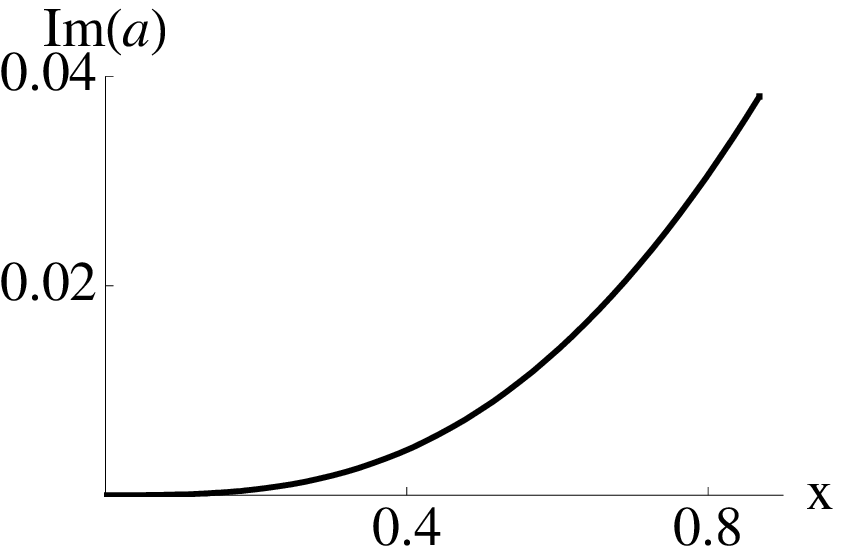} \hfill
\includegraphics[width=3.2in]{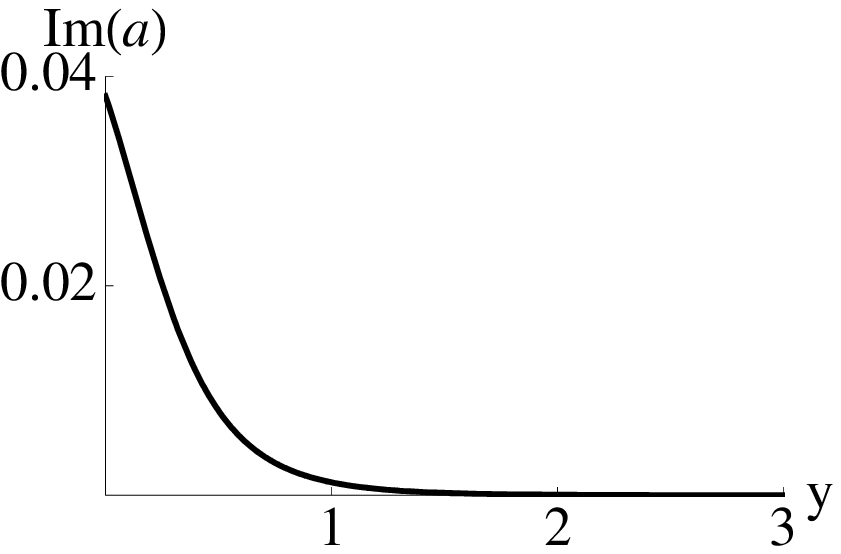} \hfill
\caption{The real and imaginary part of the scale factor for the same complex solution as in Figure \ref{ex}, along the same broken contour $C_B(X)$  in the complex $\tau$-plane.  The imaginary part of $a(\tau)$ rapidly decays to zero along the $y$-axis as a consequence of the classicality conditions, whereas the real part grows exponentially for some time.}
\label{ex2}
\end{figure}

The behavior of the scalar field $\phi (\tau)$ and the scale factor $a(\tau)$  along $C_B(X)$ are shown in Figures \ref{ex} and \ref{ex2}
for a typical complex solution that provides the semiclassical approximation to the NBWF.
The turning point $X$ is determined by the requirement that the imaginary part of the action becomes constant with increasing $y$ (cf Figure \ref{angle}). This is necessary for classicality at late times and it implies that the imaginary part of $\phi$ and $a$ decay rapidly to zero with increasing $y$. {\rf This decay can be exhibited analytically in perturbation theory, see Figure \ref{Gless}. }

The critical value $\phi_0^c$ that is present in all $\mu >3/2$ models increases slightly with $\mu$, for fixed $m$,  and tends to approximately  $1.27$ when $\Lambda \rightarrow 0$. This limiting value is  the critical value in standard scalar field models with quadratic potentials and vanishing cosmological constant (see also \cite{GR90}). We illustrate this in Fig \ref{angle} (left) where the {\sf solid} (black) curve shows $\gamma$ in a $\Lambda=0$ model with $m^2=.05$. One has $\phi_0^c=1.27$ in this model, and this is independent of the mass of the scalar field. 

It also follows from {\rf the convergence of the curves in }  Fig \ref{angle} that at large $\phi_0$, $X \sim 1/\phi_R(0)$ and $\gamma \sim -1/(\phi_R(0))^2$ independently of $\mu$. This is in agreement with the behavior \eqref{imSP} and \eqref{turningpt} for the analytic solutions \eqref{ansol}. Hence, in this regime,  only a small complex part at the SP is required to reach real $b$ and $\chi$ at late times. The metric representation of the complex geometries along the broken contours thus resemble `fuzzy instantons', with an approximately Euclidean section smoothly joined onto an approximately Lorentzian section.

\begin{figure}[t]
\begin{picture}(0,0)
\put(225,20){\Large $y$} 
\put(25,140){\Large $\vert \nabla I_R \vert^2/\vert \nabla S \vert^2$}
\end{picture}
\includegraphics[width=3.1in]{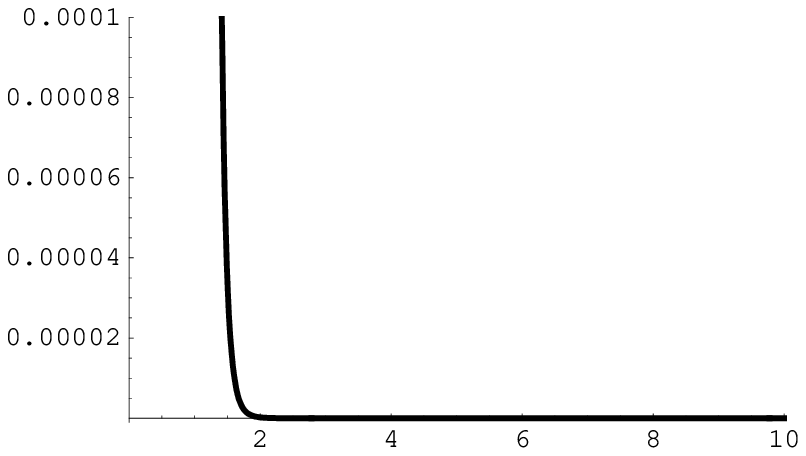}  \hfill
\includegraphics[width=3.1in]{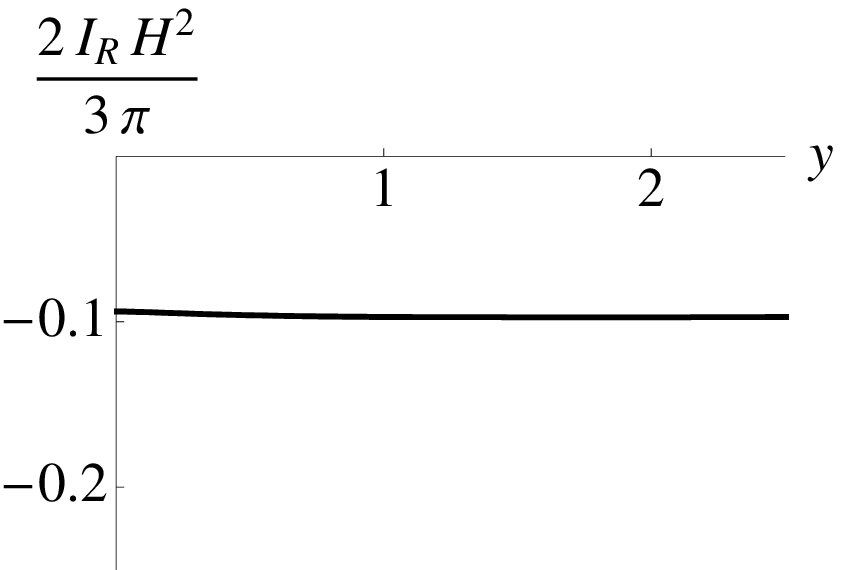}  
\caption{{\it Left panel:} The ratio of the gradient squared of the real to the imaginary part of the action {\vf plotted }along the $y$-axis, for the complex solution that behaves classically at large scale factor, with $\mu=3/4$ and $\phi_0 =2$. For $y <2$ the ratio still significantly deviates from zero, not because $I_R$ varies in the $y$-direction (as can be seen in the right panel) but because the gradient of $I_R$ in the $X$-direction is not small.  
{\it Right panel:} The real part of the action of this complex solution rapidly stabilizes along the $y$-axis.  }
\label{classconstr}
\end{figure}

Most importantly, we find numerically that in the leading semiclassical approximation there are no solutions that obey the classicality conditions \eqref{classcond} at large scale factor, other than those shown in Fig \ref{angle}.
The one-parameter set of solutions given there completely determines the ensemble of classical homogeneous and isotropic cosmologies predicted by the NBWF.  Histories other than these have zero probability in this approximaton. {\it This means that whereas the NBWF slices through the whole of phase space for $\mu <3/2$, in $\mu>3/2$ models the ensemble of possible classical histories is restricted to a surface in phase space with boundary}.

To demonstrate that the solutions given in Fig \ref{angle} satisfy the classicality conditions \eqref{classcond} we show in Figure \ref{classconstr} (left) the ratio $\vert \nabla I_R \vert^2/\vert \nabla S \vert^2$ for a typical solution for $\mu=3/4$, along the vertical line $\tau = X+iy$ where $a$ and $\phi$ tend to real functions $b(y)$ and $\chi(y)$. As discussed in Section \ref{classicality} it is necessary for classicality that this ratio tend to zero. We have also verified {\rf at a selection of points}  that the ratios of the projections of the gradients, both in the $X$ and $Y$ directions, similarly tend to zero at large $y$.
We conclude therefore that this set of solutions behaves classically at large scale factor.

\begin{figure}[t]
\includegraphics[width=5in]{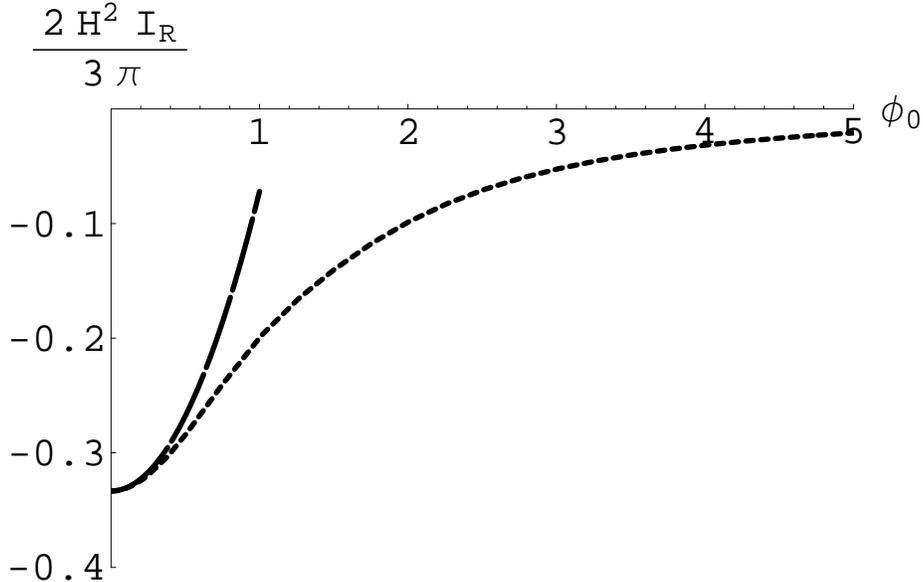}
\caption{The asymptotic value of the real part of the action of the complex solutions that behave classically at large scale factor plotted as a function of $\phi_0$ and for $\mu=3/4$. This determines the relative probabilities predicted by the NBWF for  the corresponding classical Lorentzian histories. The upper curve shows the prediction of the perturbation theory for small $\phi$ around the empty de Sitter space with cosmological constant $\Lambda$.}
\label{action-smallmu}
\end{figure}

The real part $I_R$ of the action rapidly tends to a constant along the vertical lines where the classicality condition holds. This is illustrated in Figure \ref{classconstr} (right) for a typical solution with $\mu=3/4$. This means that  along these lines the complex solutions become integral curves of $S$. {\vf They are the} classical Lorentzian histories. The value of the real part of the action provides the relative probability of the different classical histories predicted by the NBWF. 

The asymptotic value of $I_R$ is shown in Figure \ref{action-smallmu} as a function of $\phi_0$ and for $\mu=3/4$. The small $\phi_0$ solutions can be interpreted as classical perturbations of de Sitter space, and can be understood analytically in perturbation theory. The upper curve in Figure \ref{action-smallmu} shows the values for $I_R$ predicted by the perturbation theory for small $\phi$ around empty de Sitter space, discussed in Appendix A. One sees that this provides a good approximation for $\phi_0 < 3/4$. 

As mentioned earlier,  when $\mu>3/2$ there is a critical $\phi_0^c \sim {\cal O}(1)$ at which $\gamma$ diverges. Furthermore there are no solutions that obey the classicality condition \eqref{classcond} at large scale factor for $\phi_0 <\phi_0^c$. In particular, although it is still possible to tune the angle $\theta$  such that $a$ and $\phi$ are simultaneously real at some endpoint $\upsilon=X+iY$ in the complex $\tau$-plane, we find the ratio of the projection in the $X$-direction of the gradients of the real to the imaginary parts of the action is always at least of ${\cal O}(1)$. 

An analytic analysis of the complex solutions in perturbation theory for small values of $\phi$ around the empty de Sitter solution supports this conclusion: In Appendix A it is shown that for $\mu>3/2$ the real part of the action of regular complex solutions of the scalar field perturbation equation does not approach a constant along the vertical integral curves of the putative classical Lorentzian solutions. Instead it oscillates along these curves, as shown in Fig \ref{IRalongLor}. The curves along which $I_R$ is constant oscillate around some mean value $\bar X$ at large $y$ and are shown in Figure \ref{constIRXY}. Along these curves however the ratio of the gradients projected in the $\chi$ direction does {\it not} become small at large $Y$, as shown in Figure \ref{pt-cc}. Hence, at least in the semiclassical approximation, all classical histories for small scalar fields in all $\mu >3/2$ models have zero probability in the NBWF.  Even in perturbation theory the classicality condition is non-trivial.

\begin{figure}[t]
\includegraphics[width=3.2in]{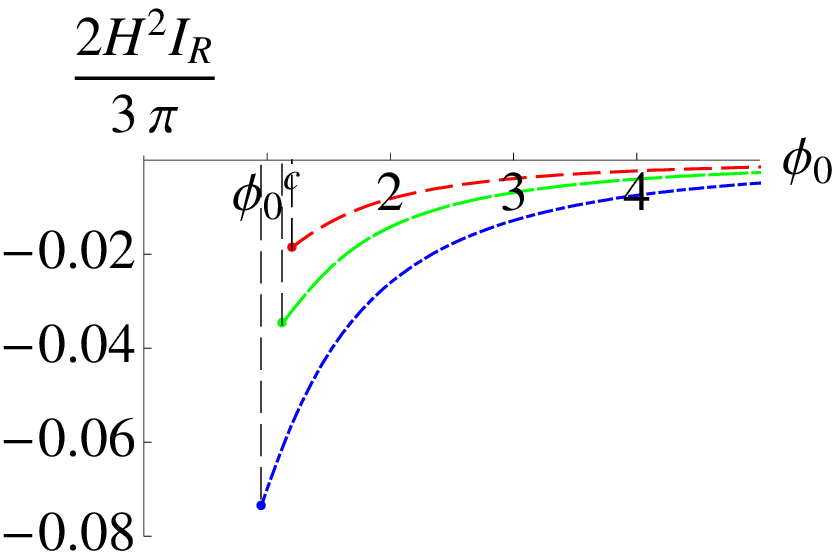}  \hfill
\includegraphics[width=3.2in]{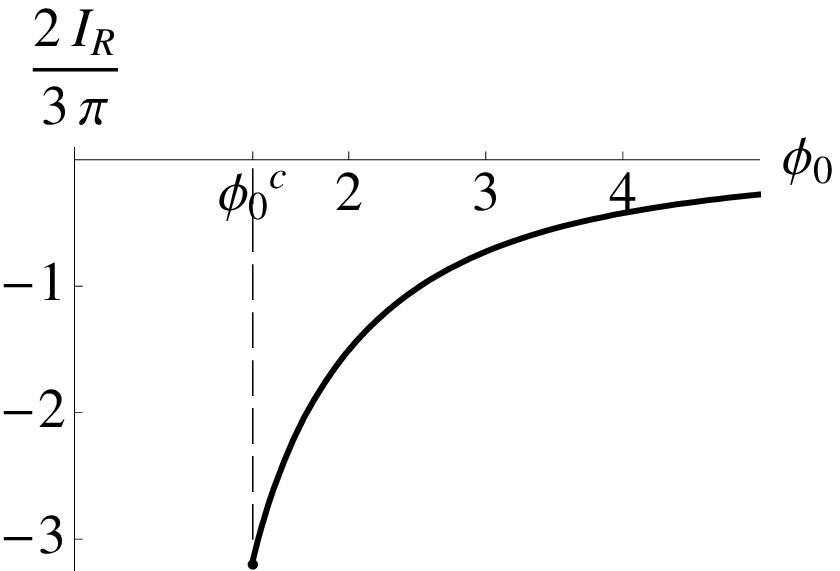}
\caption{The asymptotic value of the real part of the action of the complex solutions that behave classically at late times plotted as a function of $\phi_0$, in three $\mu >3/2$ models (left) and in a 
scalar field model with quadratic potential and $\Lambda=0$ (right). {\rf The values of $\mu$ and $m^2$} are as in Figure \ref{angle}. The action tends to a finite value at the lower bound $\phi_0^c$ that arises from the classicality condition, and it goes to zero as $\sim 1/  \phi_R^2(0)$ at large $\phi_0$. }
\label{action-largemu}
\end{figure}

For large values of the scalar field the background cosmological constant is largely irrelevant in the early universe. In this regime the {\uf complex extremizing}  solutions are qualitatively similar in all models we have considered.
The asymptotic value of the real part of the action of the complex solutions that imply classical behavior at late times is shown in Figure \ref{action-largemu} as a function of $\phi_0$  in three $\mu >3/2$ models (left) and in a $\Lambda=0$ model with a quadratic potential (right). The action tends to a finite value at the lower bound $\phi_0^c$ that arises from the classicality condition. At large $\phi_0$ it goes to zero as $ \sim 1/ \phi_R^2(0)$, in agreement with the behavior \eqref{act} that follows from the slow roll approximation. One sees that without further constraints the NBWF universally favors histories with $\phi_0$ near the lower bound $\phi_0^c$.

\section{Classical Histories}
\label{sec6}\label{classicalhistories}

The complex `fuzzy instantons' that extremize the Euclidean path integral defining the NBWF {\sf were discussed in the previous section.   They provide the probabilities for the ensemble of real Lorentzian classical histories which are the subject of this section. }

{\sf Complex extremizng solutions that predict classical behavior obey the no-boundary condition at the SP and the classicality condition \eqref{classcond} at the boundary at large scale factor. }The values of $a$ and $\phi$ together with their derivatives at the boundary provide Cauchy data for the ensemble of classical Lorentzian histories predicted by the NBWF as discussed in Section \ref{classens}. In this section we study various properties of the members of this ensemble by evolving {\vf these} Cauchy data backwards and forwards in time using the Lorentzian field equations \eqref{loreqns}. Combined with the results for the relative probabilities provided by the action of the complex solutions (e.g. Figs \ref{action-smallmu} and \ref{action-largemu})  this allows one to predict {\sf probabilities for several features of our specific universe if it is in the no-boundary state. These include the amount of inflation, whether it had an initial bounce or singularity, its future behavior, its time asymmetry if bouncing, and its consistency  with the standard cosmological model} . We will continue to label the individual classical histories in the ensemble by $\phi_0$, the absolute value of the scalar field at the SP of the corresponding complex solution.

\subsection{Inflation}

For large $\phi_0$ the complex solutions are well approximated by the analytic form  \eqref{ansol}.  {\vf Moving upwards along  the vertical contours where  $a$ and $\phi$ are real and $\phi_0 \geq 1$ one has (with $y(t)=t)$} 
\be
\hat \phi(t)= \phi(y(t) )\approx \phi_R(0) - \frac{\mu t}{3}, \qquad \hat a (t)=a(y(t))\sim e^{\mu \phi(t)  t},
\ee
until the scalar field becomes less than $\sim 1/2$. This is just like the behavior of Lorentzian {\vf slow roll}  inflationary solutions, and it {\vf shows} the classicality condition {\vf implies inflation at large 
$\phi_0$.}
\begin{figure}[t]
\begin{picture}(0,0)
\put(365,22){\Large $ \hat \phi$} 
\put(90,222){\Large $ \hat h$}
\end{picture}
\includegraphics[width=5in]{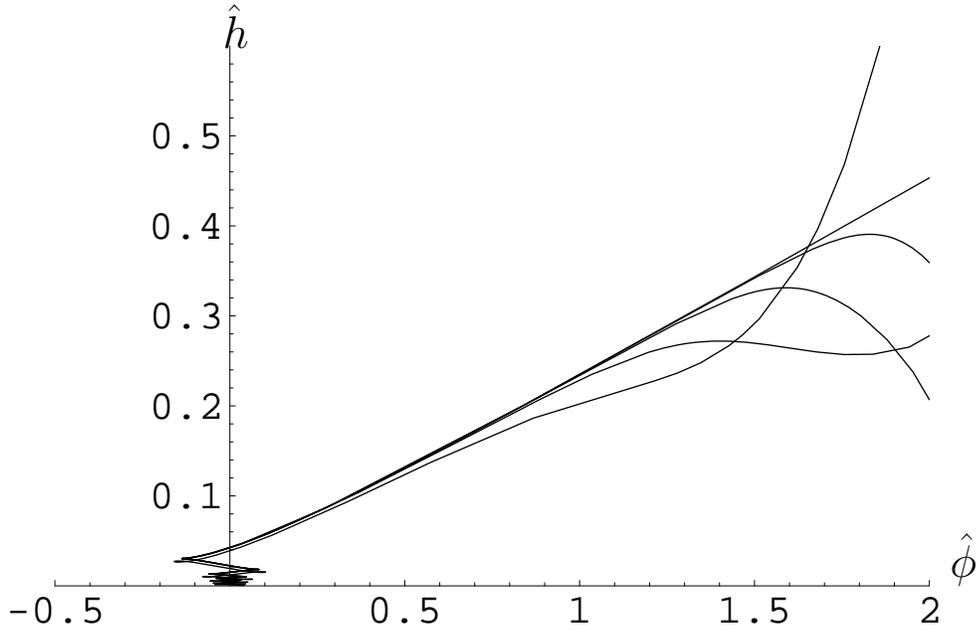} 
\caption{The no-boundary wave function predicts that all histories that behave classically at late times undergo a period of inflation at early times as shown here by the linear growth of the instantaneous Hubble constant $\hat h=\hat a_{,t}/\hat a$ in five representative classical histories for $\mu=3$ and for $\phi_0$ between 1.3 and 4.}
\label{inflation}
\end{figure}

This connection is general. Figure \ref{inflation} shows the trajectories of several {\vf numerically calculated }histories in $(\hat h,\hat \phi)$ variables, where $\hat h(t)$ is the instanteous Hubble constant $\hat h=(d\hat a/dt)/{\hat a} \equiv \hat a_{,t}/\hat a$. Five representative members of the ensemble of classical histories for $\mu=3$ and for $\phi_0$ between 1.3 and 4 are shown. When we follow the histories back in time to higher values of $\hat h$ and $\hat\phi$, they {\it all} lie within a very narrow band around $\hat h = \mu \hat \phi$. This is {\sf characteristic of Lorentzian} slow roll inflationary solutions. Furthermore, since the numerical analysis shows that there are no solutions other than those given in Fig \ref{angle}, and represented in Fig {\ref{inflation}, we conclude that  the NBWF predicts {\vf that}  a classical homogeneous and isotropic universe {\it must have}  an {\sf early} inflationary state. {\it The NBWF and  classicality at late times imply inflation at early times.} This conclusion holds for all values of $\mu$. 

{\vf  Although the classicality condition implies inflation for all values of $\mu$ the drivers of inflation are different for different values.}
For $\mu<3/2$ and small $\phi_0$ inflation is {\sf always} driven by the background cosmological constant. In all other models however, and for small $\mu$ at large $\phi_0$, inflation in the early universe is driven by the scalar field potential energy. {\sf By this we mean specifically that $\hat a_{,tt} >0$ when $\hat \phi >.5$, which we find is the minimum value of $\hat \phi$ required for inflation to occur in $\Lambda=0$ models with quadratic potentials and small $m^2$.}
\begin{figure}[t]
\includegraphics[width=3.2in]{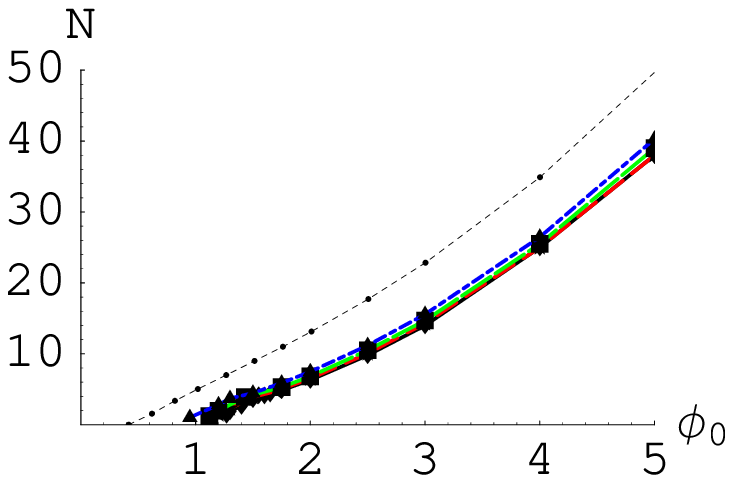} \hfill
\includegraphics[width=3.2in]{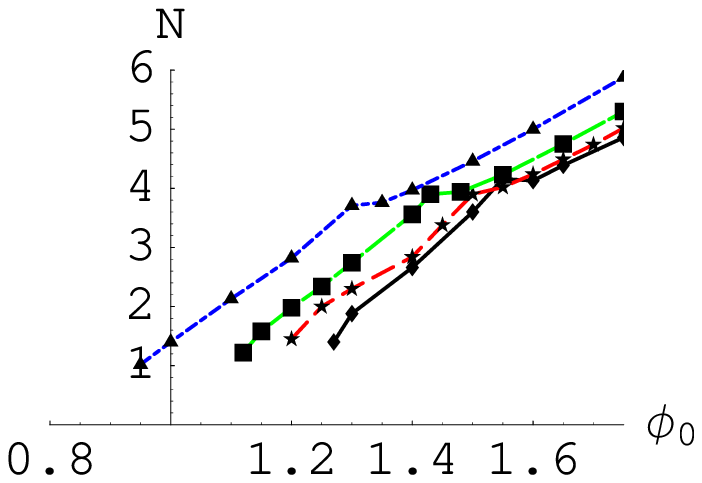}
\caption{{\it Left panel}: The number of efoldings $N$ of inflation driven by the scalar field (as opposed to the cosmological constant of the background) in the classical histories predicted by the NBWF, for five different models. From top to bottom, the different curves correspond to $\mu=3/4,\ 3,\ 9/4,\ 33/20$ and 
$\Lambda=0, m^2=.05$.
{\it Right panel:} Detail of the left panel showing the regime around the critical value $\phi_0^c$ in the $\mu>3/2$ and pure scalar field models. The lower bound on $\phi_0$ that arises from classicality implies a {\it lower bound} on the number of efoldings.}
\label{efolds}
\end{figure}

 To get a quantitative measure of the amount of  inflation predicted we calculated the number of efoldings $N \equiv \int  \hat h dt $  of scalar field driven inflation over the range of time where  $\hat a_{,tt} >0$ and $\hat \phi \geq .5$ for the members of the ensemble of Lorentzian histories predicted by the complex solutions found in Section 5. }The results are summarized in 
Figure \ref{efolds}. For $\mu >3/2$ the lower bound on $\phi_0$ arising from the classicality condition implies the number of efoldings is always greater than one\footnote{For initially singular solutions inflation generally does not begin immediately at the singularity, {\rf see e.g. Figure \ref{Lor-ex}}.}. It follows from Figure \ref{efolds}, combined with the information on the relative probability of the histories given in Fig \ref{action-smallmu} and Fig \ref{action-largemu}, that {\vf by itself} the no-boundary wave function  favors a small number of efoldings on a history by history relative probability basis. The answers to more physical questions involving probabilities {\sf conditioned on the data in our past light cone}  are obtained from the no-boundary probabilities by summing {\sf them} over {\sf those}  for classical spacetimes that contain our data at least once, and over the possible locations of our light cone in them. This {\sf sum} can significantly change the no-boundary predictions {\rm based on the wave function alone} \cite{Hawking07,HHH07}. We return to this point  in Section \ref{volumefactors}.

\subsection{Bounces and Initial Singularities}

\begin{figure}[t]
\begin{picture}(0,0)
\put(222,12){\Large $\phi_0$} 
\put(25,140){\Large$\hat a_b$}
\put(463,12){\Large$\phi_0$} 
\put(262,140){\Large$\hat \phi_b$}
\put(73,0){\Large$\phi_0^s$}
\put(315,0){\Large$\phi_0^s$}
\end{picture}
\includegraphics[width=3in]{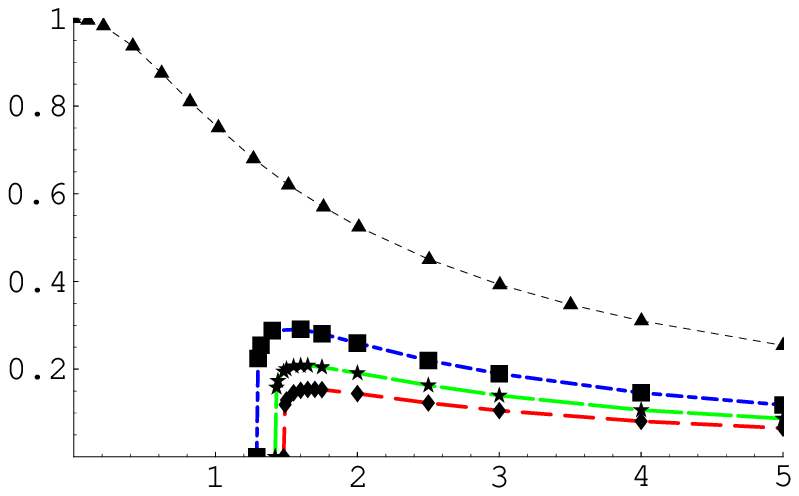} \hfill
\includegraphics[width=3in]{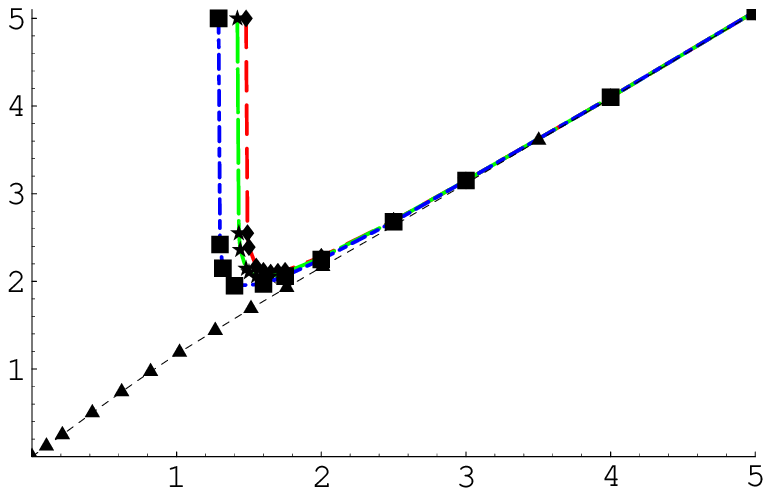}
\caption{The scale factor $\hat a_b$ (left) and the scalar field $\hat \phi_b$ (right) at the bounce of the classical histories predicted by the NBWF. {\rf The values of $\mu$ and $\Lambda$} are as in Figure \ref{efolds}. 
When $\mu <3/2$ the histories always bounce at a minimum radius in the past. By contrast for $\mu >3/2$ there is a transition from bouncing to initially singular at a critical value $\phi_0^s \approx 1.5$. {\rf Above that $\hat\phi_b \approx \phi_0$. }  }
\label{bounce}
\end{figure}

\begin{figure}[t]
\begin{picture}(0,0)
\put(65,160){\Large B} 
\put(65,145){\Large O} 
\put(65,130){\Large U} 
\put(65,115){\Large N} 
\put(65,100){\Large C} 
\put(65,85){\Large E} 
\put(130,160){\Large {classical bounce}} 
\put(130,133){\Large {initial singularity}}
\put(130,90){\Large {no classical solutions}}
\thicklines
\put(100,23){\line(0,1){170}}
\end{picture}
\includegraphics[width=5in]{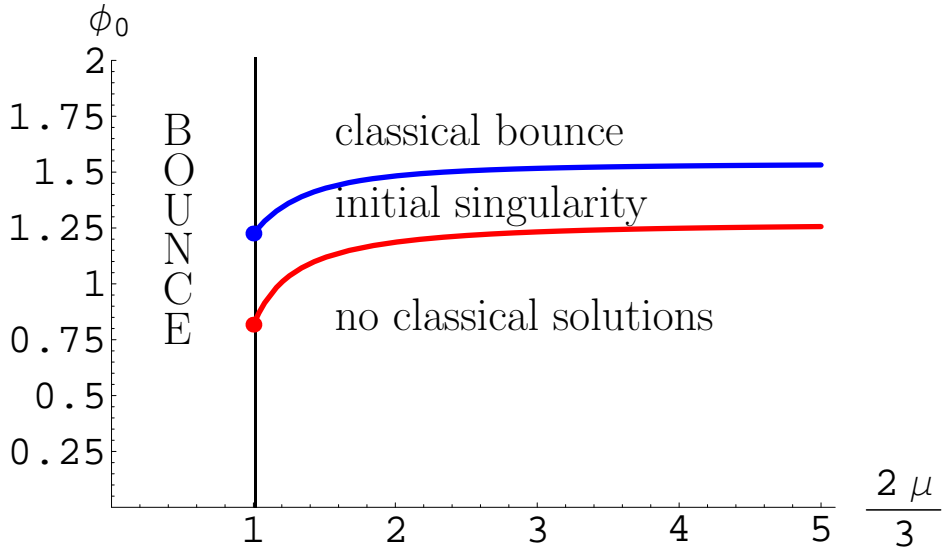} 
\caption{A `phase diagram' that summarizes some of the early universe properties of the Lorentzian histories predicted by the NBWF. For $\mu <3/2$ there is one classical history associated with each value of $\phi_0$ and the universe bounces in the past for all ranges of $\phi_0$. For $\mu >3/2$, however, there are no classical histories for $\phi_0 <\phi_0^c$ (bottom curve). Between $\phi_0^c$ and $\phi_0^s$ (top curve) the Lorentzian histories have an initial singularity. Finally, for $\phi_0 > \phi_0^s$ the Lorentzian solutions bounce at a nonzero minimum radius in the past. The limiting values of $\phi_0^c$ and $\phi_0^s$ when $\Lambda \rightarrow 0$, for fixed $m$, are  1.27 and 1.54 respectively.}
\label{phase}
\end{figure}

For $\mu>3/2$ and $\phi_0^c \leq \phi_0 \leq \phi_0^s$ the allowed classical histories of the universe are singular at an initial time $t_s$. Near the singularity both the potential and the curvature are unimportant in the Einstein equations \eqref{loreqns}, and one has $\hat a(t) \sim (t-t_s)^{1/3}$ and $\hat \phi (t) \sim \ln (t-t_s)$. But for $\phi_0 > \phi_0^s$ the histories bounce at a finite radius $\hat a_b$ in the past. The critical value $\phi_0^s$ at which there is a transition from initially singular to bouncing is determined entirely by the Einstein equations, and therefore independent of $H$. A bounce at a finite radius in the past is possible despite the singularity theorems because a scalar field and the cosmological constant violate the strong energy condition. Even though such non-singular {\rf classical} solutions form only a small subset of all scalar field gravity solutions they have significant probability in the no-boundary state. Near a bounce the universe approaches a de Sitter state with radius $\sim (H \mu \hat \phi_b)^{-1}$ {\rf where $\hat\phi_b$ is the value of the scalar field at the bounce.} For sufficiently large $\phi_0$,  $\hat \phi_b \approx \phi_0$, as shown in Fig \ref{bounce} (right panel). The scale factor at the bounce versus $\phi_0$ is shown in Fig \ref{bounce} (left panel), which clearly reveals the transition from bouncing {\rf solutions} to initially singular {\rf ones}. The critical value $\phi_0^s$ itself is shown as a function of $\mu$ in Fig \ref{phase}. One sees $\phi_0^s$ slightly increases with $\mu$ asymptoting to $\approx 1.54$ as $\Lambda \rightarrow 0$, for fixed $m$. As discussed  for $\phi_0^c$ above, this limit corresponds to the critical value that separates the bouncing {\sf from}  singular histories in pure scalar field models with quadratic potentials for any nonzero mass $m^2 <1$.


When we evolve the data provided by the complex solutions on the matching surface backwards in time we find that,  for $\mu <3/2$, {\vf all  the histories  bounce at a minimum radius in the past.}  Hence, in this  regime,  the classicality conditions at late times select a set of histories in the NBWF where either the potential energy of the scalar field or the background cosmological constant dominate the evolution at early times.
\begin{figure}[t]
\includegraphics[width=3.2in]{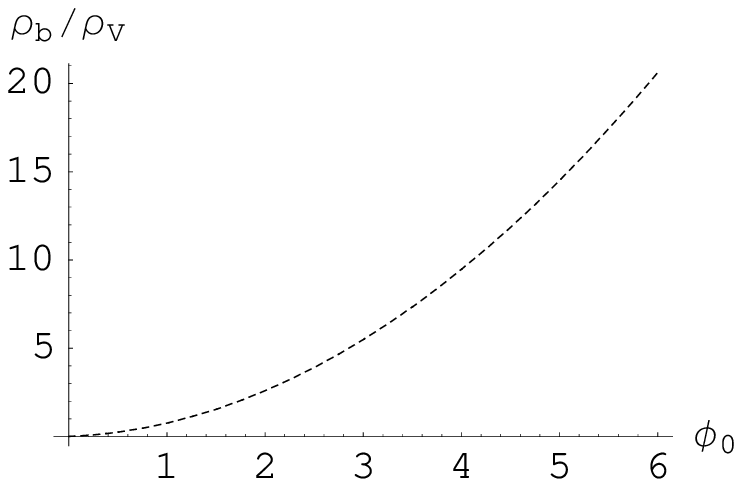} \hfill
\includegraphics[width=3.2in]{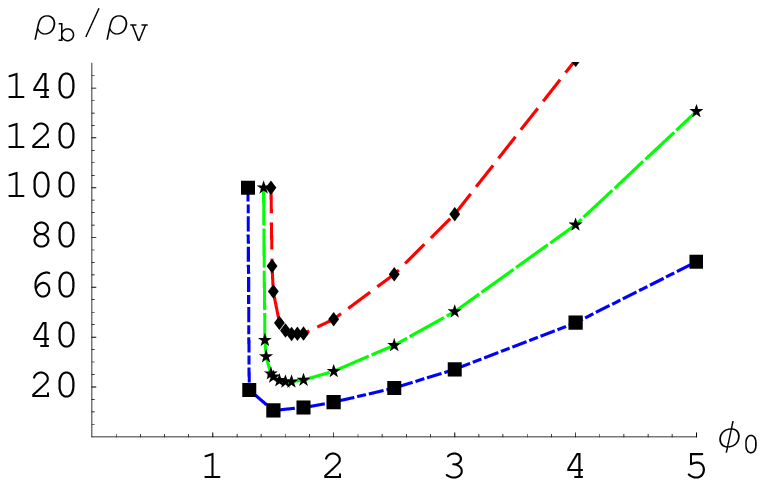}
\caption{The ratio of the energy density $\rho_b$ in the scalar field at the bounce over the vacuum energy density $\rho_V$, for the allowed Lorentzian histories for $\mu=3/4$ (left) and for (right, from top to bottom) $\mu=3,\ 9/4$ and $33/20$. When $\mu>3/2$ there is a minimum matter density needed at early times for the history to exhibit classical behavior at late times}
\label{density}
\end{figure}

Figure \ref{density} shows the ratio of the scalar field energy density \eqref{rho} over the vacuum energy density at the bounce, again as a function of $\phi_0$. Since $\hat \phi_b \approx \phi_0$ for most $\phi_0$, (Figure \ref{bounce}) this generally grows quadratically with $\phi_0$. When $\mu<3/2$ we find classical histories for all ranges of densities, whereas for $\mu >3/2$ there is a minimum matter density needed for the history to exhibit classical behavior at late times (except for the vacuum de Sitter solution). {\it For realistic values of $\Lambda$ and $\mu$, therefore, a nearly empty de Sitter solution has zero probability in the semiclassical approximation to the NBWF.}

\begin{figure}[t]
\begin{picture}(0,0)
\put(100,160){\Large {classical bounce/expansion forever}} 
\put(100,130){\Large {initial singularity/expansion forever}}
\put(160,87){\Large {initial singularity/recollapse}}
\put(100,50){\Large {no classical solutions}} 
\end{picture}
\includegraphics[width=5in]{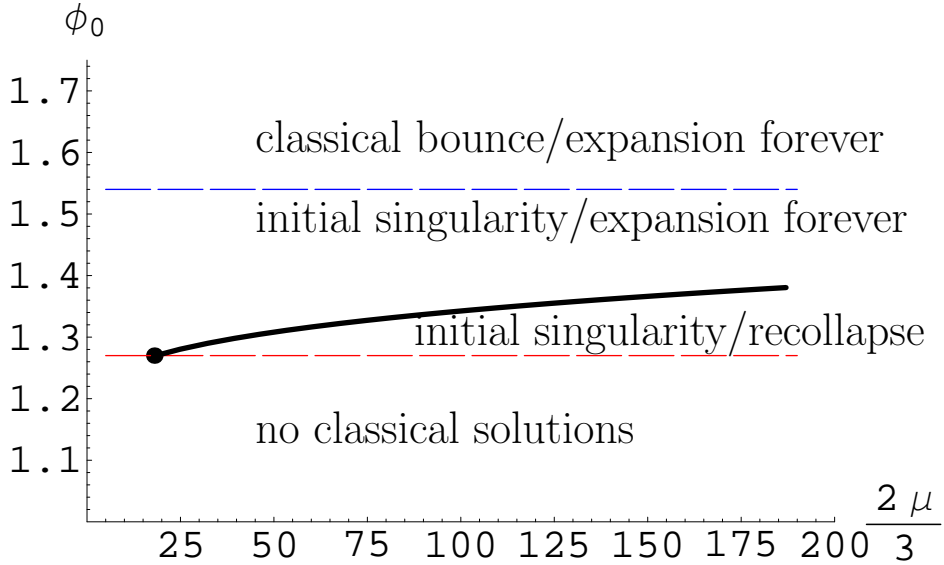}
\caption{A `phase diagram' that combines some of the early and late time properties of the Lorentzian histories predicted by the NBWF.  Below the {\sf solid} curve the homogeneous isotropic classical universes predicted by the NBWF recollapse to a Big Crunch, whereas above the curve the universes continue to expand forever.}
\label{phase2}
\end{figure}

\subsection{Eternal Expansion and Final Singularities}
Next we explore the late time properties of the classical histories. Evolving the data provided by the complex solutions on the matching surface forward in time, the universe either expands forever or recollapses again to a singularity\footnote{\rf We did not find any Lorentzian solutions with multiple bounces \cite{Staro77,Pag84}. However, our numerical search procedure identifies discrete solutions and may miss such highly fine tuned examples.}.  We find that when $\mu <18$ the histories expand forever for all ranges of $\phi_0$ that admit classical histories. In these universes, when the scalar field rolls down the potential, the cosmological constant of the background takes over to drive the exponential expansion. When $\mu >18$, however, there is critical value $\phi_0^r$, and for $\phi_0 < \phi_0^r$ the universe recollapses. We plot $\phi_0^r$ as a function of $\mu$ in Fig \ref{phase2}, which shows that  this slowly increases with $\mu$.

\begin{figure}[t]
\includegraphics[width=3.2in]{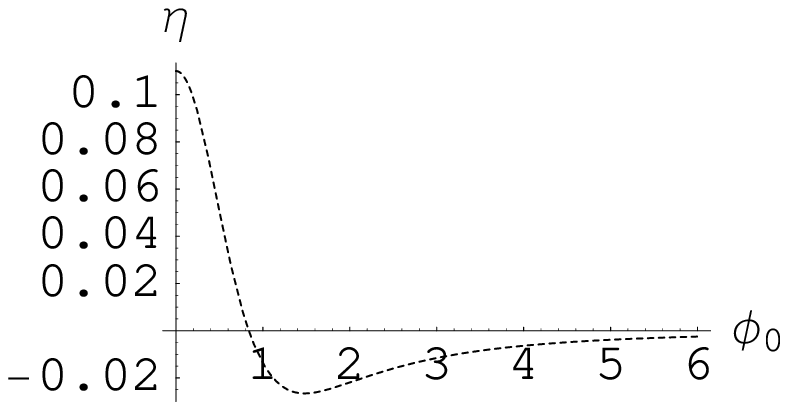} \hfill
\includegraphics[width=3.2in]{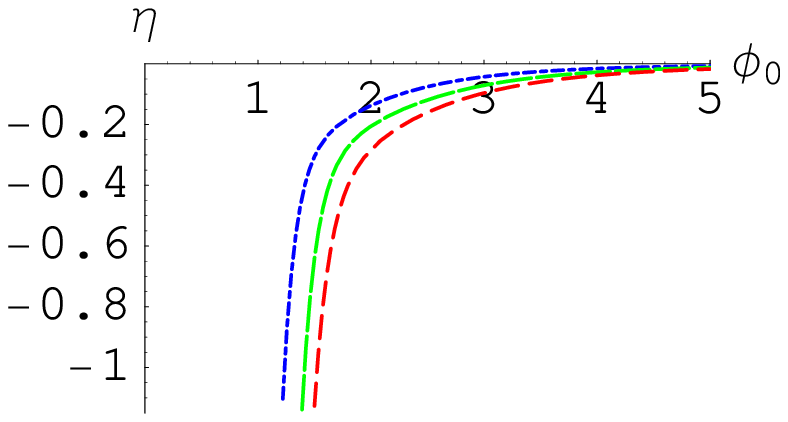}
\caption{The reality of the NBWF implies the ensemble of allowed classical histories is time-symmetric. The individual classical bouncing histories, however, are generally not time-symmetric about the bounce. A natural measure of the amount of time-asymmetry of the individual histories is provided by the quantity $\eta = (\hat \phi_{,t})_b/\hat \phi_b$. The left panel shows $\eta$, as a function of $\phi_0$, for $\mu=3/4$. In the right panel we plot $\eta$ for (from top to bottom) $\mu=33/20,\ 9/4$ and $3$.  One sees that for $\mu>3/2$, $\eta$ diverges when $\phi_0 \rightarrow \phi_0^s$ {\vf where $\phi_0^s$  separates singular from bouncing solutions.}}
\label{eta}
\end{figure}

\subsection{Time Asymmetry}

\begin{figure}[t]
\begin{picture}(0,0)
\put(192,10){\Large$t$} 
\put(20,110){\Large$\hat a$}
\put(305,115){\Large$\hat \phi$} 
\put(470,60){\Large$t$}
\put(188,-105){\Large$t$} 
\put(20,-10){\Large$\hat a$}
\put(310,-5){\Large$\hat \phi$} 
\put(468,-55){\Large$t$}
\put(187,-228){\Large$t$} 
\put(22,-125){\Large$\hat a$}
\put(293,-125){\Large$\hat \phi$} 
\put(465,-227){\Large$t$}
\put(184,-343){\Large$t$} 
\put(95,-244){\Large$\hat a$}
\put(375,-247){\Large$\hat \phi$} 
\put(467,-338){\Large$t$}
\end{picture}
\includegraphics[width=2.6in]{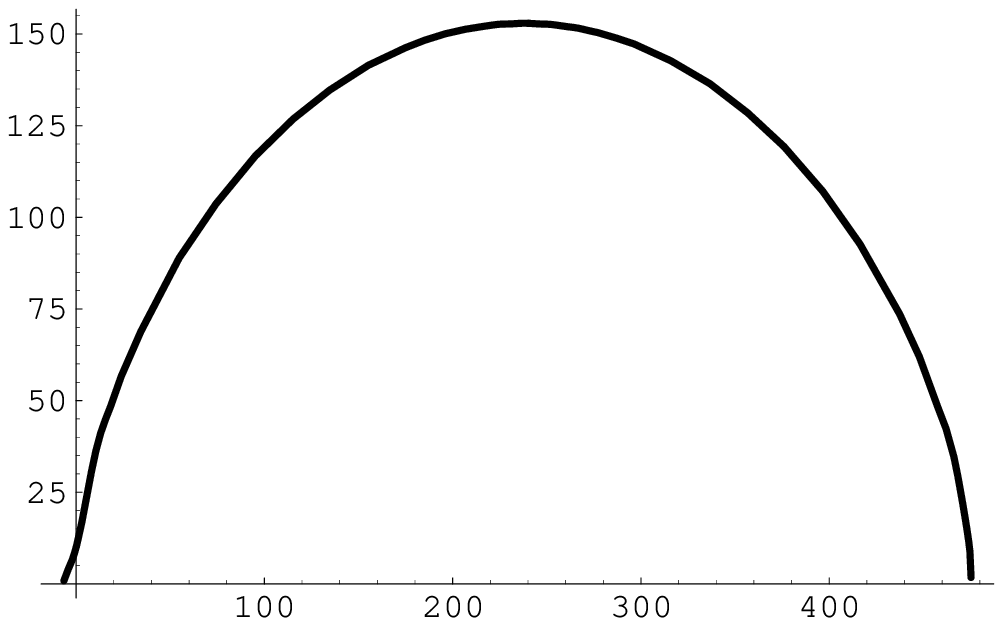} \hfill
\includegraphics[width=2.6in]{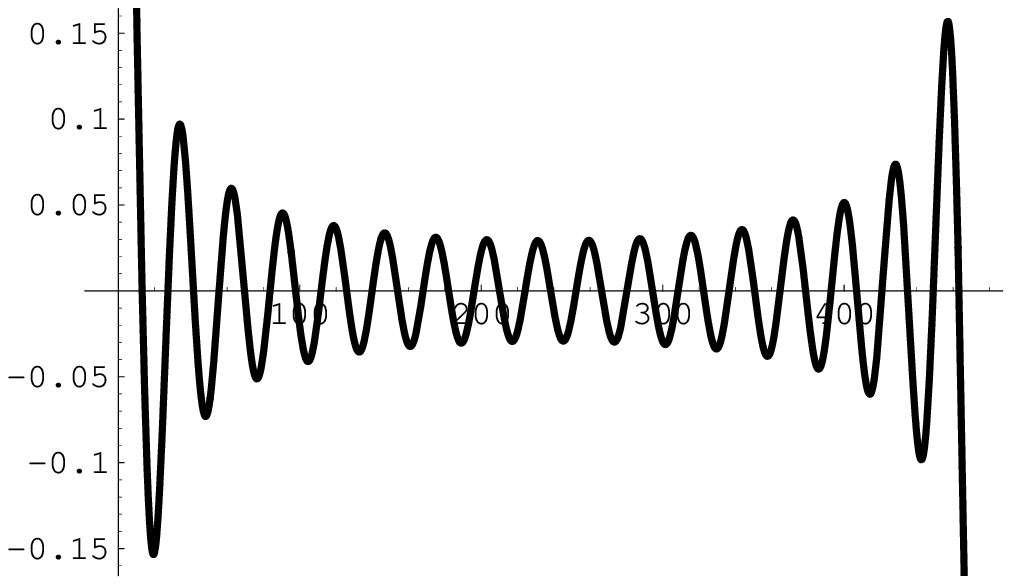} \hfill
\includegraphics[width=2.6in]{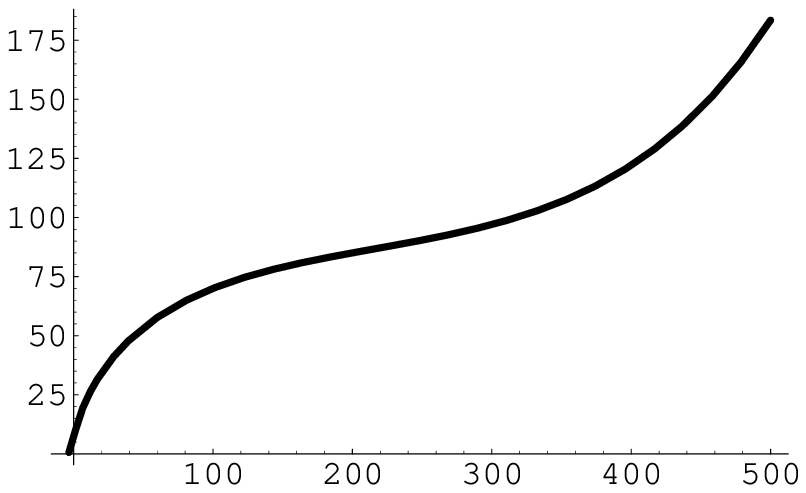} \hfill
\includegraphics[width=2.6in]{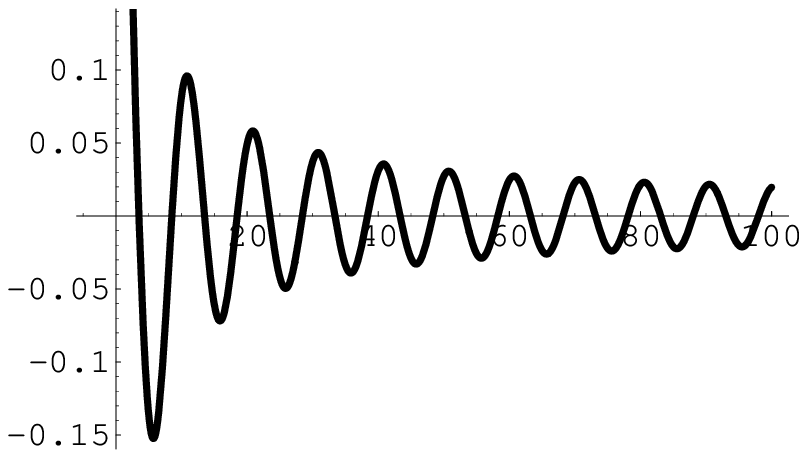} \hfill
\includegraphics[width=2.6in]{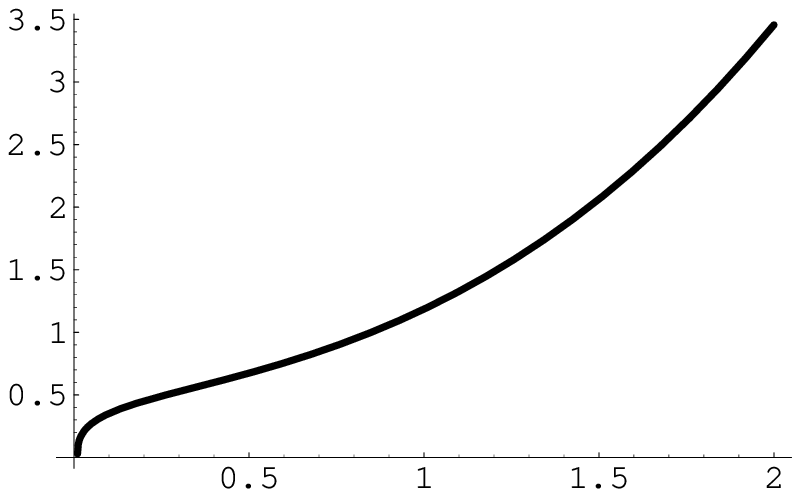} \hfill
\includegraphics[width=2.6in]{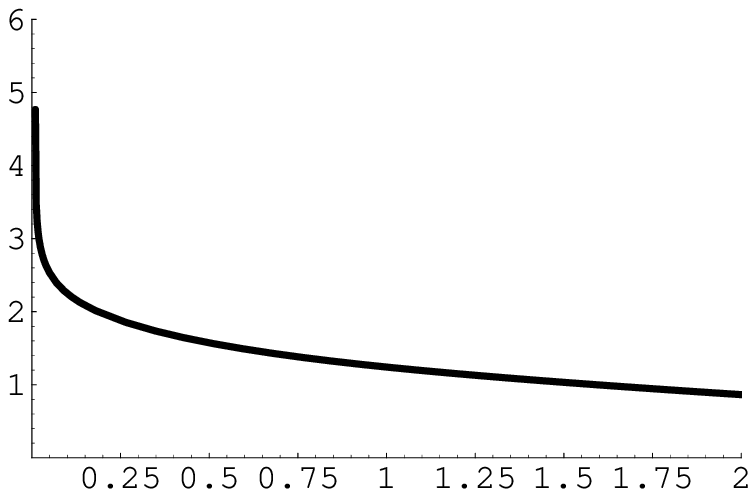} \hfill
\includegraphics[width=2.6in]{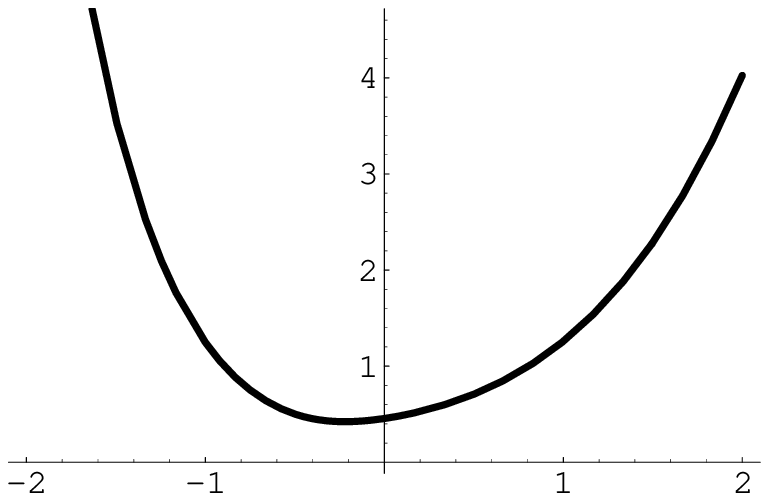} \hfill
\includegraphics[width=2.6in]{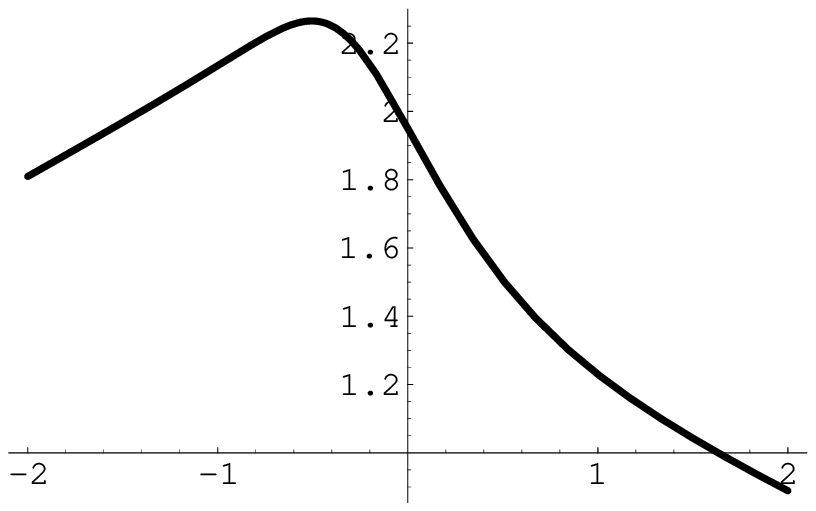}
\caption{{\rf A gallery of classical Lorentzian histories.} The scale factor $\hat a(t) $ (left) and the scalar field $\hat \phi(t)$ (right) in the classical histories labeled by $\phi_0=1.32$, for four different values of $\mu$. The value of $\mu$ decreases from top to bottom taking the values   $\mu=100, 96, 9/4, 33/20$. {\rf If $m^2$ is fixed these correspond to increasing $\Lambda$. }  These are four qualitatively different cosmologies, ranging from initially singular histories that recollapse again (top) to eternally expanding universes that bounce in the past  (bottom). At intermediate values of $\mu$ (2nd row) the NBWF is consistent with the standard picture of inflationary cosmology, consisting of a short period of inflation as the scalar field rolls down from high up the potential followed by an era of  {\rf oscillation representing particle creation and ensuing matter domination. Eventually}  the cosmological constant takes over to drive a second (future-eternal) phase of exponential expansion.}
\label{Lor-ex}
\end{figure}

 Bouncing classical histories are generally time-asymmetric about the bounce. This can be seen in perturbation theory (e.g. Fig \ref{pt_lor}) and a particular non-perturbative case is the fourth example in Figure \ref{Lor-ex}.  A natural measure of the time asymmetry of histories at the bounce is given by 
 \begin{equation}
 \eta \equiv(\hat \phi_{,t})_b/\hat \phi_b.
 \label{etadef}
 \end{equation} 
  We plot $\eta$, as a function of $\phi_0$ in Figure \ref{eta}, for $\mu=3/4$ (left) and for several values of $\mu>3/2$ (right). In the former model, the limiting value $\eta \approx .11$ for $\phi_0 \rightarrow 0$ agrees with the prediction of the perturbation theory for small $\phi$, which we obtain in Appendix A. One sees that, for $\mu >3/2$, $\eta$ diverges when $\phi_0 \rightarrow \phi_0^s$ --- the boundary between singular and bouncing histories.  For large $\phi_0$. $\eta \rightarrow 0$ in all models.

At the current level of our analysis in which we restrict attention to homogeneous isotropic minisuperspace models it is not clear whether the time-asymmetry of the bouncing classical histories in the NBWF has any physical (observable) effects. But one might expect observable signatures of the time asymmetry to show up in the spectrum of inhomogeneous perturbations. We intend to calculate these in future work. We emphasize also that although individual classical bouncing histories are not generally time-symmetric about the bounce, the reality of the NBWF implies the ensemble of allowed classical histories is time-symmetric. For every history in this ensemble, its time reversed is also a member.

\subsection{Cosmological Models}
We next turn to particular kinds of Lorentian histories. 
A gallery of {\sf qualitatively different classical histories for one value of $\phi_0$} is exhibited in Figure \ref{Lor-ex}.
A class of histories that are particularly interesting from an observational point of view are represented by points that lie just above the curve of $\phi^r_0$ vs $\phi_0$ {\vf over a range of } $\mu$  in Figure \ref{phase2}. These turn out to correspond to universes that undergo an early period of inflation that is followed by an era of oscillating scalar field {\vf that will lead to matter generation and domination.  Eventually  the} cosmological constant takes over to drive a second phase of exponential expansion that lasts forever. The NBWF, therefore, appears to be consistent with the standard picture of inflationary cosmology {\vf for our universe in which} a scalar field  rolls down from high up the potential and subsequently oscillates around the minimum losing its energy into {\vf created}  particles. An example of a history of this kind is given in Figure \ref{Lor-ex} (2nd row). 

Further increasing $\mu$ for fixed $\phi_0$ yields a qualitatively different universe. For $\mu >98$ the $\phi_0=1.32$ histories lie below the {\rf solid} curve in Fig \ref{phase2}. These universes have an initial singularity and recollapse again to a Big Crunch. An example of a universe of this kind is given in Fig \ref{Lor-ex}, 1st row. The third example in Fig \ref{Lor-ex} shows the (initially singular) $\phi_0=1.32$ history for $\mu=9/4$, where the cosmological constant immediately takes over to drive the expansion when the scalar field has rolled down its potential. Inflation never really ends in this universe.

\section{Volume Weighting}
\label{volumefactors}\label{sec8}

The NBWF gives the probabilities of entire classical histories. But we are interested in probabilities that refer to our data, which are limited to a part of our past light cone. Among these are the top-down probabilities for our past conditioned on (a subset of) our present data \cite{HH06}. Hawking \cite{Hawking07} and the current authors \cite{HHH07} have argued that  {\qf in homogeneous models } these are obtained by multiplying the NBWF probabilities for classical histories by a factor $\exp(3N)$ proportional to the volume of the hypersurface on which our data approximately lie\footnote{Anthropic reasoning has also been used as an argument to include a volume factor \cite{Page97}.}. This multiplication can be understood as resulting from a sum over the probabilities for classical spacetimes that contain our data at least once, and over the possible locations of our light cone in them \cite{HHH07}. In a large universe there are more places for our data to be. 

\begin{figure}[t]
\includegraphics[width=3in]{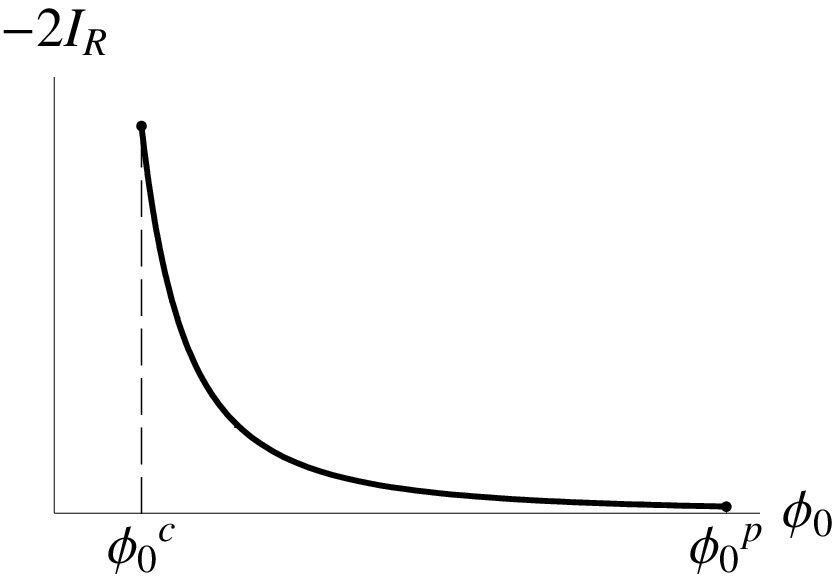} \hfill
\includegraphics[width=3in]{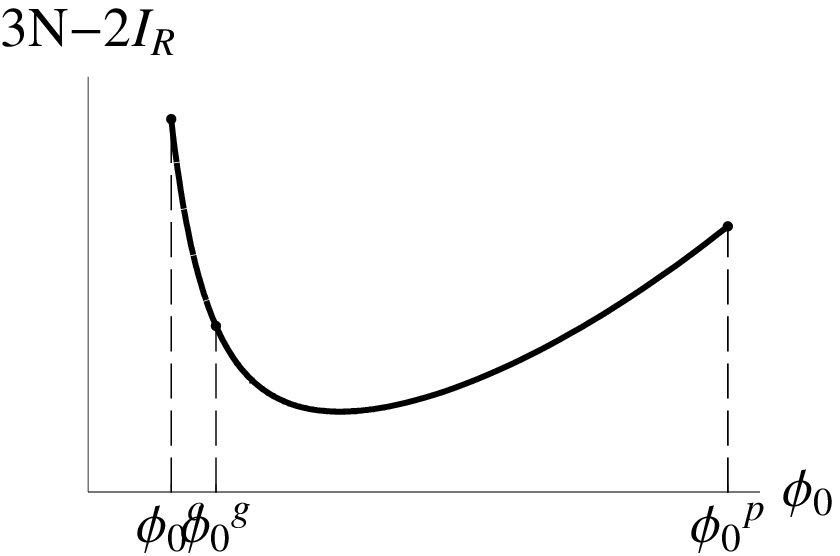}
\caption{To account for the different possible locations in the universe of the Hubble volume that contains our data, one ought to multiply the relative probabilities for classical histories coming from the NBW (left) by a volume factor, to obtain the probability (right) for what we observe in our past light cone. 
The resulting volume-weighted probability distribution favors a large number of efoldings when the ensemble is restricted to universes that last sufficiently long.}
\label{volume}
\end{figure}

In order for the volume weighted probabilities to be physically meaningful as probabilities relevant for what we observe the universe must  obviously last {\qf to the present age of 14 Gyr.} This further restricts the ensemble of histories, requiring $\phi_0$ to be larger than a critical value $\phi_0^g \sim 2$ or, equivalently, $N \geq 5$.

Figure \ref{volume} shows the {\it qualitative} effect in $m^2 \phi^2$ models of multiplying the relative probabilities for classical histories  $\exp(-2I_R)$  coming from the NBWF by a volume factor $\exp(3N)$.  Volume weighting clearly enhances the probability for a large number of efoldings. An important feature of the volume weighted probability distribution is that there is a wide region where the probability is strongly increasing with $N$. Indeed when one considers the probability distribution $ \sim \exp(3N-2I_R)$, as a function of the  {\qf  value $\phi=\phi_i$ at which inflation starts}, the gradient of this probability distribution is positive provided
\be \label{eternal}
V^3 \geq \vert V_{,\phi} \vert ^2.
\ee
For quadratic potentials this condition is satisfied well below the Planck density.

For a realistic value of $m$ Figure \ref{volume} shows qualitatively  that the two constraints of classicality and minumum age yield a restricted ensemble of histories whose volume weighted probabilities slightly favor a large number of efoldings. This can be understood analytically. Since $-I_R \sim 1/(m \phi_R(0))^2\sim 1/(m\phi_0)^2$ for the slow roll solutions predicted by the NBWF, and since $\phi_i \approx \phi_R(0)\approx \phi_0$, the volume factor $\exp(3N)$ is comparable to the no-boundary weight $\exp(-2I_R)$ for $\phi_i=1/m$, i.e. for  solutions that start inflating near the Planck density. Hence the volume weighted probability distribution is peaked both at low $\phi_0$ and for solutions that start inflating near the Planck density. The latter peak slightly dominates when the constraint that the universe lasts $\sim 14$ Gyr is taken in account (solid curve).

We expect  the effect of the volume factor on the probability distribution to be much more dramatic in the context of a landscape potential \cite{HHH07}. Indeed it appears likely that in some regions of a landscape potential and in particular around {\qf broad saddle-points of $V$}, the volume factor more than compensates for the reduction in amplitude due to the higher value of the potential. This would lead to the prediction that in a landscape potential, the most probable universe  consistent with our data had a large number of efoldings and began in an unstable de Sitter like state near a broad saddle-point of the potential. {\qf Because the dominant saddle-points are well below the Planck density we furthermore expect that the most probable histories lie entirely in the semi-classical regime \cite{Haw05, HH06}.}

\section{Arrows of Time}
\label{arrows}\label{sec7}
Suppose  that our classical universe bounced at an early time at a radius well above the Planck length.  At no time in its history were {\qf there large} quantum fluctuations in the geometry of spacetime. Could events, structures, and processes before the bounce have influenced events, structures and processes today?  Could we receive information from intelligent aliens living before the bounce encoded in gravitational waves, neutrinos, or boxes made of some durable form of matter not yet discovered by us? 

The overwhelming observational evidence for an early hot period in the universe suggests that most information from before the bounce could not get through {\vf to us}  in any accessible form.  Matter was in thermal equilibrium at least at temperatures high enough to dissociate nuclei and information encoded at lower energy scale phenomena would be wiped out. But gravitational waves whose coupling to matter is the same as that governing the expansion may not participate in this equilibrium.

But even if information could propagate from one side of the bounce to another we have to consider the thermodynamic arrow of time to discuss whether events on one side could influence events on the other. Causation is generally possible only in that direction. That, for instance, is why we remember past events but not future ones. 

In the trenchant analysis of the arrow of time by Hawking, LaFlamme and Lyons \cite{HLL93}  the thermodynamic arrow of time is taken to coincide with the time direction in which fluctuations away from homogeneity and isotropy grow. Small fluctuations grow under the action of gravitational attraction into large inhomogeneities. That is order into disorder.  
Hawking, LaFlamme and Lyons examine the evolution of fluctuations in the extremizing solutions that provide the semiclassical approximation to the NBWF.
They show that regularity conditions at the South Pole imply that the fluctuations in the extremizing solutions are small there and therefore increase away from the South Pole because they have nowhere to go but up.

The Lorentzian histories predicted by the NBWF are not the same as the extremizing solutions, but they are closely connected. In the homogeneous, isotropic case for example the curves of constant real part of the action are Lorentzian trajectories in minisuperspace when the classicality condition \eqref{classcond}  holds. 
We have not yet calculated the fluctuations to these models. But when we do it seems reasonable to suppose that regularity at the South Pole will imply that the fluctuations are small near the bounce and tend to increase away from {\qf it for a significant time}. 

Assuming this result,  the arrow of time in bouncing solutions increases away from the bounce\footnote{This has also been considered  by Carroll and Chen in a different context \cite{CCXX}}. Put differently it {\qf points}  in opposite directions on opposite sides of the bounce. It therefore seems unlikely {\qf on general thermodynamic grounds} that events on the opposite side of the bounce could influence events on this side.  To do so their influence would have to travel backward in time. Unless intelligent aliens find some way to send information backward in time over billions of years we are as unlikely to find any messages from them as we are to find ones {\qf sent by}  intelligent aliens in our own future. 
Can we say then that the other side of a bounce is `real'? It is just as real as  the {\qf pocket}  universes in an eternally inflating spacetime which also can neither communicate with us nor influence us. 

This situation is in sharp contrast with the causality in ekpyrotic cosmologies \cite{Khoury02} and in the `pre-big bang' models discussed in \cite{Veneziano90}, where one  typically starts with an ordered state in the infinite past and as the universe evolves, departures from this state grow in time. Hence in these models the arrow of time always points forward.

However, a subset of the class of histories predicted by the NBWF has a singularity in the past. We have seen (cf. Fig \ref{eta}) that the time asymmetry becomes infinitely large as we approach the regime of initially singular solutions. The NBWF does not tell us whether evolution continues past this singularity, and it has in fact not been shown rigorously whether this is possible at all in any realistic model\footnote{See however \cite{Turok07} for recent work on this.}. But if evolution continues past this singularity, it is conceivable based on Fig \ref{eta} that the arrow of time in these histories will always point in the same direction. This subset of histories may therefore represent the pre-big bang spacetimes predicted by the NBWF in which the arrow of time always points forward and information can propagate from the contracting phase to the expanding regime.

\section{Conclusions}
\label{conc}\label{sec9}

The large scale properties of our specific universe can be summarized in a short list of facts \cite{Rees97}:
Classical physics  applies on coarse-graining scales above the Planck length. The universe is expanding from a hot big bang in which light elements were synthesized. There was a period of inflation, which led to a flat universe today. Structure was seeded by Gaussian irregularities, which are relics of quantum fluctuations. The dominant matter is cold and dark, and there is dark energy which is dynamically dominant at late times. Very roughly this list of features constitutes the standard cosmological model. Quantum cosmology seeks to provide a theory of the quantum state of the universe that would predict connections between these facts. 

The first item on the list --- the wide range of time, place, and scale on which classical physics applies --- is central to all the others. This quasiclassical realm is such a manifest feature of our experience that most treatments of cosmology assume it.  But, classical behavior is not a general feature of quantum systems. Rather, it  emerges only for particular coarse-grainings in a restricted class of states.  That is especially true for the emergence of classical spacetime geometry in a quantum theory of gravity\footnote{Eternal inflation is sometimes said to vitiate the dependence of the present universe on the details of its initial quantum state. But those statements typically assume that spacetime geometry is classical.}. {\qf Any viable theory of the quantum state of our universe must predict classical spacetime over the whole of its visible part from the Planck epoch to the distant future.} Broadly speaking, this paper has mainly focussed on two issues connected with the emergence of a classical cosmological spacetime from the `no-boundary' theory of its quantum state:  (a) What is the ensemble of classical histories predicted by the NBWF and what are their probabilities?  (b) What are the implications of the classicality condition for the standard model of cosmology?  In particular,  what are the important properties of the members of the ensemble of classical histories predicted by the NBWF, and what are the resulting probabilities for what we observe in our past light cone?

We have analyzed these issues in a very simple class of homogeneous, isotropic minisuperspace models with a single scalar field moving in a quadratic potential and a cosmological constant. Our main results are as follows:

{\it Classical Prediction:}  Generalized quantum mechanics is a  clear framework for the prediction {\qf of classical behavior} from the NBWF \cite{Har95c,HHer08}.  {\qf Probabilities are predicted for an ensemble of four-dimensional classical histories of geometry and matter field.}  The complex  `fuzzy  instanton'  metrics that extremize the sum-over-histories defining the NBWF are distinct from the real Lorentizan classical metrics for which they  provide the probabilities. 

{\it The no-boundary measure of the universe:} The probabilities for histories in the NBWF classical ensemble define a measure on classical phase space.  The NBWF measure is concentrated on a surface in phase space which in realistic models has a boundary arising from the classicality condition. It is this concentration to a bounded surface in phase space that gives the NBWF predictive power. More specifically,  for given $\mu$ the NBWF generally singles out {\it at most} a one parameter subfamily from the two parameter family of classical, Lorentzian homogeneous, isotropic solutions.  For $\mu<3/2$ we found classical histories for all ranges of possible matter content. But for the more realistic case of $\mu>3/2$ we found a certain amount of matter ($\phi_0>1.27$) is necessary for classical behavior if there is  any matter at all.  For $\mu>3/2$  a  nearly empty, almost deSitter space is not the most probable Lorentzian history. A  significant amount of matter is required for classical histories.  

{\it Inflation and Classicality:}  All allowed histories that behave classically at late times inflate at early times near the bounce or the initial singularity. For $\mu<3/2$  and small $\phi_0$ the cosmological constant {\qf drives the inflation}. For $\mu>3/2$ the required matter {\qf is the driver}.  The NBWF and classicality imply inflation. {\qf This result illustrates the predictive power of the NBWF. Indeed, using a measure extending over all of phase space motivated by classical dynamics Gibbons and Turok found a negligible probability for inflation \cite{GT06}. }  

{\it  Number of  efoldings: } As Fig \ref{action-smallmu} and Fig \ref{action-largemu}, combined with Fig \ref{efolds}, show,  the NBWF on its own favors a small number of efoldings on a history by history relative probability basis.  However, we can ask the more physical, top-down,  question of what is the probability of inflation in our past conditioned on our limited present data in a Hubble volume. Then the probability for a long period of inflation is enhanced as discussed in Section \ref{sec8}. {\qf Roughly inflation leads to a larger universe with more possible locations for our Hubble volume. Requiring that the universe lasts to the age of 14 Gyr inferred from observation enhances the probability for a long period of inflation further.} 

{\it Bounces and Initial Singularities:}  Some {\qf histories of the NBWF classical ensemble bounce at a minimum radius and some are  initially singular.  The diagram in Fig \ref{phase} shows the range of parameters corresponding to each.}  On a history by history relative basis Fig \ref{action-smallmu} and Fig \ref{action-largemu} show that the NBWF prefers singular beginnings. But we can again ask the more physical, top-down question of what is the most probable origin of the universe given our limited present data in a Hubble volume. Then, as discusssed in \cite{HHH07} and in Section \ref{sec8} here, the most probable origin may be (depending on the model) a bouncing universe in which the universe was always in the semiclassical regime. 

{\it Future-eternal expansion and Final Singularities:} The NBWF predicts probabilities for classical histories and therefore for their long term fate just as much as for  their origins. 
Recollapse to a singularity and future-eternal inflation are the two possible futures for homogeneous models. Figure \ref{phase2} shows the range of parameters $\mu$ and $\phi_0$ that correspond to each. Recollapse is possible only for large $\mu$ (small $\Lambda$) and for universes that have an initial big bang singularity as well as a final singular big crunch.

{\it Singularity Resolution:} Even for classical histories that are singular at early times the NBWF unambiguously predicts probabilities for late time observables such as CMB fluctuations. That is because it predicts probabilities for histories rather than their initial data. The NBWF therefore resolves the big bang singularity, in the sense that it is no longer an obstruction to prediction. 

{\it Time Asymmetry and the Arrow of Time:}  Figure \ref{eta} shows that individual Lorentzian histories are generally  time asymmetric although the ensemble of histories is time-symmetric on general grounds. For large $\phi_0$ this asymmetry is small. {\qf The restriction to homogeneous models does not permit a conclusive discussion of the thermodynamic arrow of time. However, one possibility is that it points away from a bounce on either side.} Causality in this set of histories would be  very different from causality in ekpyrotic and pre-big bang  cosmologies, where the arrow of time always points in the same direction.

There is much to be done to extend these models to more realistic ones and to back up various theoretical assumptions that have been made. {\qf Two extensions are of particular importance:} First, relaxing the restriction to homogeneous and isotropic models would allow consideration of the evolution of quantum fluctuations whose effects could be detectable in the CMB. Further, bubble nucleation, {\qf and the arrow of time}  could be discussed.  Second, as we suggested in \cite{HHH07} in more realistic landscape potentials the classicality condition can act as a vacuum selection principle resulting in top-down probabilities that favor a bouncing universe {\vf that had} a long period inflation and was always in the semi-classical regime. 

On a technical level, going beyond the lowest semiclassical approximation that has been used here could yield more satisfactory probabilities and facilitate comparison with other measures such as the classical one developed in \cite{GT06}. A study of physically realistic coarse-grainings of spacetime geometry and the decoherence of sets of alternative histories defined by them would help back up a number of assumptions that we have made. 

These opportunities for extension, however, should not obscure the fact that our results in the simple models of this paper already demonstrate that the NBWF and classicality condition can play a central role in understanding what we observe of our quantum universe.

\begin{acknowledgments}

We are grateful for discussions or correspondence with Tim Clunan, David Coule, Gary Gibbons, Gary Horowitz, Don Marolf,  Don Page, and  Neil Turok.  Our research was carried out in part at several different places where the authors were able to meet. We are grateful for the hospitality of   the Mitchell family at their Cook's Branch Conservancy and to Chris Pope for hospitality at the Mitchell Institute at Texas A\&M University.  We thank Marc Henneaux  and the International Solvay Institutes for support.  TH thanks the KITP in Santa Barbara for support.   JH and TH thank Stephen Hawking and the Centre for Theoretical Cosmology at DAMTP for support at Cambridge University. This work was supported in part by the National Science Foundation under grants  PHY05-55669.

\end{acknowledgments}

\appendix
\section{Perturbation Theory}

When the scalar field is small  it is a perturbation on the model with only a cosmological constant. This is not a physically interesting case since we do not live in a nearly empty de Sitter space. But, it  is a case where the entire discussion can be carried out essentially analytically so as to provide a guide for the detailed numerical calculations for the physically interesting non-perturbative situations.

This appendix discusses the first two orders of perturbation theory. To avoid a d\'ebauche d'indices   we will use $a$, $\hat a$  for the leading order Euclidean and Lorentzian scale factors driven only by a cosmological constant, and $\phi$, $\hat\phi$  for the perturbations in the scalar field. 

\subsection{ No-Boundary Semiclassical  Solutions}
We first calculate the complex solutions to the equations \eqref{euceqns}  that provide the semiclassical approximation to the no-boundary wave function at given real values of $(b,\chi)$. These are solutions which are regular at the 
South Pole and match the given values at the other endpoint. 

{\it No Scalar Field:}  
When there is no scalar field our only concern is the geometry. 
There is only one solution of \eqref{eucconstraint}  that is regular at the South Pole  $\tau=0$ and that is 
\begin{equation}
a(\tau) = \sin(\tau) = \sin(x +i y).
\label{pert1}
\end{equation}
As discussed in Section \ref{complexgauge} , finding a regular solution for a given value of $b$  means finding a contour in the $\tau$-plane connecting the origin to a point $\tau=\upsilon\equiv X +iY$ where $a(\upsilon)$ is real and equal to $b$. 
The scale factor $a(\tau)=\sin(\tau)$ is real along the curves $y=0$ and along $x=\pm\pi/2, x=\pm 3\pi/2, \cdots$.
If $b<1$ there is a solution with $\upsilon$ on the real axis. If $b>1$ there are two candidate solutions corresponding to complex conjugate values of $\upsilon$ along the constant $x$ curves where $a(x+iy)$ is real with $Y=\pm \cosh^{-1}(b)$. We will argue in a moment that only $X=\pi/2$ corresponds to  a solution on the no-boundary manifold. 

If $b<1$ the contour between $\tau=0$ and $\tau=\upsilon$ can be chosen to lie on the real axis. Then the metric is real, Euclidean and corresponds to part of the Euclidean 4-sphere. For $b>1$ consider the solution with $X=\pi/2$ and $Y=\cosh^{-1}(b)$. The connecting contour can be taken to run along the real  axis to $X=\pi/2$ and then up the $y$-axis to $Y$. This corresponds to the geometry of half a unit radius round Euclidean four-sphere joined smoothly across a surface of vanishing extrinsic curvature to half of a Lorentzian de Sitter space starting at the bounce. This is the well known no-boundary instanton \cite{HH83} nucleating de Sitter space, and we will call this the NBI contour.

Proceeding along a contour of real $\tau$ to $X=3\pi/2$ and then upwards to $Y$  gives the same geometry as with $X=\pi/2$ but with an additional Euclidean four-sphere attached at the South Pole. This geometry is not strictly regular on the no-boundary manifold. We therefore exclude it and all candidate solutions with larger values of $X$. Solutions with $X=-\pi/2$ are the same as those with $X=+\pi/2$. 

The solution with $X=\pi/2$ and  $Y=-\cosh^{-1}(b)$ has the opposite sign of the imaginary part of the  action from the one with positive $Y$. We therefore count it as an independent semiclassical solution. The two solutions  make complex conjugate contributions to the wave function ensuring that it is real as discussed in Section \ref{timesym}. These two solutions dominate the semiclassical approximation to the no-boundary wave function when there is no scalar field.
 \begin{figure}[t]
\includegraphics[width=3.2in]{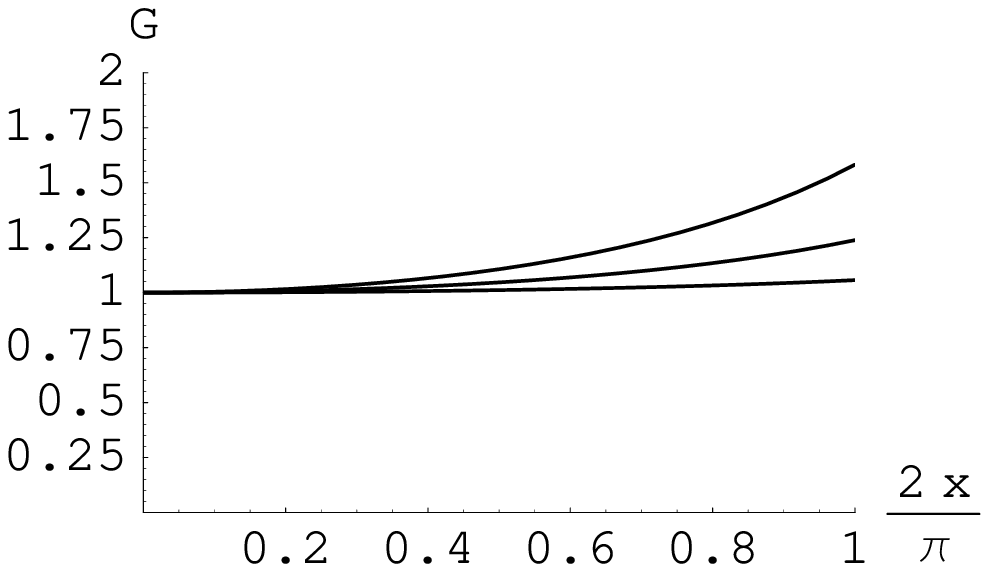} \hfill
\includegraphics[width=3.2in]{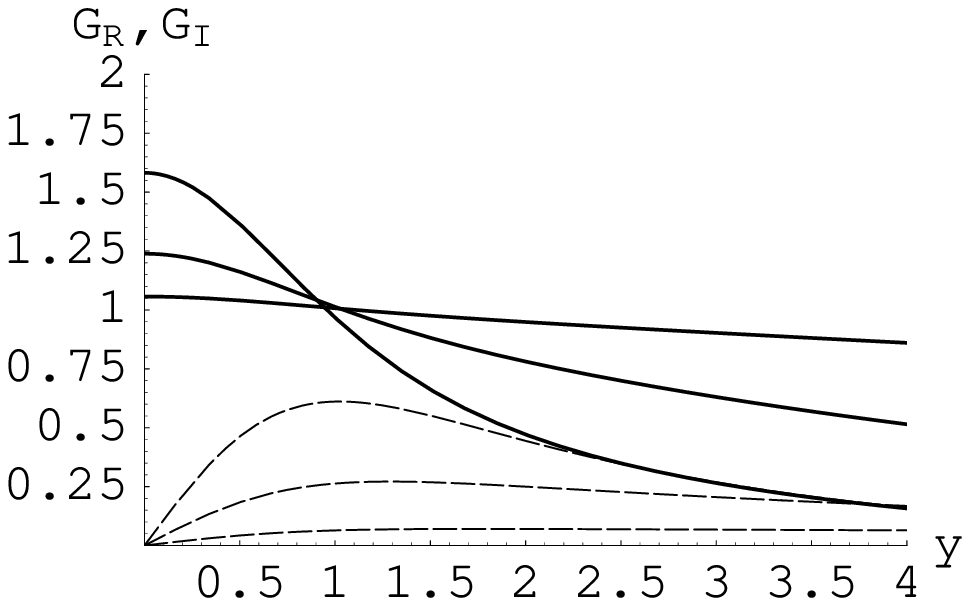}
\caption{The complex extremizing solution $G$. $G(\tau,\mb)$ is the complex solution of the linearized equation for the scalar field which is regular at the origin and equal to $1$ there. All other field extrema are multiples of $G$. Three plots are shown for $\mb$ equalling $.25$, $.50$, and $.75$. Each uses the NBI  contour in the $\tau$-plane that extends horizontally from the origin along the real axis  to $x = \pi/2$ and then vertically  in the imaginary ($y$) direction. 
Along this particular contour the unperturbed metric makes a smooth transition between a Euclidean instanton and a Lorenzian deSitter metric. $G$ is real along the real part of the contour but complex along the imaginary component. The real parts of $G$ are indicated by solid lines, the imaginary parts by dashed lines. The curves are in fact continuous along the contour with appropriate matching conditions for the derivatives reflecting the change in direction of the contour at $x=\pi/2, y=0$. (cf Figure \ref{ex}.)}
\label{Gless}
\end{figure}

{\it Perturbing Scalar Field:}
The first order scalar field perturbations to the empty extremizing solutions for the scale factor found above satisfy \eqref{eucphieqn}  with $a(\tau)$ given by \eqref{pert1}.  The boundary conditions are that $\phi(\tau)$ be regular at the South Pole $\tau=0$ (meaning that $\dot \phi$ vanishes there) and match the given $\chi$ at the boundary. 

\begin{figure}[t]
\includegraphics[width=3.2in]{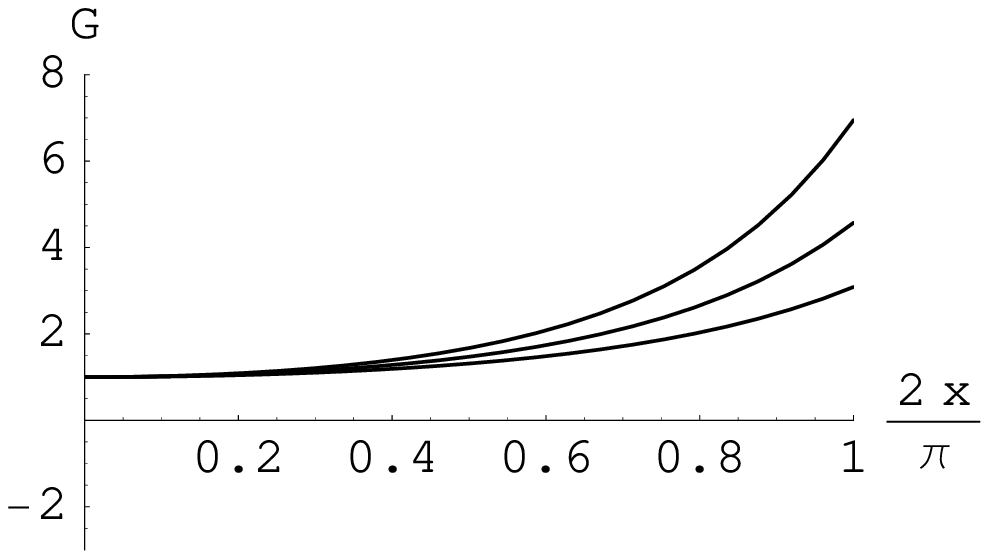} \hfill
\includegraphics[width=3.2in]{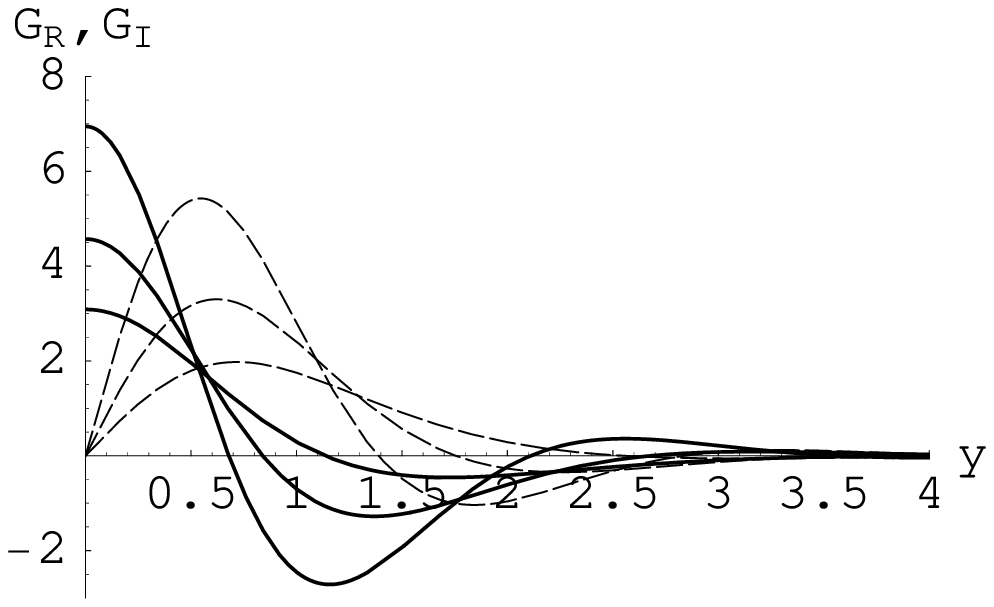}
\caption{The complex  extremizing solution $G(\tau,\mb)$ for $\mb$ equaling $1.25$, $1.50$, and $1.75$. The figure is otherwise the same as Figure \ref{Gless}.}
\label{Ggtr}
\end{figure}

We denote by $G(\tau)$ the (regular) solution to \eqref{eucphieqn}   with $G(0)=1$ and $\dot G(0)=0$. This is
\begin{equation}
G(\tau) = F(a,b,2, (1-\cos(\tau))/2) \ , 
\label{pert2}
\end{equation}
where $F(a, b, c, z)$ is the hypergeometric function, $\mb \equiv 2\mu/3$, and 
\begin{align}
a&\equiv (3/2)(1+\sqrt{1-{\bar\mu}^2}) , \nonumber \\ 
 b&\equiv (3/2)(1-\sqrt{1-{\bar\mu}^2}) \ . 
 \label{pert3}
 \end{align}
(Evidently, $G$ depends on $\bar\mu^2$ as well as $\tau$ but we will not usually indicate this explicitly.) The function $G(\tau)$ is real on the real axis for $x<\pi$  and therefore real analytic, specifically $G^*(\tau)\equiv [G(\tau^*)]^* = G(\tau)$. The function $G(\tau)$ is multi-valued and there is therefore a cut  which can be taken to extend along the real axis from $x=\pi$ to infinity. We will assume that we are considering perturbations of the no-boundary instanton solution defined by the  NBI contour discussed above and that this lies on the first sheet of $G(\tau)$.
 The behavior of $G(\tau)$ for several values of $\mb$ is illustrated in Fgures \ref{Gless} and \ref{Ggtr}. 

The general regular solution will be a complex number times $G(\tau)$. This number can be found as follows:
 The value of $b$ determines a point in the complex plane from the zeroth order calculation above. For $b<1$ this is on the real axis at $(X=\sin^{-1}(b)<\pi/2, Y=0)$. Thus the required solution is 
\begin{subequations}
\label{pertphi}
\begin{equation}
\phi(\tau; b, \chi) = \chi~  G(\tau)/G(\sin^{-1}(b)),   \quad  (b<1) \   .
\label{pert5}
\end{equation}
This is not an especially interesting case from the point of semiclassical prediction since the imaginary part of the action, $S$, will vanish. 

The interesting case $b>1$ is similar. For definiteness, focus on  the specific extremizing solution whose zeroth order approximation is labeled by the point $(X=\pi/2, Y=+\cosh^{-1}(b))$.  The regular solution for the scalar field matching $\chi$ at that value is
 \begin{equation}
\phi(\tau; b, \chi) = \chi~ G(\tau) / G(\pi/2 +i\cosh^{-1}(b)),   \quad (b>1)\  .
\label{pert6}
\end{equation}
\end{subequations}
This solution is unique once the contour defining the unperturbed solution is fixed.

For the non-pertubative case discussed in Sections \ref{sec5} and \ref{sec6}  analytic solutions are not available. Typically we start at the origin with a value for $\phi(0)$,  integrate along some contour in the $\tau$-plane, and adjust the endpoint of integration and the complex value of $\phi(0)$ to reach given real values of $b$ and $\chi$.  This connection between $\phi(0;b, \chi)$ and $(b,\chi)$ is given in perturbation theory by Eqs \eqref{pertphi}, e.g for $b>1$, 
\begin{equation}
\phi(0; b, \chi) \equiv  \phi_0(b,\chi) \exp[i\theta(b,\chi)] =  \chi  / G(\pi/2 +i\cosh^{-1}(b)).
\label{pert4}
\end{equation}
Since $\chi$ is real we have 
\begin{equation}
\theta(b,\chi)= -{\rm Arg}[G(\pi/2 +i\cosh^{-1}(b))]
\label{pert7} 
\end{equation} 
where $\rm Arg$ is the complex phase, and 
\begin{equation}
\phi_0(b,\chi) = \chi / |G(\pi/2 +i\cosh^{-1}(b))| \  .
\label{pert8}
\end{equation}
\begin{figure}[t]
\includegraphics[width=3.2in]{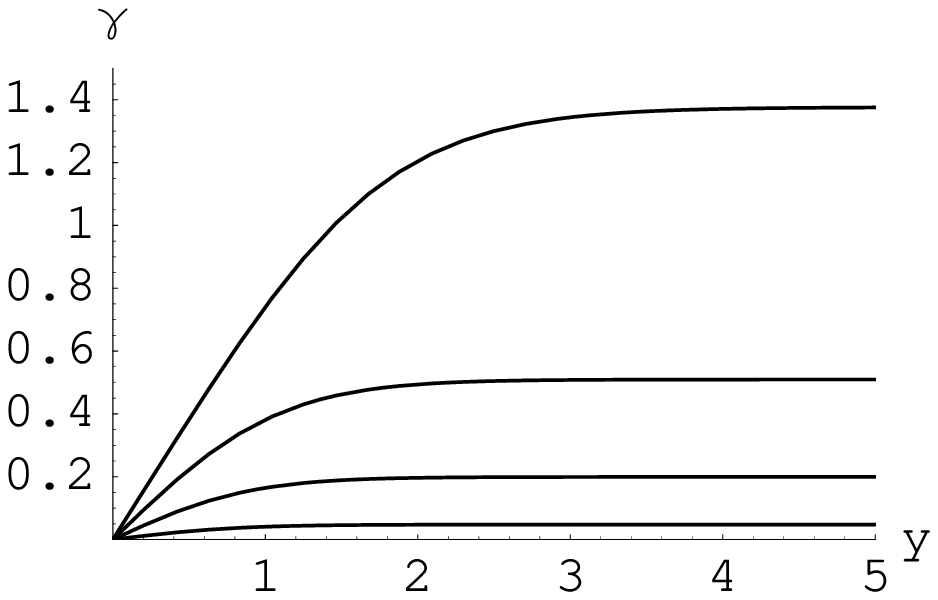} \hfill
\includegraphics[width=3.2in]{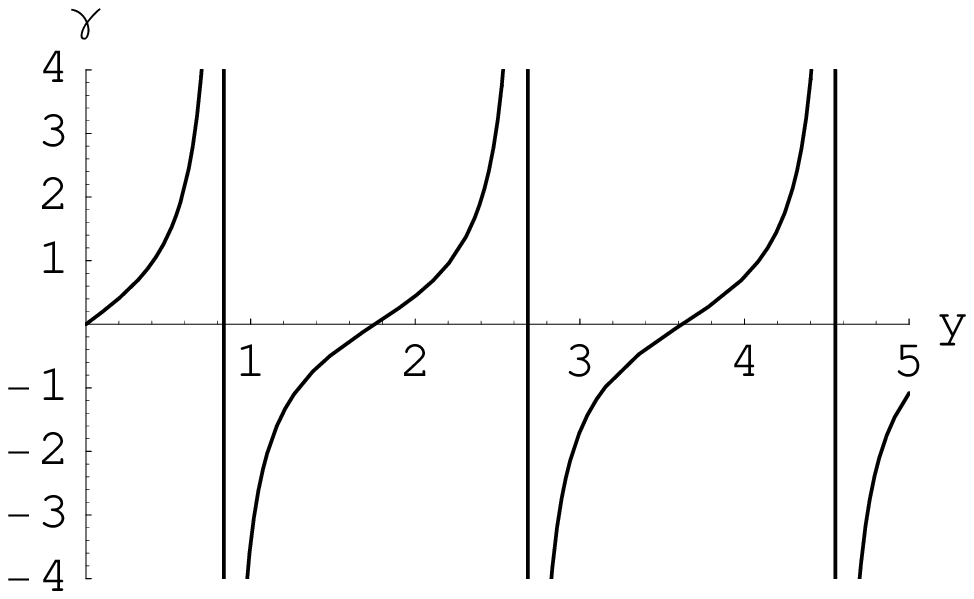}
\caption{The parameter $\gamma \equiv \Im(\phi_0)/\Re(\phi_0)$ plotted along the vertical contour $x=\pi/2,y$. This is the  ratio required to have $\phi$ real along this contour. On the left the curves  reading from bottom to top for $\mb$ equal to $.2$, $.4$, $.6$ and $.8$. The ratio becomes constant at large $y$. On the right the single value $\mb=1.5$ is plotted showing the generic lack of stabilization at large $y$.  (cf Figure \ref{angle}.) }
\label{gam}
\end{figure}
Figure \ref{gam} shows the perturbative values of $\gamma \equiv \tan(\theta)$ for various values of $\mb$.
\begin{figure}[t]
\includegraphics[width=3.2in]{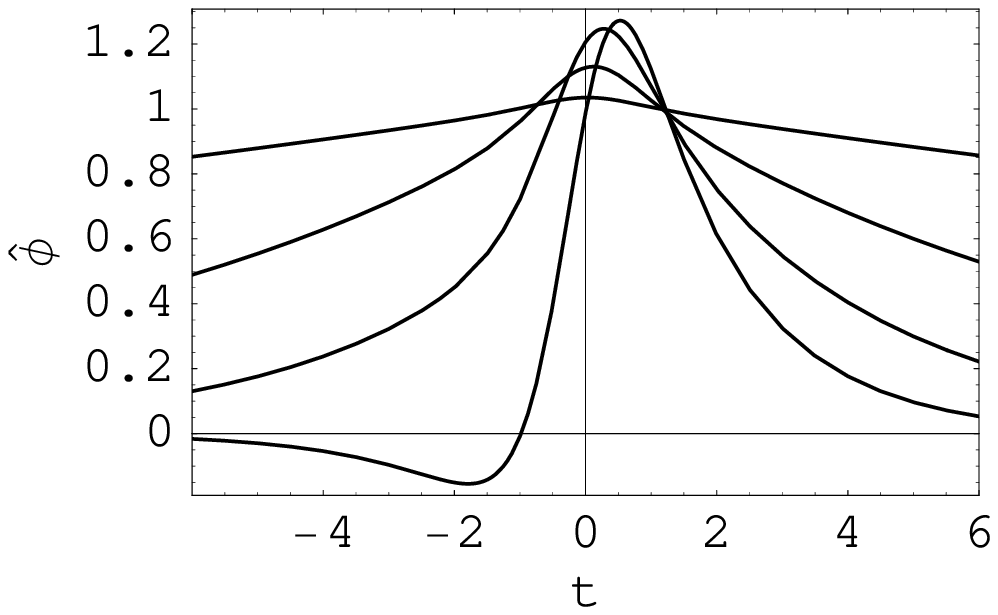} \hfill
\includegraphics[width=3.2in]{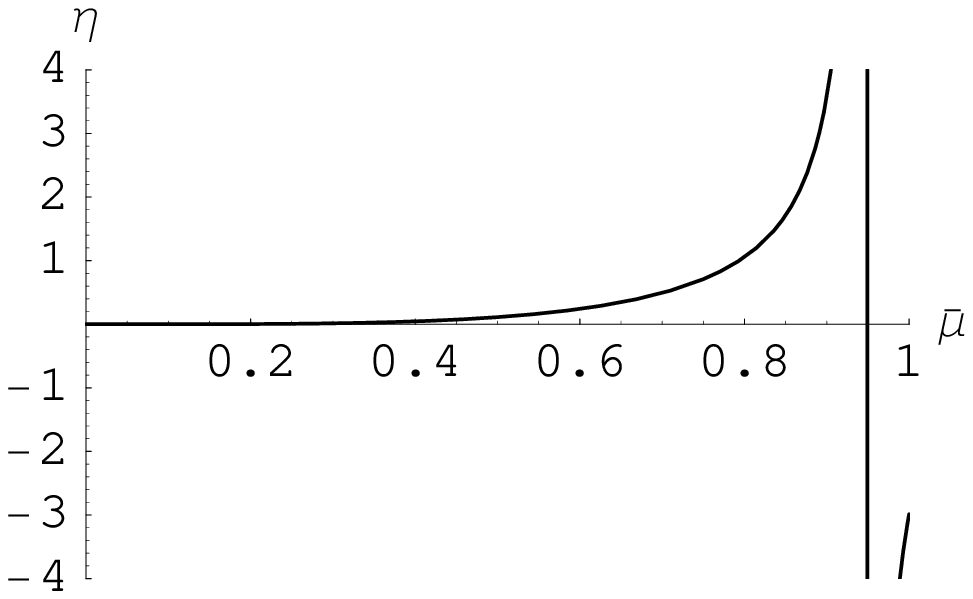}
\caption{The first figure shows Lorentzian histories of the scalar field for various values of $\mb$. The value of $\ch/\phi_0$  is plotted vertically. Starting from the top on the left and moving downward the values of $\mb$ are $.2$, $.4$, $.6$  and $.8$. The solutions are not generally time-symmetric although the ensemble of Lorentzian solutions is time-symmetric. The second figure shows the time-asymmetry parameter $\eta$ \eqref{etadef} as a function of $\mb$. (cf. Figure \ref{eta}.)}
\label{pt_lor}
\end{figure}

\subsection{The Classical Ensemble}
We next turn to the perturbative construction of the ensemble of classical Lorentzian histories predicted by the no-boundary wave function and the evaluation of their probabilities. The prescription for this is described in Section \ref{classens} and is straightforward to implement explicitly in perturbation theory. For each extremizing solution, we chose a {\vf matching} surface $b=b_*$ in minisuperspace and at each point along it labeled by $\chi=\chi_*$ we evaluate initial data for the classical Lorentzian solutions $(\bh(t), \ch(t))$ that are the integral curves of $S\equiv -\Im(I)$. 

With this data we  then integrate the Lorentzian equations to find the classical solutions labeled by $(b_*,\chi_*)$. Later we find the probabilities for these Lorentzian histories. The complete ensemble of classical predictions is the union of those from all the extremizing solutions that contribute to the semiclassical approximation to the no-boundary wave function. 

Choosing a different value of $b_*$  to implement this procedure simply means that the same  Lorentzian solution will be labeled by a different value $\chi_*$ of  the $\chi$ at which it intersects the new surface in minisuperspace. 

{\it Zeroth Order --- No scalar Field:}  
The Lorentzian equation for the scale factor $\bh(t)$ is
\begin{equation}
\frac{d\bh}{dt}= \sqrt{\bh^2-1}.
\label{lorb}
\end{equation}
There is only one solution to this equation which is the scale factor for empty de Sitter space. Choosing the origin of $t$  to be the time of the bounce this is
\begin{equation} 
\bh(t)= \cosh(t) \ . 
\label{lor0}
\end{equation} 
It is easy to verify that this solution satisfies the zeroth order versions of the Cauchy data  \eqref{match}   determined by the complex solution \eqref{pert1}, namely
\begin{equation}
\bh(t_*) = b_* , \quad \left. d\bh/dt\right |_{t_*} = -\left. \Im(\dot a)\right|_{\upsilon_*}.  
\label{data0}
\end{equation}
where  $\upsilon_*= \pi/2 +iY_*$, $t_*=Y_*$ for one extremizing solution and  $\upsilon_*= \pi/2 -iY_*$, $t_*=-Y_*$ for the other. 

{\it First Order in the Scalar Field:}
The Lorentzian equation for the scalar field is \eqref{lorphieqn},
\begin{equation}
\frac{1}{\bh^3}\frac{d}{dt}\left( \bh^3 \frac{d\ch}{dt} \right) + \mb^2 \ch =0
\label{lorchi}
\end{equation}
where $\bh(t)$ is the scale factor \eqref{lor0} determined in zeroth order. For each contributing extremizing solution the Cauchy data along the $b=b_*$ surface specified by \eqref{matchb}  is 
\begin{equation}
\ch(t_*)=\chi_* , \quad \left. d\ch/dt\right|_{t_*} = -\left. \Im(\dot\phi)\right|_{\upsilon_*},
\label{data1}
\end{equation}

Let's first consider the extremizing solution defined by  $\upsilon_*= \pi/2 +iY_*$, $t_*=Y_*$.
It is straightforward to see that the following is the Lorentizian solution with the boundary conditions \eqref{data1},
\begin{equation}
\ch(t) = \chi_* ~\Re[ G(\pi/2 +it)/G(\pi/2+it_*)]
\label{lor1}
\end{equation}
provided $t_*$ is identified with $Y_*$. Using \eqref{pert6} the Lorentzian solutions can also be parametrized by the value of $\phi_0(\infty)$ as described in the main text for the non-perturbative case.  Figure \ref{pt_lor}  shows a few examples of $\ch(t)/\phi_0(\infty)$.

Another set of extremizing solutions is defined by  $\upsilon_*= \pi/2 -iY_*$. This subensemble of Lorntzian histories is just the time $(t)$ reversed of the one defined by  $\upsilon_*= \pi/2 +iY_*$. The whole ensemble of classical Lorentzian solutions is therefore time-symmetric. 
 The individual solututions are generally not as Figure \ref{pt_lor} shows.

\subsection{Perturbative Action}

The action function $I(b,\chi)$ determines both when classical Lorentzian histories are predicted by the NBWF through \eqref{classcond}, and, if so, what their probabilities are through $\exp(-2I_R)$. In perturbation theory the action can be expanded in powers of $\chi$, viz.
\begin{equation}
I(b,\chi) = I^{(0)}(b) + I^{(2)}(b,\chi) + \cdots  
\label{expnact}
\end{equation}
where the second term is proportional to  $\chi^2$. The terms $I^{(0)}(b)$ and $ I^{(2)}(b,\chi)$ are determined by evaluating the action integral \eqref{eucact_tau} to quadratic order in $\chi$ using the perturbative complex extremizing solutions found in the first part of this Appendix. The integral in that expression is carried out along a contour from $\tau=0$ to an endpoint  $\tau=\upsilon=X+iY$ corresponding to the given value of $(b,\chi)$. In the following we carry out this perturbative evaluation focussing exclusively on the range $b>1$ needed for classical prediction. 

{\it Zeroth Order --- No Scalar Field:} The extremizing solution is given by \eqref{pert1}. The endpoint is $\upsilon=\pi/2 + i\cosh^{-1}(b)$. The result of carrying out the integral is
\begin{equation}
I^{(0)}(b)= -\frac{\pi}{2 H^2}[1-i (b^2-1)^{3/2}] \ . 
\label{zeroact}
\end{equation}
 
{\it Second Order in the Scalar Field:}
The action integral \eqref{eucact_tau} depends on $\phi(\tau)$, $a(\tau)$ and $\upsilon$. There are perturbations in all three. Only the linear order solution for $\phi(\tau)$ is needed to evaluate the field contribution to the action to quadratic order in $\chi$. 
 It turns out that the second order perturbations  of the scale factor and  of the endpoint cancel essentially as a consequence of reparametrization invariance. 
 Integrating the $\phi$ part of the action by parts and using the equation of motion for $\phi$ then gives the following simple expression for $I^{(2)}$:
\begin{equation}
I^{(2)}(b,\chi) = \frac{3\pi}{4 H^2}  b^3 \phi(\upsilon) \dot\phi(\upsilon) \ .
\label{secondact}
\end{equation}
Explicitly, using \eqref{pert6}, this is
\begin{equation}
I^{(2)}(b,\chi) = \frac{3\pi}{4 H^2}\chi^2 b^3 \frac{\dot G[\pi/2 +i \cosh^{-1}(b)]}{G[\pi/2 +i \cosh^{-1}(b)]} \equiv  \frac{3\pi}{4 H^2}\chi^2 b^3 F(b)\  .
\label{secondact_exp}
\end{equation}
It is straightforward although laborious to check that this perturbation in the action satisfies the  Hamilton-Jacobi equation \eqref{eucHJ}  expanded to second order in $\chi$ as a consequence of the equation of motion \eqref{eucphieqn}. 
\begin{figure}[t]
\includegraphics[width=3.2in]{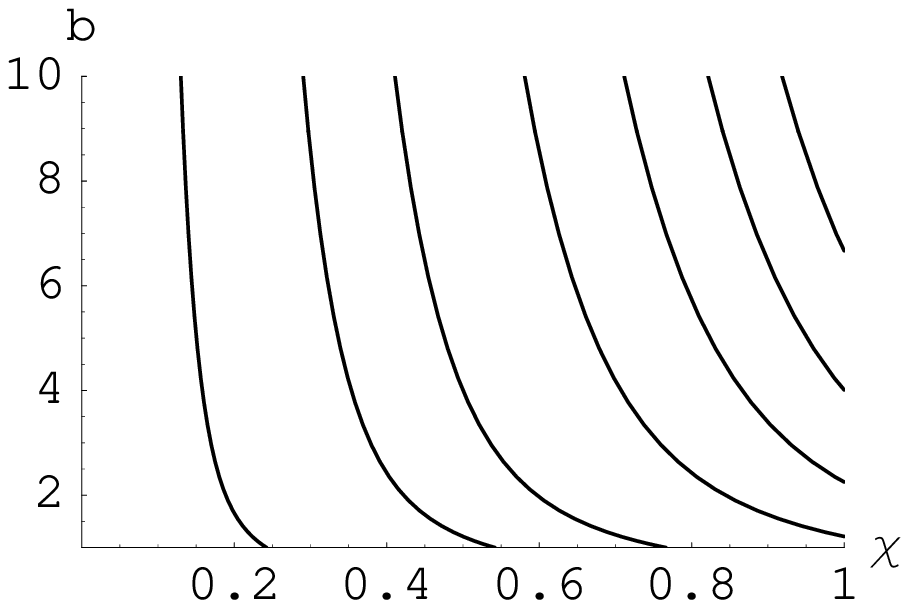} \hfill
\includegraphics[width=3.2in]{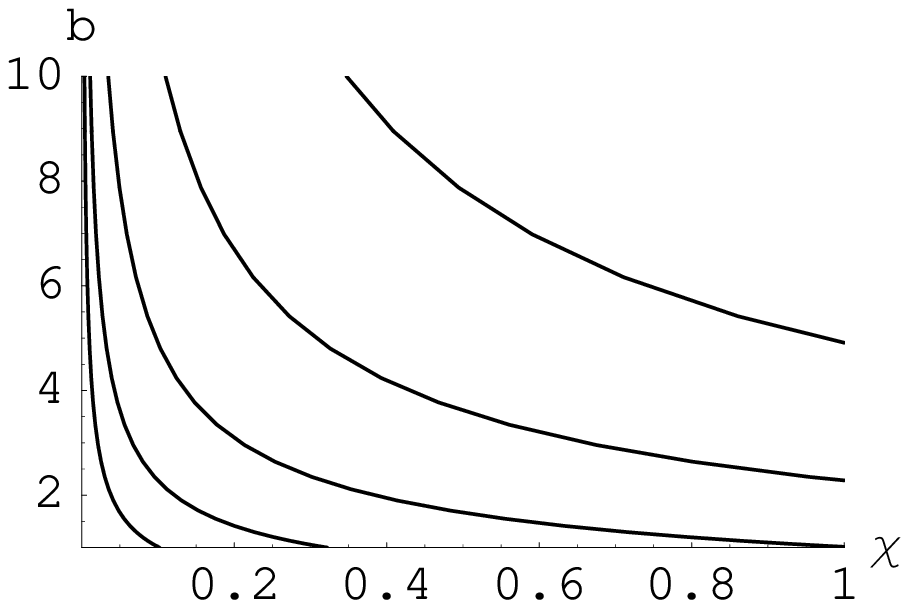}
\caption{Curves of constant real part of the action in minisuperspace. When the conditions for classicality are satisfied these will be classical Lorentzian solutions for large  $b$. That will be the case for the curves on the left with $\mb<1$, but not those on the right for which $\mb>1$.  Since the real part of the zeroth order action contributes an overall additive constant $-\pi/(2 H^2)$  it is  convenient to label the constant action curves by the value of  $(2H^2/3\pi)I^{(2)}$,  [cf \ref{secondact_exp}].  Reading from left to right the values of  $(2 H^2/3\pi)I^{(2)}$ shown are  .01, .05,  .1, .2, .3, .4, .5 and .6 for the $\mb=.5$ on the left. For $\mb=1.5$ on the right the values are .01, .1, 1, 10, 100. Note that the $b$-axis starts from $1$ which is the value of $b$ at the bounce. }
\label{constIR}
\end{figure}
\begin{figure}[t]
\includegraphics[width=3.2in]{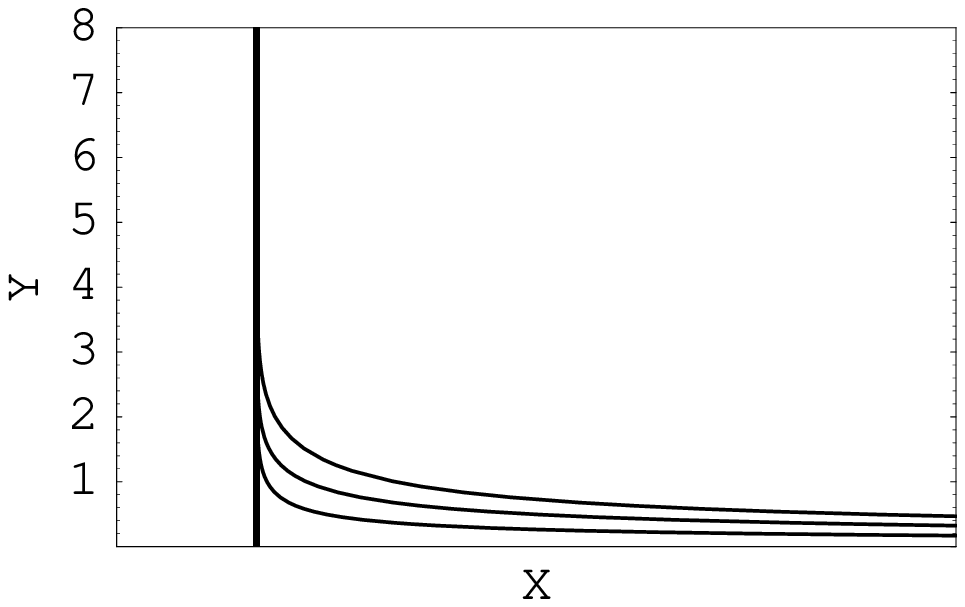} \hfill
\includegraphics[width=3.2in]{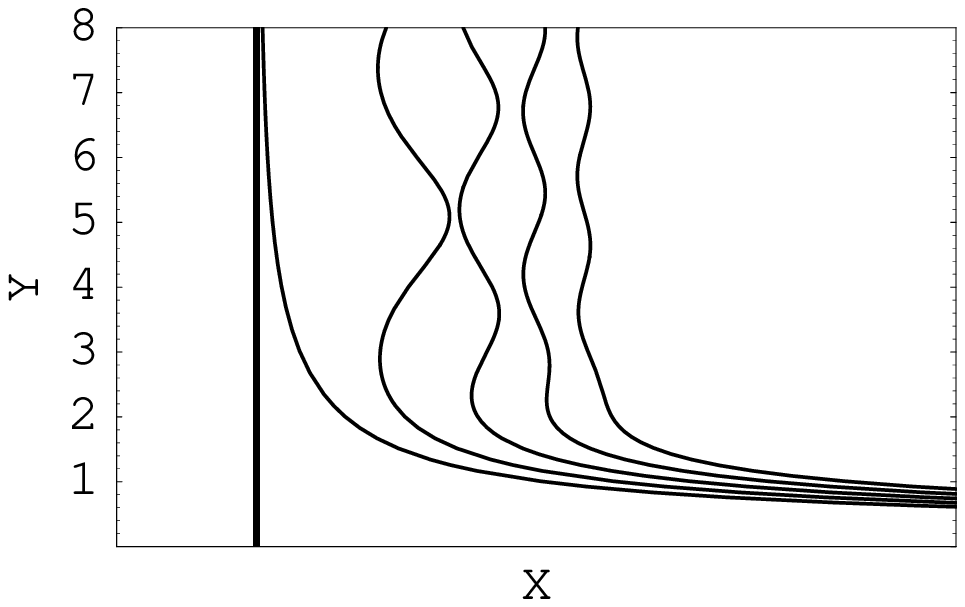}
\caption{Curves of constant real part of the action in the complex $\tau$-plane. The endpoint values $(X,Y)$ provide a set of coordinates for minisuperspace that are alternatives to $(b,\chi)$.  The unperturbed curve where $b$ is real is the vertical curve at $X=\pi/2$ along which $2 H^2I_R/3\pi=-1/3$.  Perturbations in that curve along which the perturbing matter action is constant are shown here for several values of $\mb$. On the left are curves where $\mb$ has the values $.25$, $.50$, and $.75$ reading up from the lowest to highest. On the right curves corresponding to $1.0$ to $1.4$ in steps of $.1$ reading left to right. The perturbative calculation of these curves is only valid when they remain close to the vertical line at $X=\pi/2$. For that reason no scale is indicated for the $X$-axis.   For $\mb<1$ the curves of constant $I_R$ approach classical solutions. For $\mb>1$ the oscillation at large $y$ is an indication that they do not approach classical solutions.}
\label{constIRXY}
\end{figure}

Figure \ref{constIR} and Figure \ref{constIRXY} show curves of constant real part of the action plotted in $(b,\chi)$ and $(X,Y)$ coordinates on minisuperspace for typical values of $\bar\mu$ less and greater than $1$. In regions of superspace where the classicality condition \eqref{classcond}  is satisfied these curves are integral curves of the imaginary part of the action $-S(b,\chi)$ to a good approximation and would represent the predicted classical Lorentzian histories to that approximation. In each case the curves display the decay of the scalar field with the expansion of the universe that is a property of Lorentzian solutions.
\begin{figure}[t!]
\includegraphics[width=3.2in]{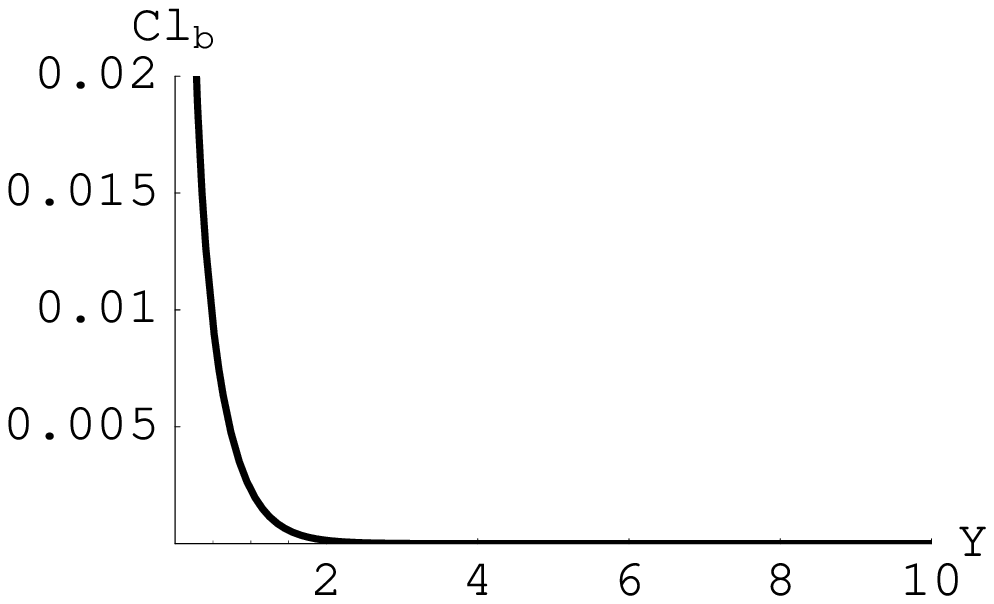} \hfill
\includegraphics[width=3.2in]{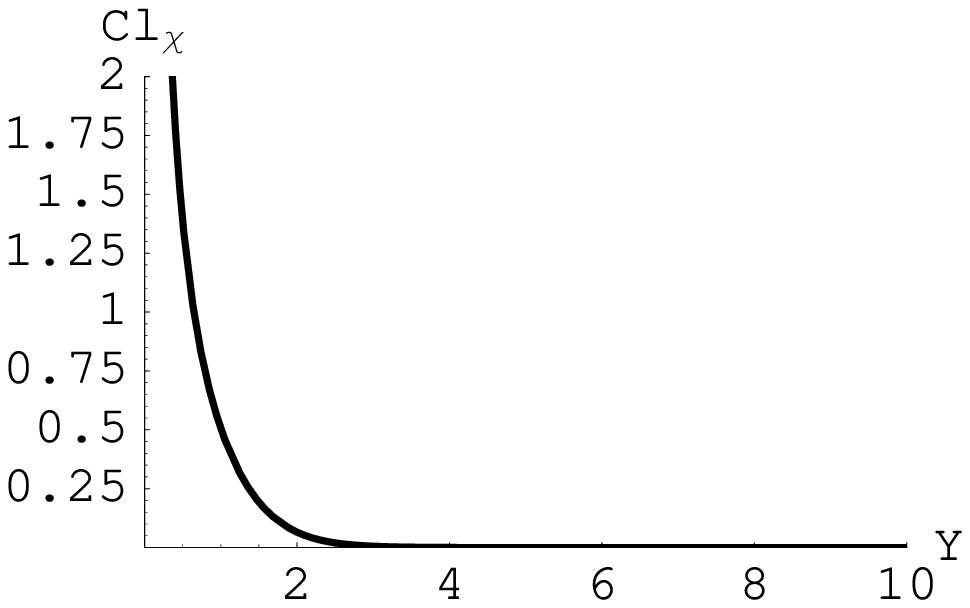} \\
\includegraphics[width=3.2in]{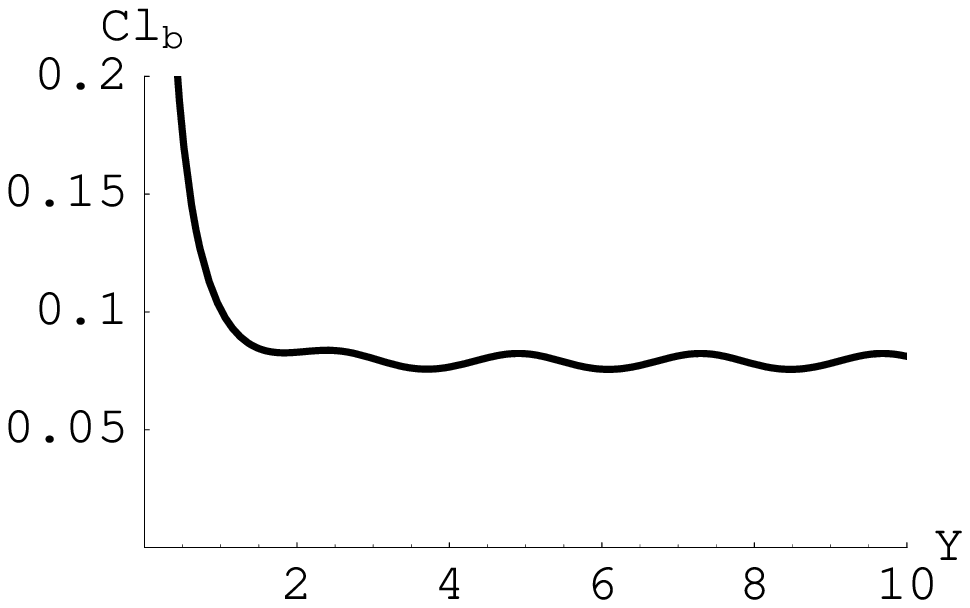} \hfill
\includegraphics[width=3.2in]{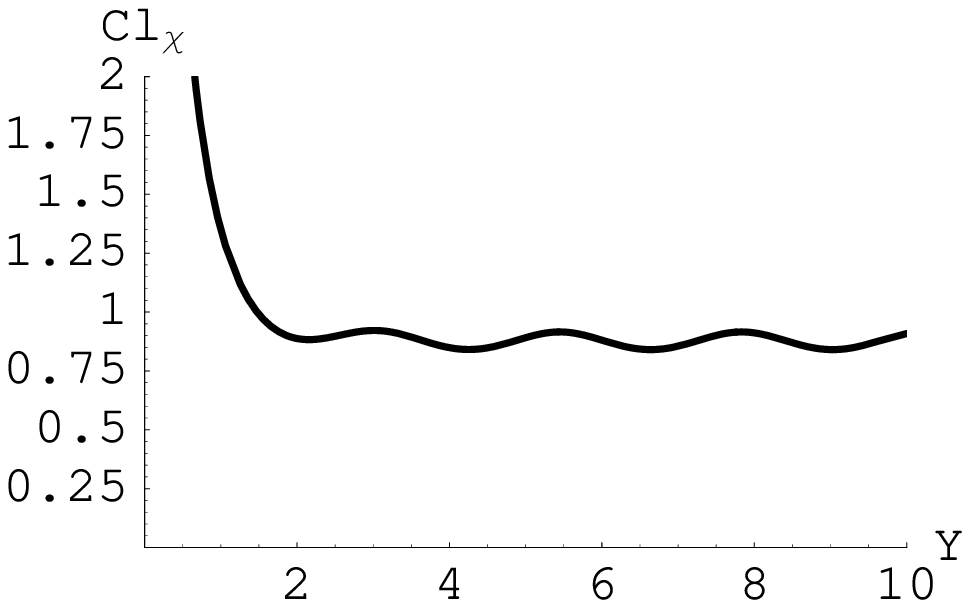} 
\caption{The classicality ratios $Cl_b$ and $Cl_\chi$ plotted for $\mu=.75$ (top pair) and $\mu=3$ (bottom pair) in minisuperspace $(b,\chi)$ where $Y\equiv \cosh^{-1}(b)$.  In lowest order perturbation theory $Cl_b$ is proportional to $\chi^2$, and the value $\chi=.1$ was used for these illustrations. $Cl_\chi$ is independent of $\chi$ in leading order. The classicality condition is well satisfied for $Y \gtrsim 2$ when $\mu=.75$. For $\mu=3$ it is not satisfied at all because $Cl_\chi$ never drops much below unity. There are no classical histories for $\mu>3/2$.}
\label{pt-cc}
\end{figure}

\subsection{Perturbative Classicality and Perturbative Probabilities}
As discussed in Section \ref{classicality}, classical Lorentzian histories are predicted when there is a region of minisuperspace where the gradients of the real part of the action $I_R(b,\chi)$ are all small compared with those of minus the imaginary part $S(b,\chi)$. A convenient measure of this classicality condition \eqref{classcond} is the  {\it classicality ratio} [cf \eqref{classcond}]
\begin{equation}
Cl_A(b,\chi) \equiv |\nabla_A I_R(b,\chi)|/|\nabla_A S(b,\chi)| . 
\label{classratio}
\end{equation}
When both these ratios are small  the classicality condition \eqref{classcond} is satisfied.  In lowest non-vanishing perturbation theory order we have
\begin{subequations}
\label{pt-classcond}
\begin{equation}
Cl_b = \frac{|\nabla_b I^{(2)}|}{|\nabla_b S^{(0)}|}\approx \frac{1}{2}\chi^2 \frac{b}{(b^2-1)^{1/2}} \left[3F_R + b \frac{dF_R}{db}\right] 
\label{classcondb}
\end{equation}
\begin{equation}
Cl_\chi = \frac{|\nabla_\chi  I^{(2)}|}{|\nabla_\chi S^{(2)}|}= \frac{F_R}{F_I} . 
\label{classcondchi}
\end{equation}
\end{subequations}
The ratio $Cl_b$ will be small for small $\chi$, but the ratio $Cl_\chi$ is independent of $\chi$ for small $\chi$. 

\begin{figure}[t]
\includegraphics[width=3.2in]{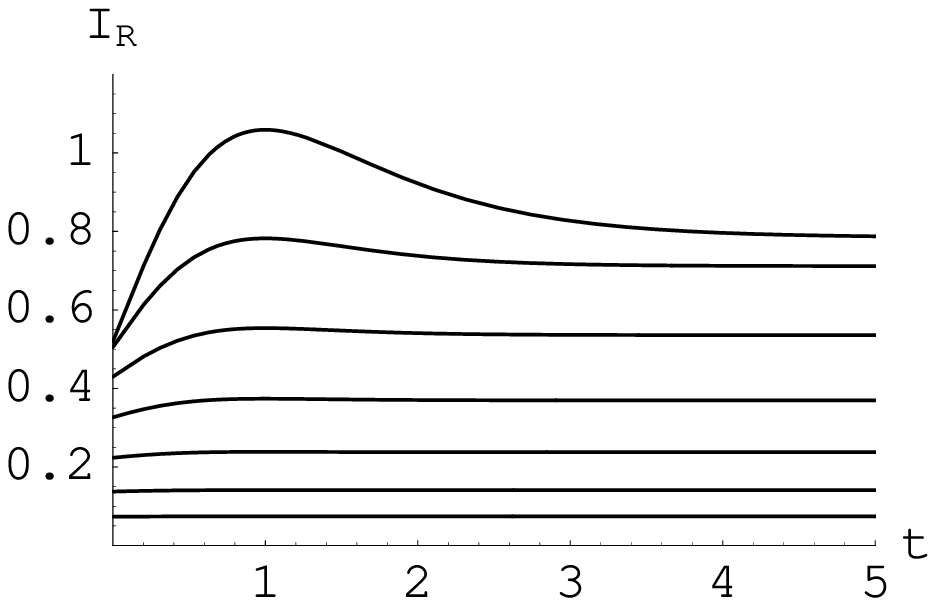} \hfill
\includegraphics[width=3.2in]{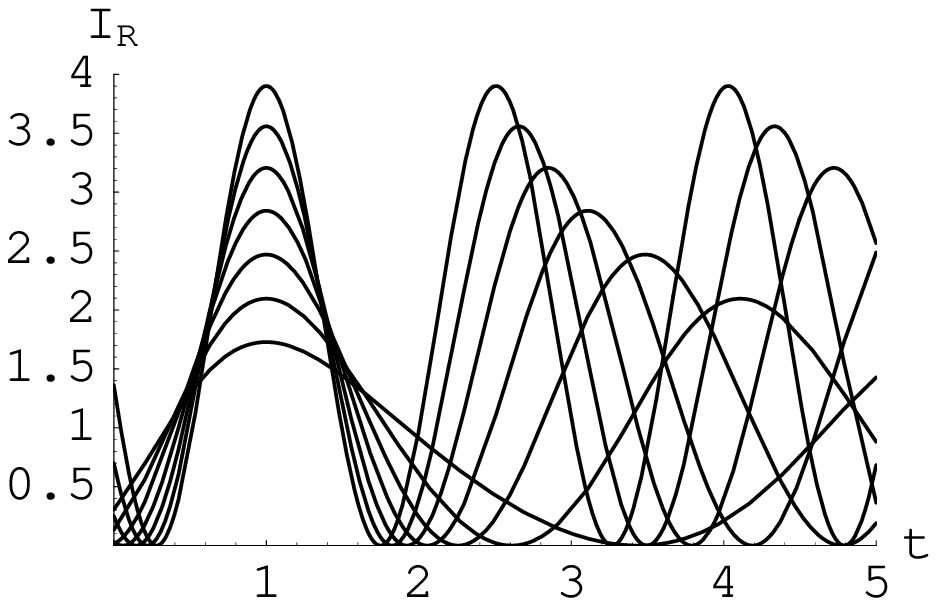}
\caption{The real part of the action along the classical Lorentzian solution with $\phi_0=1$ for several values of $\mb$. The curves on the left reading bottom top range from $\mb=.3$ to $\mb=.9$ in steps of $.1$. On the right the range from bottom to top is $\mb=1.1$ to $\mb=1.7$ in steps of $.1$. The marked qualitative difference between the $\mb<1$ and $\mb>1$ is important for classical predictions. On the  left the real part approaches a constant at large $t$. Its gradient in this direction is small compared to the gradient of $S$. The curves approach the integral curves for classical, Lorentzian solutions with probabilites proportional to $\exp(-2I_R)$. For $\mb>1$ the real part of the action does not approach a constant, its gradient remains comparable to the gradient of $S$,  and, as a consequence, classical behavior is not predicted for the scalar field.}
\label{IRalongLor} 
\end{figure}

Figure \ref{pt-cc} shows these ratios for two values of $\mu$ --- one below $3/2$ and one above.  Its evident that the classicality condition is not satisfied for $Cl_\chi$ for the larger value and this is true for all values $\mu>3/2$. By contrast, the condition is satisfied for all $\mu<3/2$. 

Figure \ref{IRalongLor} shows the situation with respect to classicality in the $\mu<3/2$ and $\mu>3/2$ regimes in a different way. The real part of the action must become  constant along any curve in superspace that is a predicted  classical history. Were $I_R$ not constant along a Lorentzian histories, then $\exp(-2I_R)$ could not be its probability. There is {\it one} probability for each history. 
\begin{figure}[t!]
\includegraphics[width=3.2in]{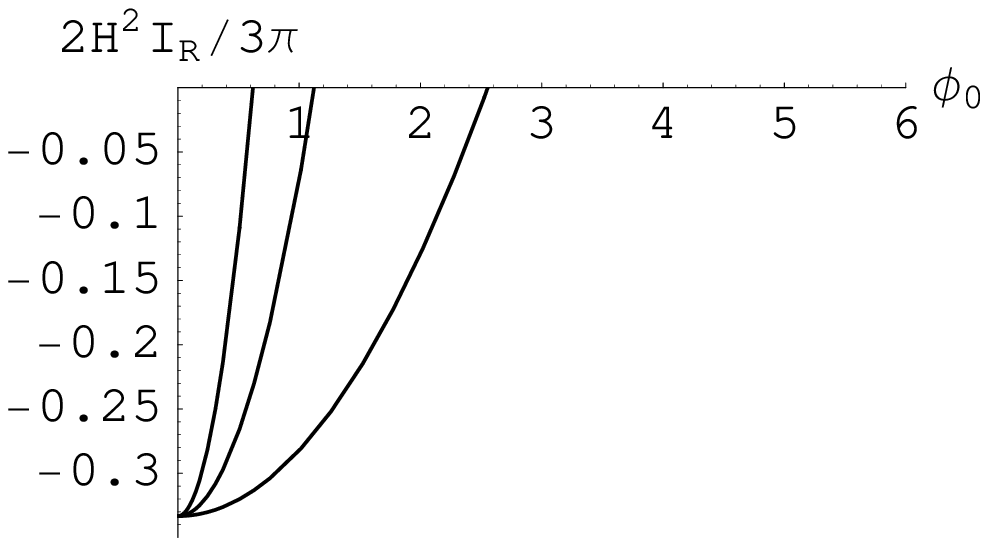} 
\caption{The real part of the perturbative Euclidean action $I_R$  \eqref{expnact} is plotted  for three values of $\mb<1$ where classical behavior is predicted.  The relative probabilities for classical Lorentzian histories labeled by different values of $\phi_0$  are  $\exp(-2I_R)$.  Reading from left to right the three values of $\mb$ are $.75$, $.50$ and $.25$. }
\label{pt-probs}
\end{figure}
In the approximations that we are using in this paper, the relative probabilities of classical  Lorentzian histories are given by $\exp(-2I_R)$  when the classicality conditions are satisfied. For small values of the scalar field this  is when $\bar\mu<1$. 
The results  for the real part of the action are shown in Fig. \ref{pt-probs} for a range of values of $\bar\mu<1$. Comparison with Figure \ref{action-smallmu}  shows that the perturbation theory is a reasonable approximation for small values of $\phi_0$. 



\end{document}